%% file: cuckooXORSAT_ICALP_submit.tex
\let\accentvec\vec 
\documentclass{llncs}
\let\vec\accentvec %

\makeatletter
    \providecommand*{\toclevel@title}{0}  
    \providecommand*{\toclevel@author}{0} 
\makeatother

\def\conf{1}    
\def\longf{0}   
\def\pageNum{1} 

\usepackage[letterpaper]{geometry}
\usepackage{times}
\ifnum\pageNum=1
\fi

\usepackage[utf8]{inputenc}
\usepackage[T1]{fontenc}

\usepackage{amsmath}
\usepackage{amssymb}
\usepackage{amsfonts}

\usepackage{graphicx}
\usepackage{subfigure}
\usepackage{xcolor}

\usepackage{url}
\usepackage{hyperref}
\definecolor{darkblue}{rgb}{0,0,.5}
\hypersetup{colorlinks=true, breaklinks=true, linkcolor=darkblue, menucolor=darkblue, urlcolor=darkblue, citecolor=darkblue}

\usepackage[ruled,vlined,linesnumberedhidden]{algorithm2e}
\usepackage[neveradjust]{paralist}

\newcommand{\figPath}{figures}
\newcommand{\imgScale}{0.58}
\newcommand{\labIncr}{\large}
\newcommand{\inFigSpace}{\vspace{-0.4cm}}
\newcommand{\outFigSpace}{\vspace{-0.4cm}}

\newcommand{\nodes}{\ensuremath{m}}
\newcommand{\edges}{\ensuremath{n}}
\newcommand{\id}[1]{\mathcal{D}(#1)}
\newcommand{\ud}[1]{\mathcal{U}(#1)}

\renewcommand{\qed}{\hfill$\square$}
\renewenvironment{proof}[1][]{\par{\noindent \bf Proof #1: }}{\qed}
\spnewtheorem{fact}{Fact}{\bfseries}{\itshape}

\newcommand{\low}{\ensuremath{l}}
\newcommand{\high}{\ensuremath{k}}
\newcommand{\keys}{\ensuremath{n}}
\newcommand{\randv}{\ensuremath{k}}
\newcommand{\mean}{\ensuremath{\kappa}}
\newcommand{\cells}{\ensuremath{m}}
\newcommand{\ve}{\ensuremath{\varepsilon}}
\newcommand{\pmf}{\ensuremath{{\rho}}}

\newcommand{\Po}{\mathrm{Po}}
\newcommand{\Bin}{\mathrm{Bin}}

\newcommand{\reals}{\mathbb{R}}
\newcommand{\integers}{\mathbb{Z}}

\newcommand{\domain}{\mathcal{D}_{m,n}} 
\newcommand{\domainCompl}{\overline{\mathcal{D}}_{m,n}} 
\newcommand{\domainGam}{\mathcal{D}_{\gamma}} 

\newcommand{\graphSet}{\mathcal{G}^k_{m,n}}
\newcommand{\cGraphSet}{\mathcal{H}^k_{m,n}}
\newcommand{\cGraphSetH}{\mathcal{H}^k_{\hat{m},\hat{n}}}

\DeclareMathOperator{\E}{E}
\DeclareMathOperator{\rank}{rank}
\DeclareMathOperator{\coeff}{coeff}
\DeclareMathOperator{\atanh}{atanh}
\DeclareMathOperator{\Hess}{Hess}

\newcommand{\blank}{\text{ }}

\begin{document}

\title{Tight Thresholds for Cuckoo Hashing via XORSAT}

\author{Martin Dietzfelbinger\thanks{Fakult\"at f\"ur Informatik und Automatisierung, Technische Universit\"at Ilmenau. 
Research supported by DFG grant DI 412/10-1. 
{\tt \{martin.dietzfelbinger,michael.rink\}@tu-ilmenau.de}
} 
\and 
Andreas Goerdt\thanks{Fakult\"at f\"ur Informatik, Technische Universit\"at Chemnitz.
{\tt goerdt@informatik.tu-chemnitz.de}
} 
\and 
Michael Mitzenmacher\thanks{Harvard University, School of Engineering and Applied Sciences.  
Part of this work was done while visiting Microsoft Research New England. {\tt michaelm@eecs.harvard.edu}
} \and 
\\
Andrea Montanari\thanks{Department of Electrical Engineering and Department of Statistics, Stanford University.
Part of this work was done while visiting Microsoft Research New England.  {\tt montanar@stanford.edu}} 
\and 
Rasmus Pagh\thanks{Efficient Computation group, IT University of Copenhagen. {\tt pagh@itu.dk}
} 
\and 
Michael Rink${}^\star$}
\tocauthor{M. Dietzfelbinger, A. Goerdt, M. Mitzenmacher, A. Montanari, R. Pagh, M. Rink}

\date{}
\institute{}

\maketitle

\ifnum\pageNum=1
\thispagestyle{plain}
\fi

\ifnum\conf=1
\vspace{-0.27 in}
\fi

\begin{abstract}
We settle the question of tight thresholds for offline cuckoo hashing.
The problem can be stated as follows: we have $n$ keys to be hashed
into $m$ buckets each capable of holding a single key.
Each key has $k \geq 3 $ (distinct) associated buckets
chosen uniformly at random and independently of the choices
of other keys.  A hash table can be constructed successfully if each
key can be placed into one of its buckets.  We seek thresholds $c_k$
such that, as $n$ goes to infinity, if $n/m \leq c$ for some $ c < c_k$ then a hash
table can be constructed successfully with high probability, and if
$n/m \geq c$ for some $c > c_k$ a hash table cannot be constructed successfully with
high probability.  Here we are considering the offline version of the
problem, where all keys and hash values are given, so the problem is
equivalent to previous models of multiple-choice hashing. We find the
thresholds for all values of $k > 2$ by showing that they are in
fact the same as the previously known thresholds for the random
$k$-XORSAT problem.  We then extend these results to the setting where
keys can have differing number of choices, and provide evidence in
the form of an algorithm for a conjecture extending this result to
cuckoo hash tables that store multiple keys in a bucket.
\end{abstract}

\ifnum\conf=1
\vspace{-0.28in}
\fi

\section{Introduction}

Consider a hashing scheme with $n$ keys to be hashed into $m$ buckets each
capable of holding a single key.  Each key has $k \geq 3$ (distinct)
associated buckets chosen uniformly at random and independently of the
choices of other keys.  A hash table can be constructed successfully
if each key can be placed into one of its buckets.  This setting
describes the offline load balancing problem corresponding to multiple
choice hashing~\cite{ABKU} and cuckoo hashing~\cite{FPSS:2005,cuckoo1}
with $k \geq 3$ choices.  An open question in the literature (see, for
example, the discussion in~\cite{MMsurvey}) is to determine a tight
threshold $c_k$ such that 
if $n/m \leq c$ for some $ c < c_k$ then a hash
table can be constructed successfully with high probability, and if
$n/m \geq c$ for some $c > c_k$ a hash table cannot be constructed successfully with
high probability. In this paper, we provide these thresholds.

We note that, in parallel with this work, two other papers have
similarly provided means for determining the 
thresholds~\cite{Fountoulakis:2009,Frieze:2009}.  
Our work differs from these works in substantial ways.  Perhaps the most
substantial is our argument that, somewhat surprisingly, the
thresholds we seek were actually essentially already known.  We show that tight
thresholds follow from known results in the literature, and in fact
correspond exactly to the known thresholds for the random $k$-XORSAT
problem.  We describe the $k$-XORSAT problem and the means for computing
its thresholds 
in more detail in the following sections.  Our argument
is somewhat indirect, although all of the arguments appear to rely
intrinsically on the analysis of corresponding random hypergraphs, and
hence the alternative arguments of~\cite{Fountoulakis:2009,Frieze:2009} provide additional insight
that may prove useful in further explorations.

With this starting point, we extend our study of the cuckoo hashing
problem in two ways.  First, we consider {\em irregular cuckoo
hashing}, where the number of choices corresponding to a key is not
a fixed constant $k$ but itself a random variable depending on the
key.  Our motivations for studying this variant include past work on
irregular low-density parity-check codes~\cite{ldpc} and recent work
on alternative hashing schemes that have been said to behave like
cuckoo hashing with ``3.5 choices''~\cite{cuckoo35}.  Beyond finding
thresholds, we show how to optimize irregular cuckoo hashing schemes
with a specified average number of choices per key; for example, with
an average of 3.5 choices per key, the optimal scheme is the natural
one where half of the keys obtain 3 choices, and the other half
obtain 4.  Second, we consider the generalization to the setting where
a bucket can hold 
more than one key.  
We provide a conjecture regarding
the appropriate threshold behavior for this setting, and provide a
simple algorithm that, experimentally, appears to perform remarkably
close to the thresholds predicted by our conjecture.

\ifnum\longf=1
\subsection{Paper overview}
\fi

Section~\ref{sect:facts:cores} presents an exposition of known results
on cores of random hypergraphs. Readers familiar with this material
may want to skip directly to Section~\ref{sec:main}, which provides our
proof that the thresholds for $k$-XORSAT and $k$-ary cuckoo hashing
are identical. In Section~\ref{sec:non-integer} we extend the
discussion of thresholds to the case where $k$ is any real number
greater than 2. Finally, Section~\ref{subsec:selfless:algorithm}
presents our simple algorithm to construct hash tables and presents
experimental evidence that it is able to achieve load factors close to
the thresholds.  
\ifnum\conf=1
Further details appear in the appendices.
\fi

\section{Technical background on cores}\label{sect:facts:cores}

The key to our analysis will be the behavior of 
cores in random hypergraphs.  
We therefore begin by providing a review of this
subject.  To be clear, the results of this section are not new; the
reader is encouraged to see~\cite[Ch.~18]{MezMontBook:2009}, as well
as references~\cite{Coo:2004,DubMan:2002,Molloy:RSA:2005} for more background.

We consider the set of all $k$-uniform hypergraphs with $m$ nodes and $n$ hyperedges $\graphSet$. 
More precisely, each hypergraph $G$ from $\graphSet$ consists of $n$ (labeled) hyperedges 
of a fixed size $k\ge2$, chosen independently at random, with repetition, 
from the $\binom{m}{k}$ subsets of $\{1,\ldots,m\}$ of size $k$.
This model will be regarded as a probability space.
We always assume $k$ is fixed, $m$ is sufficiently large, and 
$n=c m$ for a constant $c$.

For $\ell\ge2$, the $\ell${\emph{-core}} of a hypergraph $G$ is
defined as the largest induced sub-hypergraph that has minimum degree
$\ell$ or larger.  It is well known that the $\ell$-core can be
obtained by the following iterative ``peeling process'': While there
are nodes with degree smaller than $\ell$, delete them and their
incident hyperedges.  By pursuing this process backwards one sees that
the $\ell$-core, conditioned on the number of nodes and hyperedges it contains, is a
uniform random hypergraph that satisfies the degree constraint.

The fate of a fixed node $a$ after a fixed number of $h$ iterations of
the peeling procedure is determined by the $h$-neighborhood of $a$,
where the $h$-neighborhood of $a$ is the sub-hypergraph induced on
the nodes at distance at most $h$ from $a$.  For example, the
$1$-neighborhood contains all hyperedges containing $a$.  In our
setting where $n$ is linear in $m$ the $h$-neighborhood of 
node $a$ is a hypertree of low degree (at most $\log \log m$) with
high probability. We assume this in the discussion to come.

We can see whether a node $a$ is removed from the hypergraph in the
course of $h$ iterations of the peeling process in the following way.
Consider the hypertree rooted from $a$ (so the children are nodes that
share a hyperedge with $a$, and similarly the children of a node share
a hyperedge with that node down the tree).  First, consider the nodes
at distance $h-1$ from $a$ and delete them if they have at most $\ell-2$ child
hyperedges; that is,  their degree is at most $\ell-1.$  Second, treat the nodes at
distance $h-2 $ in the same way, and so on, down to distance $1$, the
children of $a$. Finally, $a$ is deleted if its degree is at most $\ell-1.$

The analysis of such random processes on trees has been well-studied
in the literature.  (See, for example,
\cite{BroderFriezeUpfal,AndOrTrees} for similar analyses.)  We wish to
determine the probability $q_h$ that node $a$ is deleted after $h$
rounds of the peeling process.  For $j < h$ let $p_j$ be the
probability that a node at distance $h-j$ from $a$ is deleted after
$j$ rounds of the peeling process.  The discussion becomes easier for
the binomial random hypergraph with an expected number of $cm$ hyperedges:
Each hyperedge is present with probability $k! \cdot c /m^{k-1}$ independently. It is
well known that $\graphSet$ and the binomial hypergraph are
equivalent as far as asymptotic behavior of cores are concerned when $c$ is a constant.

Let $\Bin(N,p)$ denote a random variable with a binomial distribution, 
and $\Po(\beta)$ a random variable with a Poisson distribution. 
Below we make use of the Poisson approximation of the binomial distribution and
the fact that the number of child hyperedges of a node in the hypertree asymptotically 
follows  the binomial distribution. 
This results in additive terms that tend to zero as $m$ goes to infinity. We have $p_0 = 0$,
\begin{eqnarray*} 
p_1 & =& \Pr\bigg[\Bin\left( \binom{m-1}{k-1} \,\,,\,\,k! 
\cdot \frac{c}{m^{k-1}} \right) \le \ell-2 \bigg]\,  \\ & & \\ 
  & = & \Pr[\Po(k c ) \le \ell-2] \pm o(1),     \\  \mbox{ } & \\
p_{j+1} &= & \Pr\bigg[\Bin\left(\binom{m-1}{k-1} \,\,
, \,\, k! \cdot \frac{c}{m^{k-1}} \cdot (1- p_j)^{k-1}\right)\, \le \ell-2\bigg]\, \\ & & \\
&=&
\Pr[ \Po(kc(1-p_j)^{k-1}) \le \ell-2]  \pm o(1), \mbox{ for } j=1,\dots,h-2.
\end{eqnarray*} 
The probability $q_h$ that $a$ itself is deleted 
is given by the following different formula:
\begin{equation}  \label{del}
q_h = \Pr[\Po(kc(1-p_{h-1})^{k-1} )\le  \ell-1] \pm o(1). 
\end{equation} 

The $p_j$ are monotonically increasing and $0 \le p_j \le 1$, so $p=
\lim p_j$ is well-defined.  The probability that $a$ is deleted
approaches $p$ from below as $h$ grows. %
Continuity of the functions involved implies that
$p$ is the smallest non-negative solution of
\begin{eqnarray*}
p & = & \Pr[ \Po(kc(1-p)^{k-1}) \le \ell-2].
\end{eqnarray*}
Observe that $1$ is always a solution. 
Equivalently, applying the monotone function 
$t\mapsto kc(1-t)^{k-1}$ 
to both sides of the equation, $p$ is the smallest solution of 
\begin{equation}  \label{eq:def:p}
kc(1-p)^{k-1} = kc \left( 1 - \Pr
[ \Po(kc(1-p)^{k-1}) \le \ell-2] \right)^{k-1}.
\end{equation}

Let $\beta = kc (1-p)^{k-1}$.  It is helpful to think of $\beta$ with the following
interpretation:

Given a node in the hypertree, the number of child hyperedges
(before deletion)
follows the distribution $\Po(kc)$.
Asymptotically, a given child  hyperedge is not deleted
with probability $(1-p)^{k-1}$, independently for all children.
Hence the number of child hyperedges after deletion
follows the distribution $\Po(kc(1-p)^{k-1}).$  And $\beta$ is the key parameter
for the node giving the expected number of hyperedges
containing it
that could contribute to keeping it in the core.

Note that~(\ref{eq:def:p}) 
is equivalent to 
$$c = \frac{1}{k}\cdot\frac{\beta}{\left (\Pr[\Po(\beta ) \,\ge \ell-1]\right)^{k-1}}.$$
\noindent
This motivates considering the function
\begin{equation} 
g_{k,\ell}(\beta)
=\frac{1}{k}\cdot\frac{\beta}{(\Pr[\Po(\beta) \ge \ell-1])^{k-1}},
\label{eq:definition:g}
\end{equation} 
which has the following properties in the range $(0,\infty)$: It tends to infinity
for $\beta\to0$, as well as for $\beta\to\infty$. Since it is convex there is exactly one global minimum. Let
$\beta^*_{k,\ell}=\arg\min_{\beta}g_{k,\ell}(\beta)$ and $c^*_{k,\ell}=\min g_{k,\ell}(\beta)$. 
For $\beta > \beta^*_{k,\ell}$ the function $g_{k,\ell}$ is monotonically increasing. 
For each $c>c^*_{k,\ell}$ let $\beta(c) \,= \, \beta_{k,\ell}(c)$ denote the unique $\beta>\beta^*_{k,\ell}$ 
such that $g_{k,\ell}(\beta)=c$. 

Coming back to the fate of $a$ under the peeling process, Equation  (\ref{del}) shows that
$a$ is deleted with probability approaching $\Pr[\Po(\beta(c))\le \ell-1].$ This probability is smaller than $1$
if and only if $c > c^*_{k, \ell}$, 
which implies that the expected number of nodes that are {\it not} deleted    
is linear in $n$.  As the  $h$-neighborhoods of two
nodes $a$ and $b$ are disjoint with high probability,
by making use of the second moment we can show that in this case 
a linear number of nodes survive with high probability. 
(The sophisticated reader would use Azuma's inequality to obtain concentration bounds.)

Following this line of reasoning, we obtain the following results,
the full proof of which is in~\cite{Molloy:RSA:2005}. 
(See also the related argument of \cite[Ch.~18]{MezMontBook:2009}.)
Note the restriction to the case $k + \ell > 4$, which means that the result does not
apply to $2$-cores in standard graphs; 
since the analysis of standard cuckoo hashing is simple, using direct arguments, 
this case is ignored in the analysis henceforth.
\begin{proposition}\label{prop:one}
Let $k+\ell >4$ and $G$ be a random hypergraph from $\graphSet$. Then $c^*_{k,\ell}$ is the threshold for the \emph{appearance} of an $\ell$-core in $G$.
That is, for constant $c$ and $m\to \infty$, 
\begin{itemize}
	\item[{\rm(a)}]
	   if  $n/m = c < c^*_{k,\ell}$, then
	   $G$ has an empty $\ell$-core with probability $1-o(1)$.
	\item[{\rm(b)}]
	   if  $n/m = c > c^*_{k,\ell}$, then
	   $G$  has an $\ell$-core  of linear size with probability $1-o(1)$.
\end{itemize}
\end{proposition}

In the following we assume  $c > c^*_{k,\ell}$. Therefore
$\beta(c)>\beta^*_{k,\ell}$ exists.
Let $\hat m$ be the number of nodes in the  $\ell$-core and 
$\hat n$ be the number of hyperedges in the $\ell$-core. 
We will find it useful in what follows to consider the {\em edge density} of the 
$\ell$-core, which is simply the ratio of the number of hyperedges to the number of nodes.

\ifnum\longf=1
\begin{proposition}\label{prop:two}
Let  $c>c^*_{k,\ell}$ and  $n/m = c\, (1\pm o(1))$.  Then with high probability
in $\graphSet$

$$\hat m = \Pr[\Po({\beta(c))}\ge \ell]\cdot m \pm o(m)$$ 
and 
$$\hat n = (\Pr[\Po({\beta(c))}\ge \ell-1])^k
\cdot n \pm o(m).$$
\end{proposition}
\fi

\ifnum\conf=1
\begin{proposition}\label{prop:two}
Let  $c>c^*_{k,\ell}$ and  $n/m = c\, (1\pm o(1))$.  Then with high probability
in $\graphSet$

$$\hat m = \Pr[\Po({\beta(c))}\ge \ell]\cdot m \pm o(m)
\mbox{  and  }
\hat n = (\Pr[\Po({\beta(c))}\ge \ell-1])^k
\cdot n \pm o(m).$$
\end{proposition}
\fi

The bound for $\hat m$ follows from the concentration of the expected
number of nodes 
surviving when we plug in the limit $p$ for $p_h$ in equation~(\ref{del}).  
The result for $\hat n$ follows similar lines:
Consider a fixed hyperedge $e$ that we assume is present
in the random hypergraph. For each node of this hyperedge
we consider its $h$-neighborhood modified in that $e$ itself 
does not belong to this  $h$-neighborhood. 
We have $k$ disjoint trees with high probability.
Therefore each of the $k$ nodes of $e$ survives 
$h$ iterations of the peeling procedure independently with probability
$\Pr[\Po(\beta(c)) \ge \ell-1]$. Note that we use
$\ell-1$ here (instead of $\ell$) because  the nodes belong to  $e.$ 
Then $e$ itself survives with $ (\Pr[\Po(\beta(c)) \ge \ell-1])^k.$
Concentration of the number of surviving 
hyperedges again follows from second moment calculations
or Azuma's inequality.

With this we have the information needed regarding the edge density
of the $\ell$-core.
\begin{proposition}\label{prop:three}
If $c>c^*_{k,\ell}$ and  $n/m = c\, (1\pm o(1))$ then with high probability 
the edge density of the $\ell$-core of a random hypergraph from $\graphSet$ is
$$
\frac{%
\beta(c)\cdot \Pr[\Po(\beta(c))\ge \ell-1]}{%
k\cdot\Pr[\Po(\beta(c))\ge \ell]}\pm o(1).
$$
\end{proposition}
\noindent
This follows directly from Proposition~\ref{prop:two}, where we have
also used 
equation (\ref{eq:definition:g}) to simplify the expression for $\hat n$.  

We define
$c_{k,\ell}$ as the unique $c$ that satisfies
\begin{equation} \label{ckll+1}
\frac{\beta(c)\cdot \Pr[\Po(\beta(c))\ge \ell-1]} 
{k\cdot\Pr[\Po(\beta(c))\ge  \ell]}\,= \, \ell-1.
\end{equation}
The values $c_{k,\ell}$ will prove important in the work to come;
in particular, we next show that $c_{k, 2}$ is the threshold
for $k$-ary cuckoo hashing for $k > 2$.  We also conjecture
that $c_{k,\ell+1}$ is the threshold for $k$-ary cuckoo hashing
when a bucket can hold $\ell$ keys instead of a single key.  

The following table contains numerical values of $c_{k,\ell}$ for $\ell=2, \ldots, 7$
and $k=2,\ldots,7$ (rounded to 10 decimal places).
Some of these numbers are found or referred to in other works, such as~\cite[Sect.~5]{Coo:2004},
\cite[Sect.\,4.4]{MezRicZec:2003},
\cite[p.\,423]{MezMontBook:2009},~\cite{FerRam:2007}, and~\cite{CaiSanWor:2007}.

\begin{table}[ht]
\begin{center}
\begin{tabular}{c|cccccc}
$\ell \backslash k$ & 2 & 3 & 4 & 5 & 6 & 7\\\hline
2 & $-$          & 0.9179352767 & 0.9767701649 & 0.9924383913 & 0.9973795528 & 0.9990637588\\
3 & 1.7940237365 & 1.9764028279 & 1.9964829679 & 1.9994487201 & 1.9999137473 & 1.9999866878\\
4 & 2.8774628058 & 2.9918572178 & 2.9993854302 & 2.9999554360 & 2.9999969384 & 2.9999997987\\
5 & 3.9214790971 & 3.9970126256 & 3.9998882644 & 3.9999962949 & 3.9999998884 & 3.9999999969\\
6 & 4.9477568093 & 4.9988732941 & 4.9999793407 & 4.9999996871 & 4.9999999959 & 5.0000000000\\
7 & 5.9644362395 & 5.9995688805 & 5.9999961417 & 5.9999999733 & 5.9999999998 & 6.0000000000\\
\end{tabular}
\end{center}
\end{table}

\ifnum\conf=1
\vspace{-0.4in}
\fi

\section{Equality of thresholds for random $k$-XORSAT and $k$-ary cuckoo hashing}\label{sec:main}

We now recall the random $k$-XORSAT problem and describe its relationship to
cores of random hypergraphs and cuckoo hashing.  The $k$-XORSAT problem is a variant of
the satisfiability problem in which every clause has $k$ literals and
the clause is satisfied if the XOR of values of the literals is 1.
Equivalently, since XORs correspond to addition modulo 2, and the negation of
$X_i$ is just $1$ XOR $X_i$, 
an instance of the $k$-XORSAT problem 
corresponds to a system of linear equations modulo 2, with
each equation having $k$ variables (none of which is negated), and randomly chosen right hand sides.  (In what follows we simply use
the addition operator where it is understood we are working modulo 2 from context.)

For a random $k$-XORSAT problem, 
let ${{\Phi}^k_{m,n}}$ be the set of all sequences of $n$ linear equations
over $m$ variables $x_1,\ldots,x_m,$ where an equation is
$$
x_{j_1}+\cdots+x_{j_k}=b_j,
$$
where $b_j\in\{0,1\}$ and $\{j_1,\ldots,j_k\}$ is a subset of $\{1,\ldots,m\}$ with $k$ elements.
We consider ${{\Phi}^k_{m,n}}$ as a probability space with the uniform distribution.

Given a $k$-XORSAT formula $F$, it is clear that $F$ is satisfiable if
and only if the formula obtained from $F$ by repeatedly deleting variables that
occur only once (and equations containing them) is satisfiable.  Now
consider the $k$-XORSAT formula as a hypergraph, with nodes
representing variables and hyperedges representing equations.  (The
values $b_j$ of the equations are not represented.)  The process of repeatedly
deleting all variables that occur only once, and the corresponding
equations, is exactly equivalent to the peeling process on the
hypergraph.  Hence, after the peeling process, we obtain the 2-core
of the hypergraph.

This motivates the following definition.  Let $\Psi^k_{m,n}$ be the
set of all sequences of $n$ equations 
such that each variable appears at least twice. We consider $\Psi^k_{m,n}$
as a probability space with the uniform distribution.  

Recall that if we start with a uniformly chosen random $k$-XORSAT formula, and perform the peeling
process, then conditioned on the remaining number of equations and variables
($\hat n$ and $\hat m$), 
we are in fact left with a uniform random formula from
$\Psi^k_{\hat m,\hat n}$.  
Hence, the imperative question is 
when a random formula from $\Psi^k_{\hat m,\hat n}$ will be satisfiable.  
In~\cite{DubMan:2002},
it was shown that this depends entirely on the edge density of the corresponding hypergraph.  If the
edge density is smaller than 1, so that there are more variables than
equations, the formula is likely to be satisfiable, and naturally,
if there are more equations than variables, the formula is likely
to be unsatisfiable.  Specifically, we have the following theorem
from~\cite{DubMan:2002}.

\begin{theorem} \label{thresh}
\label{DM:Theorem}
Let $k > 2$ be fixed. For $n/m = \gamma$ and $m\to\infty$,
	\begin{itemize}
	\item[{\rm(a)}] if $\gamma>1$
	then a random formula from $\Psi^k_{m,n}$
	is unsatisfiable with high probability.
	\item[{\rm(b)}] if $\gamma<1$
	then a random formula from $\Psi^k_{m,n}$
	is satisfiable with high probability.
\end{itemize}
\end{theorem}

The proof of Theorem \ref{thresh} in Section 3 of~\cite{DubMan:2002}
uses a 
first moment method argument for the simple direction (part (a)).
Part (b) is significantly more complicated, and is based on the second 
moment method.  Essentially the same problem has also arisen in 
coding theoretic settings;  analysis and techniques can be found in
for example \cite{MMU}.
\ifnum\conf=1
It has been suggested by various readers of earlier drafts of this paper that
previous proofs of Theorem \ref{thresh} have been insufficiently complete, 
particularly for $k > 3$.  We therefore provide a detailed proof in Appendix~C for completeness.
\fi

We have shown that the edge density is concentrated around a specific
value depending on the initial ratio $c$ of hyperedges (equations) to nodes (variables).
Let $c_{k,2}$ be the value of $c$ such that the resulting edge density is concentrated
around 1.  Then Proposition \ref{prop:three} and
Theorem \ref{thresh} together with the preceding consideration implies: 

\begin{corollary} \label{corthresh}
Let $k > 2$ and consider ${{\Phi}^k_{m,n}}.$  
The satisfiability threshold with respect to the edge density $c=n/m$ is  $c_{k, 2}$. 
\end{corollary}

Again, up to this point, everything we have stated was known from previous work.
We now provide the connection to cuckoo hashing, to show that we obtain the same
threshold values for the success of cuckoo hashing.  That is, we argue the following:

\begin{theorem}\label{thm:equivalence:variant}
For $k > 2$,
$c_{k,2}$ is the threshold for $k$-ary cuckoo hashing to work. 
That is, and with $n$ keys to be stored and 
$m$ buckets, with $c=n/m$ fixed and $m\to\infty$,  
\begin{itemize}
	\item[{\rm(a)}]
	   if  $c > c_{k,2}$, then
	   $k$-ary cuckoo hashing does not work with high probability.
	\item[{\rm(b)}]
	   if  $c < c_{k,2}$, then
	   $k$-ary cuckoo hashing works with high probability.
\end{itemize}
\end{theorem}
	   
\begin{proof}
Assume a set of  $n$ keys $S$   is given, and for each $x \in S$ 
a random set $A_x\subseteq\{1,\ldots,m \}$ 
of size $k$  of possible buckets is chosen. 

To prove part (a), note that the sets $A_x$ for $x\in S$
can be represented by a random hypergraph from $\graphSet$.	   
If $n/m = c > c_{k,2}$ and $m\to\infty$, then
with high probability 
the edge density in the 2-core is greater than 1.
The hyperedges in the 2-core correspond to a set of keys, and
the nodes in the 2-core to the buckets available for these keys. 
Obviously, then, cuckoo hashing does not work. 

To prove part (b), consider the case where $n/m = c < c_{k,2}$ and $m \to\infty$.
Picking for each $x$ a random $b_x \in \{0, 1\}$, the sets $A_x$, $x\in S$,
induce a random system of equations from  ${{\Phi}^k_{m,n}}.$
Specifically,  $A_x = \{ j_1 , \dots,  j_k\}$ 
induces the equation  $x_{j_1} + \dots + x_{j_k} = b_x.$ 
 
By Corollary \ref{corthresh} a random system of equations from ${{\Phi}^k_{m,n}}$ is satisfiable  with high 
probability.  This implies that the
the matrix $M$ made up from the left-hand sides of these
equations consists of linearly independent rows
with high probability. This is because a given set of left-hand 
sides with  dependent rows is only satisfiable 
with probability at most $1/2$ when we pick the $b_x$ at random. 

Therefore we have an $n \times n$-submatrix in $M$ with 
a nonzero determinant. The expansion of the determinant of this submatrix
as a sum of products by the Leibniz formula must contain a product with all 
factors being variables $x_{i_j}$ (as opposed to 0).
This product term corresponds to a permutation mapping 
keys to buckets, showing that cuckoo hashing is indeed possible.
\end{proof}


We make some additional remarks.  We note that the idea of using the
rank of the key-bucket matrix to obtain lower bounds on the cuckoo hashing
threshold is not new either;  it appears in~\cite{DP:2008}.  
There the authors use a result bounding the rank by Calkin~\cite{Calkin} to obtain a
lower bound on the threshold, but this bound is not tight in this context.
More details can be found by reviewing
\cite[Theorem 1.2]{Calkin} and~\cite[Exercise 18.6]{MezMontBook:2009}.  
Also, Batu et al.~\cite{BBC} note that 2-core thresholds provide
an upper bound on the threshold for cuckoo hashing, but fail to note the connection
to work on the $k$-XORSAT problems. 

\section{Non-integer choices}\label{sec:non-integer}

The analysis of $k$-cores in Section~\ref{sec:main} and the correspondence to
$k$-XORSAT problems extends nicely to the setting where the number of
choices for a key is not necessarily a fixed number $k$.  This
can be naturally accomplished in the following way: when a key
$x$ is to be inserted in the cuckoo hash table, the number of choices
of location for the key is itself determined by some hash
function; then the appropriate number of choices for each key $x$
can also be found when performing a lookup.  Hence, it is possible to
ask about for example cuckoo hashing with 3.5 choices, by which we
would mean an average of 3.5 choices.  Similarly, even if we decide to
have an average of $k$ choices per key, for an integer $k$, it is not immediately
obvious whether the success probability in $k$-ary cuckoo
hashing could be improved if we do not fix the number of possible
positions for a key but rather choose it at random from a cleverly
selected distribution.

Let us consider a more general setting where for each $x\in U$ the set $A_x$ is
chosen uniformly at random from the set of all $\randv_x$-element
subsets of $[\cells]$, where $\randv_x$ follows some probability mass
function $\pmf_x$ 
on $\{2,\ldots,\cells\}$.\footnote{We could in principle also consider the possibility of keys having only a single choice. However, this is generally not very interesting since even a small number of keys with a single choice would make an assignment impossible whp., by the birthday paradox. Hence, we restrict our attention to at least two choices.}
Let $\mean_x=E(\randv_x)$ and
$\mean^*=\frac{1}{\keys}\sum_{x\in S}\mean_x$. Note that $\mean^*$ is
the average (over all $x \in S$) worst case lookup time for successful
searches.  We keep $\mean^*$ fixed and study which sequence
$(\pmf_x)_{x\in S}$ maximizes the probability that cuckoo hashing is
successful. 

We fix the sequence of the expected number of choices per key
$(\mean_x)_{x\in S}$ and therefore~$\mean^*$. Furthermore we assume
$\mean_x\leq n-2$, for all $x \in S$; obviously this does not exclude
interesting cases. For compactness reasons, there is a system of
probability mass functions~$\pmf_x$ that maximizes the success
probability.  We will show the following:

\begin{proposition}
\label{prop:constant_expected_degree}
Let $(\pmf_x)_{x \in S}$ be an optimal sequence. Then we have, for all $x \in S$:
 $$\pmf_x( \lfloor \mean_x \rfloor )=1-(\mean_x-\lfloor \mean_x\rfloor), \text{ and }\pmf_x( \lfloor \mean_x \rfloor+1 )=\mean_x-\lfloor \mean_x\rfloor.$$
\end{proposition}
That is, the success probability is maximized if for each $x\in S$ the
number of choices $\randv_x$ is concentrated on $\lfloor
\mean_x\rfloor$ and $\lfloor \mean_x\rfloor +1$ (when the number of
choices is non-integral). Further, in the natural case where all keys $x$ have the same
expected number $\mean^*$ of choices, the optimal
assignment is concentrated on $\lfloor \mean^* \rfloor$ and $\lfloor
\mean^*\rfloor +1$.  Also, if $\mean_x$ is an integer,
then a fixed degree $\randv_x=\mean_x$ is optimal.  This is very different
from other similar scenarios, such as erasure- and error-correcting codes,
where irregular distributions have proven beneficial~\cite{ldpc}. 

\ifnum\conf=1
The proof is given in Appendix~A.
\fi

\ifnum\longf=1
\begin{proof}
We consider a random bipartite graph $G_S$ with left node set
$S$, right node set $[\cells]$ and an edge between two nodes $x\in S$
and $a\in [\cells]$ if and only if $a\in A_x$. Let the sequence
$(\mean_x)_{x\in S}$ be fixed. For each $x \in S$ we want to obtain a
distribution $\pmf_x$ for the degree $\randv_x$ 
(or, equivalently, the cardinality of
$A_x$), such that we have $E(\randv_x)=\mean_x$ and the
following quantity is maximized:
\begin{equation}
\label{eq:success_prob}
\Pr(\text{``success''}):=\Pr( (A_x)_{x\in S} \text{ admits a left-perfect matching}\footnote{In the following ``matching'' and ``left-perfect matching'' are used synonymously.}\text{ in $G_S$})\blank.
\end{equation}
We study the sequence $(\pmf_x)_{x\in S}$ that realizes the maximum.
Let $z$ be an arbitrary but fixed element of $S$ with probability mass
function $\pmf_z$. To prove Proposition
\ref{prop:constant_expected_degree} it is sufficient to show that if
there exist two numbers $\low$ and $\high$ with $\low <\mean_z <
\high$ and $\high-\low\geq2$ as well as $\pmf_z(\low)>0$ and
$\pmf_z(\high)>0$ then \eqref{eq:success_prob} cannot be maximal. 

We start by fixing $\randv_x$ and $A_x$ for each $x\in S-\{z\}$ and consider the corresponding bipartite graph $G_{S-\{z\}}$. Let $B\subseteq [\cells]$ be the set of right nodes in $G_{S-\{z\}}$ that are matched in every matching. Then there is a matching for the whole key set $S$ in $G_S$ if and only if $A_z \not \subseteq B$. Note that $0\leq |B|<\cells$, i.\,e., there must be at least one right node that is not matched. Let $p=\min\{ \pmf_z(\low), \pmf_z(\high) \}>0$ and $|B|=b$. We will show that changing $\pmf_z$ to 
\begin{align*}
\pmf_z'(\low)&:=\pmf_z(\low)-p     &\pmf_z'(\high)&:=\pmf_z(\high)-p \\
\pmf_z'(\low+1)&:=\pmf_z(\low+1)+p &\pmf_z'(\high-1)&:=\pmf_z(\high-1)+p,
\end{align*}
with $\pmf_z'(j)=\pmf_z(j)$ for $j\notin\{\low,\high\}$, 
increases \eqref{eq:success_prob}, while leaving $\mean_z$ unchanged. This is the case if and only if
\begin{equation}
\label{eq:first_ineq}
p\cdot \frac{\binom{b}{\low}}{\binom{\cells}{\low}} +  p\cdot \frac{\binom{b}{\high}}{\binom{\cells}{\high}} 
\geq
p\cdot \frac{\binom{b}{\low+1}}{\binom{\cells}{\low+1}} + p\cdot \frac{\binom{b}{\high-1}}{\binom{\cells}{\high-1}}, 
\end{equation}
and the strict inequality holds for at least one value $b$ that occurs with positive probability.
The left sum of \eqref{eq:first_ineq} is the $2\cdot p$ fraction of the failure probability (by $\pmf_z(\low)$ and $\pmf_z(\high)$) before the change of $\pmf_z$ under the condition that $B$ has cardinality $b$; the right sum is the corresponding fraction of the failure probability after the change. Depending on $b$ we have to distinguish several cases.
\setlength{\leftmargini}{10pt} 
\begin{compactenum}[\text{Case} 1:]
\item $b=\cells-1$. In this case both sides of \eqref{eq:first_ineq} are equal, i.\,e., the modification we do to $\pmf_z$ will not change the success probability. 
\item $\high \leq b < \cells-1$. Canceling $p$ and subtracting ${\binom{b}{\high}}/{\binom{\cells}{\high}}$ and ${\binom{b}{\low+1}}/{\binom{\cells}{\low+1}}$ from both sides of \eqref{eq:first_ineq} shows that the strict inequality holds if and only if
\begin{equation}
\label{eq:intermediate}
\begin{split}
\frac{b\cdots(b-\low+1)}{\cells\cdots(\cells-\low+1)}-\frac{b\cdots(b-\low)}{\cells\cdots(\cells-\low)}>\frac{b\cdots(b-\high+2)}{\cells\cdots(\cells-\high+2)}-\frac{b\cdots(b-\high+1)}{\cells\cdots(\cells-\high+1)} \blank. 
\end{split}
\end{equation}
Factoring out $\frac{b\cdots(b-\low+1)}{\cells\cdots(\cells-\low+1)}$ on the left side and $\frac{b\cdots(b-\high+2)}{\cells\cdots(\cells-\high+2)}$ on the right side gives
\begin{equation}
 \frac{b\cdots (b-\low+1)}{\cells\cdots (\cells-\low)}>\frac{b\cdots (b-\high+2)}{\cells\cdots(\cells-\high+1)}
 \Leftrightarrow
 \frac{\cells\cdots(\cells-\high+1)}{\cells\cdots (\cells-\low)}>\frac{b\cdots (b-\high+2)}{b\cdots (b-\low+1)}\blank.
\end{equation}
Since $\low \leq \high -2$, this is equivalent to 
\begin{equation}
(\cells-\low+1)\cdot(\cells-\low)\cdots(\cells-\high+1)>(b-\low)\cdot(b-\low-1)\cdots(b-\high+2)\blank,
\end{equation}
which is true for $m-1>b$.
\item $\low\leq b<\high$.  Calculations along the lines of case 2 show that the strict inequality of \eqref{eq:first_ineq} also holds in this case. Note that $\binom{b}{k}$, $\binom{b}{k-1}$ and $\binom{b}{l+1}$ can be zero.
\item $0\leq b<\low$. In this case both sides of \eqref{eq:first_ineq} are zero, i.\,e., the modifications we do to $\pmf_z$ will not change the success probability.
\end{compactenum}

Since in cases 1 and 4 above there was no change in the success probability, to show
that \eqref{eq:success_prob} cannot be maximal when $\high-\low\geq2$ as we are considering, 
it remains to show that at least one of the Cases 2 and 3 occurs with positive probability.
We construct a situation in which one of these cases applies, and which occurs with positive
probability. 

Choose degrees $\randv_x$ for all elements $x \in S-\{z\}$ such that $\randv_x\leq \mean_x$ and $\pmf_x(\randv_x)>0$.
Consider a permutation of the elements $x \in S-\{z\}$ such that these degrees are ordered, i.\,e., $\randv_{x_1}\leq \randv_{x_2}\leq \ldots \leq \randv_{x_{\keys-1}}$. Choose the first element $x_i$ with  $i \geq \low$ and $\randv_{x_i}\leq i$. Such an element must exist, since we assume $\randv_x\leq \mean_x\leq \keys-2$, in particular we have $l<\keys-2$. Arrange that $A_{x_j}\subseteq [i],1\leq j \leq i,$ such that there is a matching in $G_{\{x_1,\ldots,x_i\}}$. This implies $b\geq \low$. Then arrange that $|A_{x_j}-\bigcup_{1\leq {j'}<j} A_{x_{j'}}|=1$, for all $i<j\leq n-2$, as well as $|A_{x_{n-1}}-\bigcup_{1\leq {j'}<n-1} A_{x_{j'}}|=2$, which implies $b<m-1$. This finishes the proof of Proposition \ref{prop:constant_expected_degree}. 
\end{proof}
\fi

\subsection{Thresholds for non-integral degree distributions}

We now describe how to extend our previous analysis to derive thresholds for the case of a non-integral number of choices per key;  
equivalently, we are making use of thresholds for XORSAT problems with an irregular number of literals per clause.  

Following notation that is frequently used in the coding literature,
we let
$\Lambda_k$ be the probability that a key obtains $k$ choices, and
define $\Lambda(x) = \sum_k \Lambda_k x^k$.  
Clearly, then, $\Lambda'(x) = \sum_k \Lambda_k k x^{k-1}$,
and $\Lambda'(1)=\kappa^*$. 
(We assume henceforth
that $\Lambda_0 = \Lambda_1 = 0$ and $\Lambda_k = 0$ for all $k$
sufficiently large for technical convenience.)

We now follow our previous analysis from Section~\ref{sect:facts:cores};  
to see if a node $a$ is deleted after $h$
rounds of the peeling process, we let $p_j$ be the
probability that a node at distance $h-j$ from $a$ is deleted after
$j$ rounds%
.
We must now account for the differing 
degrees of hyperedges.  Here, the appropriate asymptotics is given by a mixture of binomial hypergraphs,
with each hyperedge of degree $k$ present with probability $k! \cdot c \Lambda_k /m^{k-1}$ independently. 

The corresponding equations are then given by $p_0 = 0$,
\begin{eqnarray*} 
p_1 & =& \Pr\bigg[\sum_k \Bin\left( \binom{m-1}{k-1} \,\,,\,\,k! 
\cdot \frac{c\Lambda_k}{m^{k-1}} \right) \le \ell-2 \bigg]\,  \\
  & = & \Pr\bigg[\sum_k \Po(k c \Lambda_k) \le \ell-2\bigg] \pm o(1),     \\  
  & = & \Pr[\Po(c \Lambda'(1)) \le \ell-2] \pm o(1),     \\  \mbox{ } & \\
p_{j+1} &= & \Pr\bigg[\sum_k \Bin\left(\binom{m-1}{k-1} \,\,
, \,\, k! \cdot \frac{c \Lambda_k}{m^{k-1}} \cdot (1- p_j)^{k-1}\right)\, \le \ell-2\bigg]\, \\ 
&=&
\Pr\bigg[ \sum_k \Po(kc \Lambda_k (1-p_j)^{k-1}) \le \ell-2\bigg]  \pm o(1), \mbox{ for } j=1,\dots,h-2, \\
&=&
\Pr[ \Po(c \Lambda'(1-p_j)) \le \ell-2]  \pm o(1), \mbox{ for } j=1,\dots,h-2.
\end{eqnarray*} 
Note that we have used the standard fact that the sum of Poisson random variables is itself Poisson, which
allows us to conveniently express everything in terms of the generating function $\Lambda(x)$
and its derivative. %
As before we find $p= \lim p_j$, which is now given by 
the smallest non-negative solution of
\begin{eqnarray*}
p & = & \Pr[ \Po(c\Lambda'(1-p)) \le \ell-2].
\end{eqnarray*}

When given a degree distribution $(\Lambda_k)_k$,
we can proceed as before  
to find the threshold load that allows
that the edge density of the 2-core remains greater than 1; using 
\ifnum\longf=1
a second moment argument,
\fi
\ifnum\conf=1
the approach of Appendix~C, 
\fi
this can again be shown to be the required property for the corresponding XORSAT problem to have a solution, and hence
for there to be a permutation successfully mapping keys to buckets.  Notice that this argument works for
all degree distributions (subject to the restrictions given above), but in particular we have already shown
that the optimal thresholds are to be found by the simple degree distributions that have all weight on two
values, $\lfloor\mean^*\rfloor$ and $\lfloor\mean^*\rfloor + 1$.  
Abusing notation slightly, let $c_{\mean^*,2}$ be the unique $c$ such that
the edge density of the 2-core of the corresponding mixture is equal to 1, following the same form as in
Proposition~\ref{prop:three} and equation~(\ref{ckll+1}).  The corresponding extension to Theorem~\ref{thm:equivalence:variant}
is the following:

\begin{theorem}\label{thm:equivalence:variant2}
For $\mean^* > 2$,
$c_{\mean^*,2}$ is the threshold for cuckoo hashing with an average of $\mean^*$ choices per key to work.
That is, with $n$ keys to be stored and 
$m$ buckets, with $c=n/m$ fixed and $m\to\infty$,  
\begin{itemize}
	\item[{\rm(a)}]
	   if  $c > c_{\mean^*,2}$, for any distribution on the number
           of choices per key with mean $\mean^*$, cuckoo hashing does not work with high probability.
	\item[{\rm(b)}]
	   if  $c < c_{\mean^*,2}$, then
	   cuckoo hashing works with high probability when the distribution on the number of choices per key
           is given by
           $\pmf_x( \lfloor \mean_x \rfloor )=1-(\mean_x-\lfloor \mean_x\rfloor)$ and $\pmf_x( \lfloor \mean_x \rfloor+1 )=\mean_x-\lfloor \mean_x\rfloor$, for all $x \in S$.
\end{itemize}
\end{theorem}

\begin{figure}[ht]
\begin{center}
	\begin{minipage}{7.3cm}
	\includegraphics[width=\linewidth]{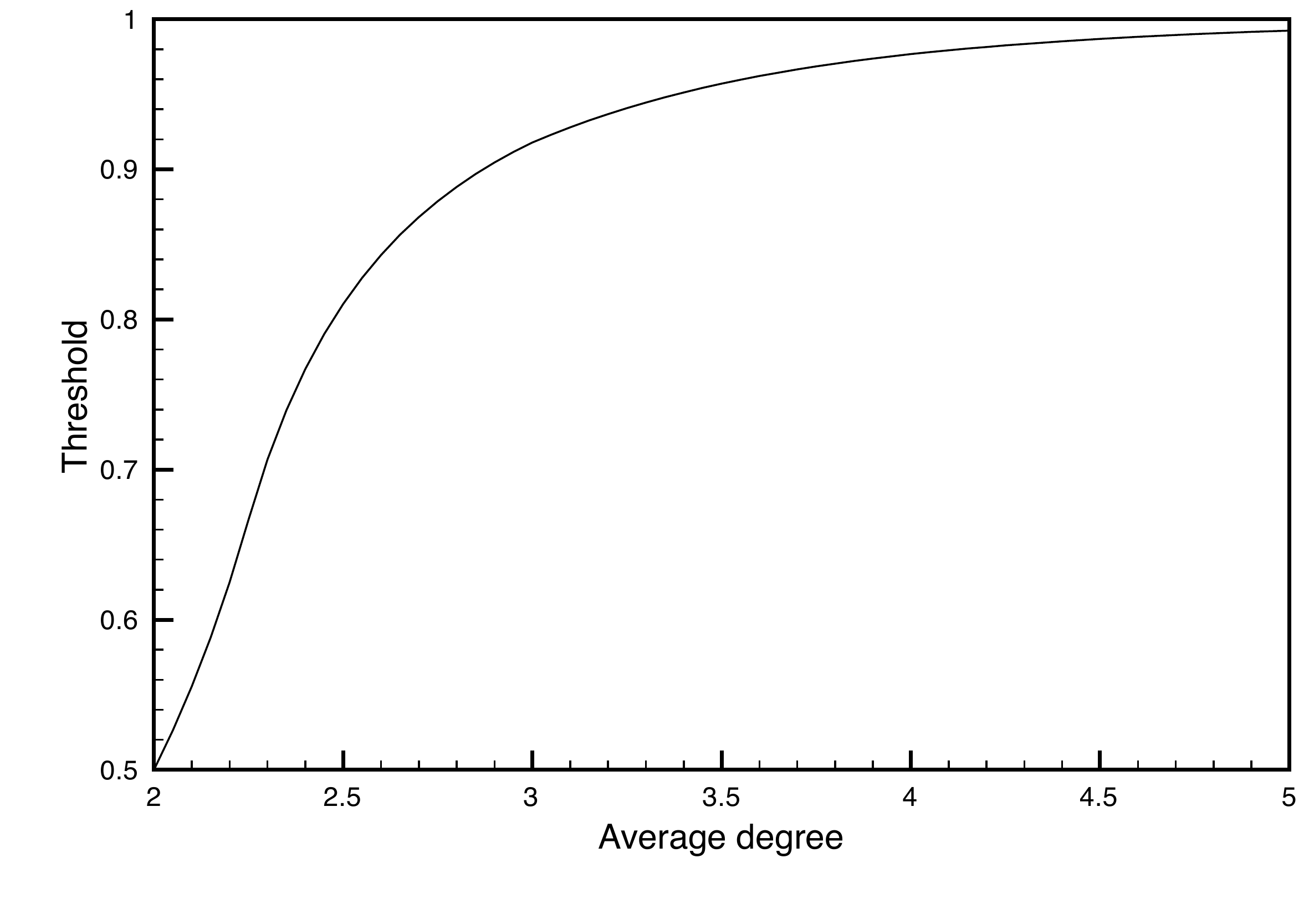}
	\end{minipage}
\begin{minipage}{0.46\textwidth}

\begin{tabular}{c|c}
 $\mean^*$ & $c_{\mean^*,2}$\\
\hline
2.25 & 0.6666666667\\ 
2.50 & 0.8103423635\\ 
2.75 & 0.8788457372\\ 
3.00 & 0.9179352767\\ 
3.25 & 0.9408047937\\ 
3.50 & 0.9570796377\\ 
3.75 & 0.9685811888\\ 
4.00 & 0.9767701649 
\end{tabular}
\hspace{0.5cm}
\begin{tabular}{c|c}
$\mean^*$ & $c_{\mean^*,2}$\\
\hline
4.25 & 0.9825693463\\
4.50 & 0.9868637629\\
4.75 & 0.9900548807\\
5.00 & 0.9924383913\\
5.25 & 0.9942189481\\
5.50 & 0.9955692011\\
5.75 & 0.9965961383\\
6.00 & 0.9973795528
\end{tabular}
\end{minipage}
\inFigSpace
\caption{Thresholds for non-integral $\mean^*$-ary cuckoo hashing, with optimal degree distribution. The values in the tables are rounded to the nearest multiple of $10^{-10}$.}\label{fig:non-integral}
\end{center}
\vspace{-0.2in}
\end{figure}
We have determined the thresholds numerically for a range of values
of~$\mean^*$. The results are shown in
Figure~\ref{fig:non-integral}. One somewhat surprising finding is that
the threshold for $\mean^*\leq 2.25$ appears to simply be given by $c =
0.5/(3-\mean^*)$. 
Consequently, in place of using 2 hash functions per key,
simply by using a mix of 2 or 3 hash functions for a key, we can increase the space utilization by adding
33\% more keys with the same (asymptotic) amount of memory.

\section{Algorithm for computing a placement}\label{subsec:selfless:algorithm}
In this section, we describe an algorithm for finding a placement for
the keys using $k$-ary cuckoo hashing when the set $S$ of
keys is given an advance.  The algorithm is an adaptation of the  
``selfless algorithm'' proposed by~Sanders~\cite{Sanders:SOFSEM:2004}, for the case $k=2$,  and analyzed
in~\cite{CaiSanWor:2007}, for orienting standard undirected random graphs so
that all edges are directed and the maximum indegree of all nodes is
at most $\ell$, for some fixed $\ell \geq 2$.  We generalize this algorithm
to hypergraphs, including hypergraphs where hyperedges can have varying degrees.

Of course, maximum matching algorithms can solve this problem
perfectly.  However, there are multiple motivations for considering
our algorithms. First, it seems in preliminary experiments that the
running times of standard matching algorithms like the Hopcroft-Karp
algorithm~\cite{HK:1973} will tend to increase significantly as the
edge density approaches the threshold (the details of this effect are
not yet understood), while our algorithm has linear running time which
does not change in the neighborhood of the threshold.  This proves
useful in our experimental evaluation of thresholds.  Second, we
believe that algorithms of this form may prove easier to analyze for
some variations of the problem.

We first describe the generalized selfless algorithm for 
bucket size $\ell=1$. 
A description in pseudocode
is given as Algorithm~\ref{algo:GeneralizedSelfless}.
The algorithm can deal with arbitrary hypergraphs, 
uniform or not. The aim is to ``orient'' the
hyperedges of the hypergraph $G$, i.\,e., associate a node $v\in e$ 
to each hyperedge $e$ so that at most one hyperedge is directed towards
any one node $v$. Initially, all hyperedges are unoriented.
Nodes that have an hyperedge directed 
towards them are saturated and are not considered further,
and similarly hyperedges once oriented are fixed.
At each step, if there is a node $v$ that is incident to 
only one undirected hyperedge $e$, we direct $v$ to $e$, breaking ties arbitrarily. 
(In the pseudocode, this is realized by giving such nodes the highest 
\emph{priority}, which is 0. Note that this rule entails that the algorithm starts
by carrying out the peeling process for the $2$-core. But the rule
is also applied when hyperedges from the 2-core have already been treated.)
If there are no such nodes, 
every unoriented hyperedge is assigned as its \emph{weight}
the number of unsaturated nodes it contains. (Intuitively, a
smaller weight means a higher need to direct the hyperedge.)
The priority of a node $v$ then is the sum of the inverses
of the weights of the hyperedges that contain~$v$. 
This corresponds to the expected
number of hyperedges $v$ would have directed toward it if all
its unoriented hyperedges were directed to 
one of their nodes at random. 
Now a vertex $v$ of smallest (highest) priority is chosen,
again breaking ties at random.
If this priority is larger than 1, then the algorithm stops
and reports ``failure''.  This is because 
the sum of all priorities is the number of undirected hyperedges,
so if the smallest priority is bigger than 1, the number of
undirected hyperedges is larger than the number of unsaturated nodes, 
and it is impossible to complete the process of directing the hyperedges.
Otherwise the algorithm 
directs the minimum weight incident hyperedge of $v$ toward $v$, 
breaking ties randomly. (Intuitively, this means that
the algorithm tries to continue the peeling process ``on average''.)
This step is repeated until all hyperedges have been oriented or failure occurs.

\begin{algorithm}
 \SetKwInput{KwPurpose}{Purpose}

 \SetKw{KwStop}{stop}
 \SetKw{KwFail}{failure}
 \SetKw{KwInput}{Input:}
 \SetKw{KwElse}{else}

 \KwIn{Hypergraph $G=(V,E)$ with $\nodes$ nodes and $\edges$ hyperedges.}
 \KwPurpose{Direct all hyperedges such that the maximum indegree is at most $1$.}
 \tcp{ $\ud{v}$ set of undirected hyperedges incident to node $v$}
 \For{$t\leftarrow1$ \KwTo $\edges$ }
 {
        $V_0\leftarrow \{v \in V : |\ud{v}|>0\}$; $E_0\leftarrow \{e\in E:e\text{ undirected}\}$\;%
        \ForAll(\tcp*[h]{calculate edge \emph{weight} $\omega(e)$}){$e\in E_0$}
        {
         $\omega(e)\leftarrow |\{v\in e : \text{no hyperedge is directed towards $v$}\}|$\;
        }
        \ForAll(\tcp*[h]{calculate \emph{priority} $\pi(v)$}){$v\in V_0$}
        {
                \lIf{a hyperedge is directed towards $v$}
                {
                        $\pi(v)\leftarrow 2$ \tcp{saturated}
                }\;
                \Else
                {
                        \lIf{$|\ud{v}|=1$}{$\pi(v)\leftarrow0$}
                        \lElse{$\pi(v)\leftarrow \sum_{e\in \ud{v}}\frac{1}{\omega(e)}$\;}\;
                }
         }
        find $v \in V_0$ with smallest priority (break ties by randomization)\;
        \lIf{$\pi(v) > 1$}{\KwRet\KwFail}\;
        choose minimum weight hyperedge $e\in \ud{v}$ (break ties by randomization)\;
        direct $e$ towards $v$
 }
  \caption{\label{algo:GeneralizedSelfless}$(k,1)$-Generalized Selfless}
\end{algorithm}

\ifnum\conf=1
We ran the generalized selfless algorithm for hypergraphs with $10^5$ and $10^6$ nodes
and tabulated the failure rate around the theoretical threshold values $c_{k,2}$ for $k=3,4,5$.
Results demonstrate that the generalized selfless algorithm achieves results quite near the
threshold;  more details and figures are given in Appendix~B.
\fi

\ifnum\longf=1
We ran the generalized selfless algorithm for hypergraphs with $10^5$ and $10^6$ nodes
and tabulated the failure rate around the theoretical threshold values $c_{k,2}$ for $k=3,4,5$. For each pair $(\nodes,k)$ we considered $81$ edge densities $c=\frac{\edges}{\nodes}$, spaced apart by $0.0001$, 
thus covering an interval of length $0.008$, which encloses the theoretical threshold value 
for the particular parameter pair $(\nodes,k)$.
The hyperedges of the hypergraphs were randomly chosen via pseudo random number generator MT19937 ``Mersenne Twister'' of the GNU Scientific Library~\cite{GNU_Scientific}. We measured the average failure rate of the algorithm over $100$ random hypergraphs for each combination $(\nodes,\edges,k)$ within the parameter space. 
To get an estimation of the threshold, i.\,e., the rate $c$ where the algorithm switches from success to failure, we fit the sigmoid function
\begin{equation}
\label{eq:fit_function}
 \sigma(c;a,b)=\frac{1}{1+e^{-(c-a)/b}}
\end{equation}
to the measured failure rate (via gnuplot\footnote{gnuplot, an interactive plotting program, version 4.2, \url{http://www.gnuplot.info}}), using the method of least squares. We determined the parameters $a,b$ that lead to a (local) minimum of the sum of squares of the $81$ residuals, denoted by $\sum_{res}$. The parameter $a$ is the inflection point of \eqref{eq:fit_function} and therefore the approximation of the threshold of the generalized selfless algorithm. Figures \ref{fig:gen_selfless_edge_size_3},
\ref{fig:gen_selfless_edge_size_4} and \ref{fig:gen_selfless_edge_size_5} show the results of the experiments.

\begin{figure}[th]
\centering
 \subfigure[$\nodes=10^5$]{\scalebox{\imgScale}{\input{\figPath/xGS_n=10E5_d=3_k=1.tex}}}
 \subfigure[$\nodes=10^6$]{\scalebox{\imgScale}{\input{\figPath/xGS_n=10E6_d=3_k=1.tex}}}
 \caption{\label{fig:gen_selfless_edge_size_3}edge size $k=3$; theoretical threshold $c_{k,2} \approx 0.91794$}
\end{figure}
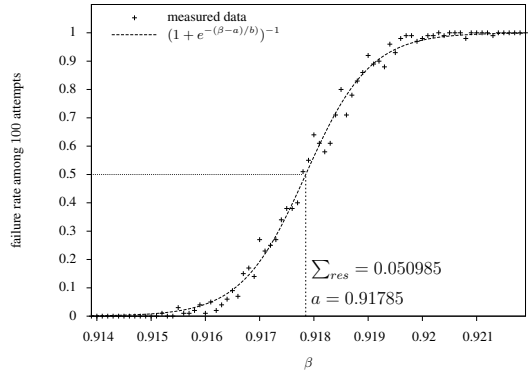
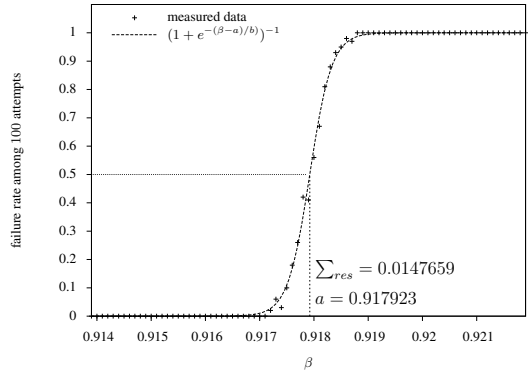
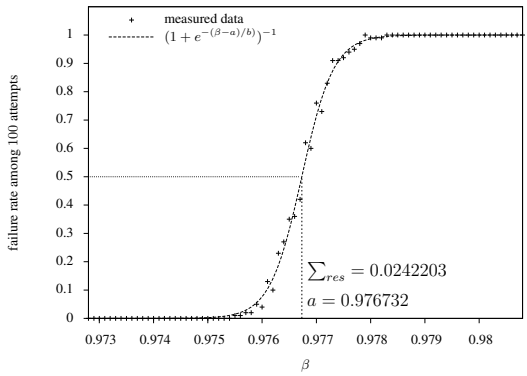
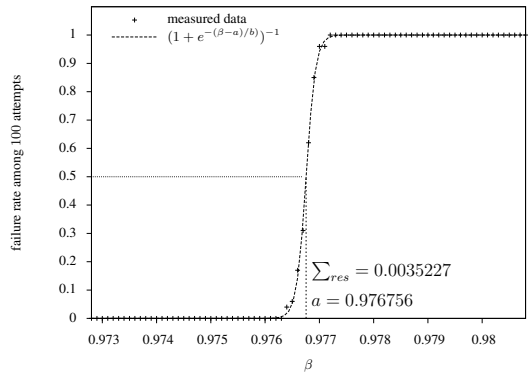
\begin{figure}[th]
\centering
 \subfigure[$\nodes=10^5$]{\scalebox{\imgScale}{\input{\figPath/xkthulhu_MTW_n=10E5_d=4_k=1.tex}}}
 \subfigure[$\nodes=10^6$]{\scalebox{\imgScale}{\input{\figPath/xkthulhu_MTW_n=10E6_d=4_k=1.tex}}}
 \caption{\label{fig:gen_selfless_edge_size_4}edge size $k=4$; theoretical threshold $c_{k,2} \approx 0.97677$}
\end{figure}
\begin{figure}[th]
\centering
 \subfigure[$\nodes=10^5$]{\scalebox{\imgScale}{\input{\figPath/xkthulhu_MTW_n=10E5_d=5_k=1.tex}}}
 \subfigure[$\nodes=10^6$]{\scalebox{\imgScale}{\input{\figPath/xkthulhu_MTW_n=10E6_d=5_k=1.tex}}}
 \caption{\label{fig:gen_selfless_edge_size_5}edge size $k=5$; theoretical threshold $c_{k,2} \approx 0.99244$}
\end{figure}
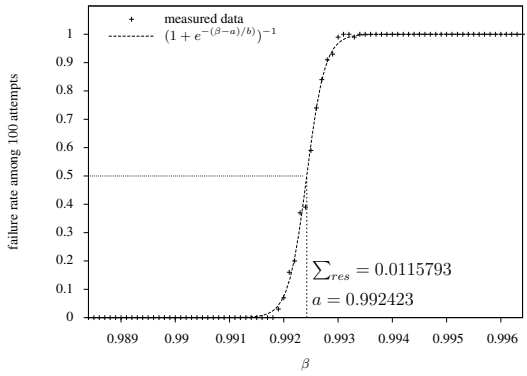
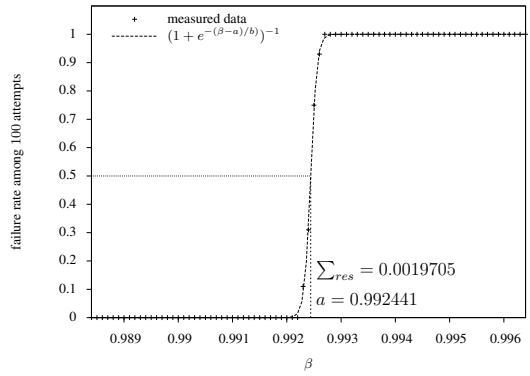
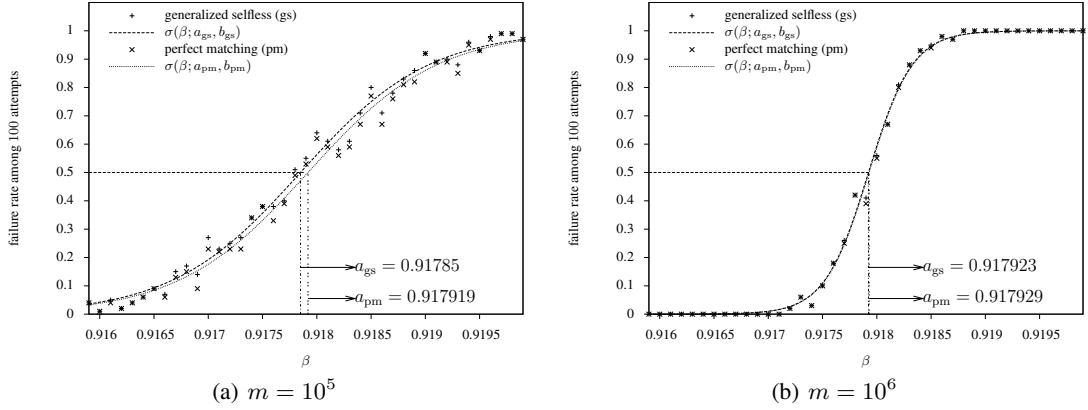
\begin{figure}[h!]
\centering
 \subfigure[$\nodes=10^5$]{\scalebox{\imgScale}{\input{\figPath/xGSvsHC_n=10E5_d=3_k=1.tex}}}
 \subfigure[$\nodes=10^6$]{\scalebox{\imgScale}{\input{\figPath/xGSvsHC_n=10E6_d=3_k=1.tex}}}
 \caption{\label{fig:selfless_vs_perfect_matching}$k=3$; theoretical threshold $c_{k,2} \approx 0.91794$, interval size $0.004$}
\end{figure}

One observes that this simple algorithm is able to construct the placements for edge densities quite close to the calculated thresholds $c_{k,2}$. The slope of the sigmoid curve increases and $\sum_{res}$ decreases with growing $\nodes$ and $k$, leading to a sharp transition from total success to total failure.
Clearly the algorithm can fail on hypergraphs that admit a matching. Experimental comparisons with a perfect matching algorithm \cite{HK:1973} showed that this is very unlikely for random hypergraphs. An example is given in Figure \ref{fig:selfless_vs_perfect_matching}, which shows the failure rate of perfect matching in comparison to the generalized selfless algorithm. Note that the plot shows an interval of size $0.004$, i.\,e., 41 data points instead of $81$. The differences in the failure rates of the algorithms become very small as~$\nodes$ grows.
\fi

\subsection{A conjecture, with evidence from a generalized selfless algorithm }\label{subsec:conjecture}
Now consider a situation in which buckets have a capacity of $\ell>1$ keys. 
There is as yet no rigorous analysis of the appropriate thresholds for cuckoo hashing for the cases $k > 2$ and $\ell>1$. 
However, our results of Section~\ref{sect:facts:cores} suggest a natural conjecture:
\begin{conjecture}
\label{conj:threshold}
For $k$-ary cuckoo hashing with bucket size $\ell$, it is \emph{conjectured} that cuckoo hashing works with high probability if 
$n/m = c >  c_{k,\ell+1}$, and does not work if $n/m = c <  c_{k,\ell+1}$,
i.\,e., that the threshold is at the point where the $(\ell+1)$-core of the cuckoo hypergraph starts having edge density larger than $\ell$.
\end{conjecture}

In order to provide evidence for this conjecture, 
we generalize our algorithm further 
so that it can deal with bucket size $\ell>1$. 
The pseudocode is given as Algorithm~\ref{algo:GeneralizedSelfless2}.
In hypergraph language, 
we are now looking for an orientation of the hyperedges of $G$ so 
that every node has at most $\ell$ hyperedges directed toward it.
Now a node is saturated if it has $\ell$ edges pointing to it. 
As long as there are nodes $v$ such that
the number of hyperedges directed toward $v$ and 
the number of undirected hyperedges containing $v$ taken together
does not exceed $\ell$, one such node is chosen and 
its undirected edges are directed toward it. 
Again, the effect of this rule is
that the algorithm starts by carrying out 
the peeling process that finds the $(\ell+1)$-core.   
Otherwise, the algorithm assigned weights and priorities as before,
and if all priorities exceed $\ell$, the algorithm stops and reports failure.
If the smallest (highest) priority is at most $\ell$,
a vertex of smallest priority is chosen and one of the incident undirected hyperedges of minimum weight
is directed toward it. 
The process is carried out until all hyperedges have
been directed or failure occurs. 

\begin{algorithm}
 \SetKwInput{KwPurpose}{Purpose}

 \SetKw{KwAnd}{and}
 \SetKw{KwStop}{stop}
 \SetKw{KwFail}{failure}
 \SetKw{KwInput}{Input:}
 \KwIn{Hypergraph $G=(V,E)$ with $\nodes$ nodes and $\edges$ hyperedges.}
 \KwPurpose{Direct all hyperedges such that the maximum indegree is at most $\ell$.}
 \tcp{ $\id{v}$ set of hyperedges directed towards node $v$\\$\ud{v}$ set of undirected hyperedges incident to node $v$}

  \For{$t\leftarrow1$ \KwTo $\edges$ }
  {
        $V_0\leftarrow \{v \in V : |\ud{v}|>0\}$; $E_0\leftarrow \{e\in E:e\text{ undirected}\}$\;
        \ForAll(\tcp*[h]{calculate edge \emph{weight} $\omega(e)$}){$e\in E_0$}
        {
         $\omega(e)\leftarrow |\{ v\in e: |\id{v}|< \ell\}|$\;
        }
        \ForAll(\tcp*[h]{calculate \emph{priority} $\pi(v)$}){$v\in V_0$}
        {
                \lIf{$|\ud{v}|+|\id{v}|\leq \ell$}{$\pi(v)\leftarrow0$}\;
                \lElse{$\pi(v)\leftarrow \sum_{e\in \ud{v}}\frac{1}{\omega(e)} +|\id{v}|$}\;
        }
         find $v \in V_0$ with smallest priority (break ties by randomization)\;
         \lIf{$\pi(v) > \ell$}{\KwRet\KwFail}\;
         choose minimum weight hyperedge $e\in \ud{v}$ (break ties by randomization)\;
         direct $e$ towards $v$
 }
 \caption{\label{algo:GeneralizedSelfless2} $(k,\ell)$-Generalized Selfless}
\end{algorithm}
Experiments with Algorithm~\ref{algo:GeneralizedSelfless2} corroborate Conjecture \ref{conj:threshold}, 
in that they show that the failure rate of the algorithm changes from 0 to 1 very close to
the possible threshold values suggested in Conjecture~\ref{conj:threshold}.
\ifnum\conf=1
Again, numerical results are given in Appendix~B.  
\fi

\ifnum\longf=1
(For an example see Figure~\ref{fig:gen_selfless_edge_size_3_bucket_size=2}.)
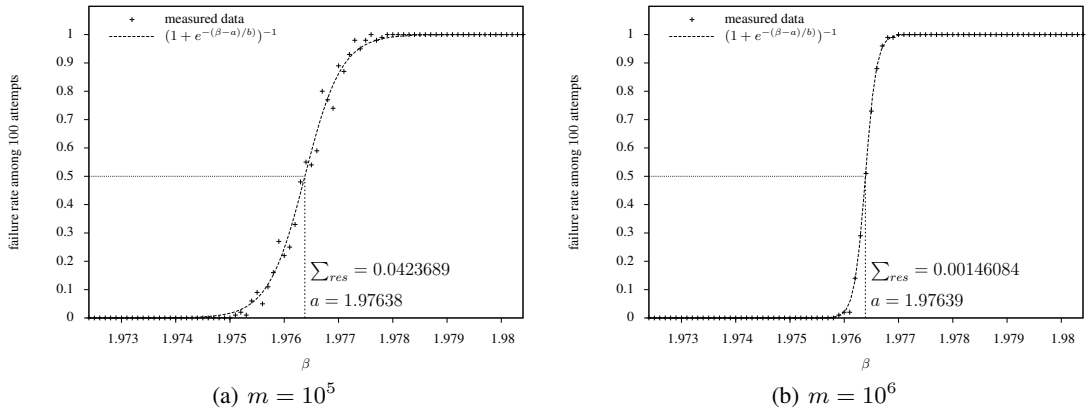
\begin{figure}[th]
\centering
 \subfigure[$\nodes=10^5$]{\scalebox{\imgScale}{\input{\figPath/xkthulhu_MTW_n=10E5_d=3_k=2.tex}}}
 \subfigure[$\nodes=10^6$]{\scalebox{\imgScale}{\input{\figPath/xkthulhu_MTW_n=10E6_d=3_k=2.tex}}}
 \caption{\label{fig:gen_selfless_edge_size_3_bucket_size=2}$k=3$, $\ell=2$; 
 \emph{conjectured} threshold value $c_{k,\ell+1} \approx 1.97640$}
\end{figure}
\fi

\section{Conclusion}

We have found tight thresholds for cuckoo hashing with 1 key per bucket,
by showing that the thresholds are in fact the same for the previous studied
$k$-XORSAT problem.  We have generalized the result to irregular cuckoo hashing
where keys may have differing numbers of choices, and have conjectured  
thresholds for the case where buckets have size larger than 1 based on an
extrapolation of our results.

\ifnum\conf=1

\appendix

\section{Optimality of degree distribution}

We present here the proof of Proposition~\ref{prop:constant_expected_degree}.
Specifically, we show that if $(\pmf_x)_{x \in S}$ is an optimal sequence, then for all $x \in S$:
 $$\pmf_x( \lfloor \mean_x \rfloor )=1-(\mean_x-\lfloor \mean_x\rfloor), \text{ and }\pmf_x( \lfloor \mean_x \rfloor+1 )=\mean_x-\lfloor \mean_x\rfloor.$$

\begin{proof} 
We consider a random bipartite graph $G_S$ with left node set
$S$, right node set $[\cells]$ and an edge between two nodes $x\in S$
and $a\in [\cells]$ if and only if $a\in A_x$. Let the sequence
$(\mean_x)_{x\in S}$ be fixed. For each $x \in S$ we want to obtain a
distribution $\pmf_x$ for the degree $\randv_x$ 
(or, equivalently, the cardinality of
$A_x$), such that we have $E(\randv_x)=\mean_x$ and the
following quantity is maximized:
\begin{equation}
\label{eq:success_prob}
\Pr(\text{``success''}):=\Pr( (A_x)_{x\in S} \text{ admits a left-perfect matching}\footnote{In the following ``matching'' and ``left-perfect matching'' are used synonymously.}\text{ in $G_S$})\blank.
\end{equation}
We study the sequence $(\pmf_x)_{x\in S}$ that realizes the maximum.
Let $z$ be an arbitrary but fixed element of $S$ with probability mass
function $\pmf_z$. To prove Proposition
\ref{prop:constant_expected_degree} it is sufficient to show that if
there exist two numbers $\low$ and $\high$ with $\low <\mean_z <
\high$ and $\high-\low\geq2$ as well as $\pmf_z(\low)>0$ and
$\pmf_z(\high)>0$ then \eqref{eq:success_prob} cannot be maximal. 

We start by fixing $\randv_x$ and $A_x$ for each $x\in S-\{z\}$ and consider the corresponding bipartite graph $G_{S-\{z\}}$. Let $B\subseteq [\cells]$ be the set of right nodes in $G_{S-\{z\}}$ that are matched in every matching. Then there is a matching for the whole key set $S$ in $G_S$ if and only if $A_z \not \subseteq B$. Note that $0\leq |B|<\cells$, i.\,e., there must be at least one right node that is not matched. Let $p=\min\{ \pmf_z(\low), \pmf_z(\high) \}>0$ and $|B|=b$. We will show that changing $\pmf_z$ to 
\begin{align*}
\pmf_z'(\low)&:=\pmf_z(\low)-p     &\pmf_z'(\high)&:=\pmf_z(\high)-p \\
\pmf_z'(\low+1)&:=\pmf_z(\low+1)+p &\pmf_z'(\high-1)&:=\pmf_z(\high-1)+p,
\end{align*}
with $\pmf_z'(j)=\pmf_z(j)$ for $j\notin\{\low,\high\}$, 
increases \eqref{eq:success_prob}, while leaving $\mean_z$ unchanged. This is the case if and only if
\begin{equation}
\label{eq:first_ineq}
p\cdot \frac{\binom{b}{\low}}{\binom{\cells}{\low}} +  p\cdot \frac{\binom{b}{\high}}{\binom{\cells}{\high}} 
\geq
p\cdot \frac{\binom{b}{\low+1}}{\binom{\cells}{\low+1}} + p\cdot \frac{\binom{b}{\high-1}}{\binom{\cells}{\high-1}}, 
\end{equation}
and the strict inequality holds for at least one value $b$ that occurs with positive probability.
The left sum of \eqref{eq:first_ineq} is the $2\cdot p$ fraction of the failure probability (by $\pmf_z(\low)$ and $\pmf_z(\high)$) before the change of $\pmf_z$ under the condition that $B$ has cardinality $b$; the right sum is the corresponding fraction of the failure probability after the change. Depending on $b$ we have to distinguish several cases.
\setlength{\leftmargini}{10pt} 
\begin{compactenum}[\text{Case} 1:]
\item $b=\cells-1$. In this case both sides of \eqref{eq:first_ineq} are equal, i.\,e., the modification we do to $\pmf_z$ will not change the success probability. 
\item $\high \leq b < \cells-1$. Canceling $p$ and subtracting ${\binom{b}{\high}}/{\binom{\cells}{\high}}$ and ${\binom{b}{\low+1}}/{\binom{\cells}{\low+1}}$ from both sides of \eqref{eq:first_ineq} shows that the strict inequality holds if and only if
\begin{equation}
\label{eq:intermediate}
\begin{split}
\frac{b\cdots(b-\low+1)}{\cells\cdots(\cells-\low+1)}-\frac{b\cdots(b-\low)}{\cells\cdots(\cells-\low)}>\frac{b\cdots(b-\high+2)}{\cells\cdots(\cells-\high+2)}-\frac{b\cdots(b-\high+1)}{\cells\cdots(\cells-\high+1)} \blank. 
\end{split}
\end{equation}
Factoring out $\frac{b\cdots(b-\low+1)}{\cells\cdots(\cells-\low+1)}$ on the left side and $\frac{b\cdots(b-\high+2)}{\cells\cdots(\cells-\high+2)}$ on the right side gives
\begin{equation}
 \frac{b\cdots (b-\low+1)}{\cells\cdots (\cells-\low)}>\frac{b\cdots (b-\high+2)}{\cells\cdots(\cells-\high+1)}
 \Leftrightarrow
 \frac{\cells\cdots(\cells-\high+1)}{\cells\cdots (\cells-\low)}>\frac{b\cdots (b-\high+2)}{b\cdots (b-\low+1)}\blank.
\end{equation}
Since $\low \leq \high -2$, this is equivalent to 
\begin{equation}
(\cells-\low+1)\cdot(\cells-\low)\cdots(\cells-\high+1)>(b-\low)\cdot(b-\low-1)\cdots(b-\high+2)\blank,
\end{equation}
which is true for $m-1>b$.
\item $\low\leq b<\high$.  Calculations along the lines of case 2 show that the strict inequality of \eqref{eq:first_ineq} also holds in this case. Note that $\binom{b}{k}$, $\binom{b}{k-1}$ and $\binom{b}{l+1}$ can be zero.
\item $0\leq b<\low$. In this case both sides of \eqref{eq:first_ineq} are zero, i.\,e., the modifications we do to $\pmf_z$ will not change the success probability.
\end{compactenum}

Since in cases 1 and 4 above there was no change in the success probability, to show
that \eqref{eq:success_prob} cannot be maximal when $\high-\low\geq2$ as we are considering, 
it remains to show that at least one of the Cases 2 and 3 occurs with positive probability.
We construct a situation in which one of these cases applies, and which occurs with positive
probability. 

Choose degrees $\randv_x$ for all elements $x \in S-\{z\}$ such that $\randv_x\leq \mean_x$ and $\pmf_x(\randv_x)>0$.
Consider a permutation of the elements $x \in S-\{z\}$ such that these degrees are ordered, i.\,e., $\randv_{x_1}\leq \randv_{x_2}\leq \ldots \leq \randv_{x_{\keys-1}}$. Choose the first element $x_i$ with  $i \geq \low$ and $\randv_{x_i}\leq i$. Such an element must exist, since we assume $\randv_x\leq \mean_x\leq \keys-2$, in particular we have $l<\keys-2$. Arrange that $A_{x_j}\subseteq [i],1\leq j \leq i,$ such that there is a matching in $G_{\{x_1,\ldots,x_i\}}$. This implies $b\geq \low$. Then arrange that $|A_{x_j}-\bigcup_{1\leq {j'}<j} A_{x_{j'}}|=1$, for all $i<j\leq n-2$, as well as $|A_{x_{n-1}}-\bigcup_{1\leq {j'}<n-1} A_{x_{j'}}|=2$, which implies $b<m-1$. This finishes the proof of Proposition \ref{prop:constant_expected_degree}. 
\end{proof}

\section{Performance results for the generalized selfless algorithm}

We present some performance results for the generalized selfless algorithm
We ran the generalized selfless algorithm for hypergraphs with $10^5$ and $10^6$ nodes
and tabulated the failure rate around the theoretical threshold values $c_{k,2}$ for $k=3,4,5$. For each pair $(\nodes,k)$ we considered $81$ edge densities $c=\frac{\edges}{\nodes}$, spaced apart by $0.0001$, 
thus covering an interval of length $0.008$, which encloses the theoretical threshold value 
for the particular parameter pair $(\nodes,k)$.
The hyperedges of the hypergraphs were randomly chosen via pseudo random number generator MT19937 ``Mersenne Twister'' of the GNU Scientific Library~\cite{GNU_Scientific}. We measured the average failure rate of the algorithm over $100$ random hypergraphs for each combination $(\nodes,\edges,k)$ within the parameter space. 
To get an estimation of the threshold, i.\,e., the rate $c$ where the algorithm switches from success to failure, we fit the sigmoid function
\begin{equation}
\label{eq:fit_function}
 \sigma(c;a,b)=\frac{1}{1+e^{-(c-a)/b}}
\end{equation}
to the measured failure rate (via gnuplot\footnote{gnuplot, an interactive plotting program, version 4.2, \url{http://www.gnuplot.info}}), using the method of least squares. We determined the parameters $a,b$ that lead to a (local) minimum of the sum of squares of the $81$ residuals, denoted by $\sum_{res}$. The parameter $a$ is the inflection point of \eqref{eq:fit_function} and therefore the approximation of the threshold of the generalized selfless algorithm. Figures \ref{fig:gen_selfless_edge_size_3},
\ref{fig:gen_selfless_edge_size_4} and \ref{fig:gen_selfless_edge_size_5} show the results of the experiments.

\begin{figure}[!ht]
\centering
 \subfigure[$\nodes=10^5$]{\scalebox{\imgScale}{\input{\figPath/xGS_n=10E5_d=3_k=1.tex}}}
 \subfigure[$\nodes=10^6$]{\scalebox{\imgScale}{\input{\figPath/xGS_n=10E6_d=3_k=1.tex}}}
 \inFigSpace
 \caption{\label{fig:gen_selfless_edge_size_3}edge size $k=3$; theoretical threshold $c_{k,2} \approx 0.91794$}
 \outFigSpace
\end{figure}

\begin{figure}[!ht]
\centering
 \subfigure[$\nodes=10^5$]{\scalebox{\imgScale}{\input{\figPath/xkthulhu_MTW_n=10E5_d=4_k=1.tex}}}
 \subfigure[$\nodes=10^6$]{\scalebox{\imgScale}{\input{\figPath/xkthulhu_MTW_n=10E6_d=4_k=1.tex}}}
 \inFigSpace
 \caption{\label{fig:gen_selfless_edge_size_4}edge size $k=4$; theoretical threshold $c_{k,2} \approx 0.97677$}
 \outFigSpace
\end{figure}
\begin{figure}[!ht]
\centering
 \subfigure[$\nodes=10^5$]{\scalebox{\imgScale}{\input{\figPath/xkthulhu_MTW_n=10E5_d=5_k=1.tex}}}
 \subfigure[$\nodes=10^6$]{\scalebox{\imgScale}{\input{\figPath/xkthulhu_MTW_n=10E6_d=5_k=1.tex}}}
 \inFigSpace
 \caption{\label{fig:gen_selfless_edge_size_5}edge size $k=5$; theoretical threshold $c_{k,2} \approx 0.99244$}
 \outFigSpace
\end{figure}
\begin{figure}[!ht]
\centering
 \subfigure[$\nodes=10^5$]{\scalebox{\imgScale}{\input{\figPath/xGSvsHC_n=10E5_d=3_k=1.tex}}}
 \subfigure[$\nodes=10^6$]{\scalebox{\imgScale}{\input{\figPath/xGSvsHC_n=10E6_d=3_k=1.tex}}}
 \inFigSpace
 \caption{\label{fig:selfless_vs_perfect_matching}$k=3$; theoretical threshold $c_{k,2} \approx 0.91794$, interval size $0.004$}
 \outFigSpace
\end{figure}
\begin{figure}[!ht]
\centering
 \subfigure[$\nodes=10^5$]{\scalebox{\imgScale}{\input{\figPath/xkthulhu_MTW_n=10E5_d=3_k=2.tex}}}
 \subfigure[$\nodes=10^6$]{\scalebox{\imgScale}{\input{\figPath/xkthulhu_MTW_n=10E6_d=3_k=2.tex}}}
 \inFigSpace
 \caption{\label{fig:gen_selfless_edge_size_3_bucket_size=2}$k=3$, $\ell=2$; 
 \emph{conjectured} threshold value $c_{k,\ell+1} \approx 1.97640$}
 \outFigSpace
\end{figure}

One observes that this simple algorithm is able to construct the placements for edge densities quite close to the calculated thresholds $c_{k,2}$. The slope of the sigmoid curve increases and $\sum_{res}$ decreases with growing $\nodes$ and $k$, leading to a sharp transition from total success to total failure.
Clearly the algorithm can fail on hypergraphs that admit a matching. Experimental comparisons with a perfect matching algorithm \cite{HK:1973} showed that this is very unlikely for random hypergraphs. An example is given in Figure \ref{fig:selfless_vs_perfect_matching}, which shows the failure rate of perfect matching in comparison to the generalized selfless algorithm. Note that the plot shows an interval of size $0.004$, i.\,e., 41 data points instead of $81$. The differences in the failure rates of the algorithms become very small as~$\nodes$ grows.

Similarly, we find our generalized algorithm for the case where the bucket size $\ell$
is greater than 1 has similar behavior.   For an example see Figure~\ref{fig:gen_selfless_edge_size_3_bucket_size=2}.

\newpage

\section{Proof of the threshold for $k$-XORSAT}

In this section we give a full proof of the
threshold for $k$-XORSAT (Corollary \ref{corthresh}).
The proof employs the notation and facts developed in Sections \ref{sect:facts:cores} and \ref{sec:main},
especially Propositions \ref{prop:two} and \ref{prop:three}, and the following fact
(known as ``Friedgut's Theorem'' for $k$-XORSAT \cite{CD03,CD09}).
\begin{fact}
\label{fact:Friedgut}
For every $k\geq 3$ there exists a function $c_k(m)\leq 1$ such that, for every $\ve>0$ and a
random formula $F$ (system $Ax=b$ of equations) from $\Phi^k_{m,n}$ we have the following:
\begin{displaymath}
 \lim_{m\to\infty} \Pr[\text{$F$ is satisfiable}]=
\begin{cases}
1, \text{ if } n=c_k(m)(1-\ve)m\\
0, \text{ if } n=c_k(m)(1+\ve)m \blank.\\
\end{cases}
\end{displaymath}
\end{fact}
Recall from Section \ref{sec:main} that $\Phi^k_{m,n}$ can be regarded as a probability space
whose elements are pairs $(A,b)$ where $A$ is an $n\times m$ matrix with entries in $\{0,1\}$,
each row containing $k$ 1's, and $b\in\{0,1\}^n$.
Alternatively, $A$ can be regarded as a node-edge incidence matrix $A_G$ of a $k$-uniform hypergraph $G\in \graphSet$.
Via the obvious correspondence we identify $\graphSet$ with the set of bipartite graphs $G$ with $n$ left nodes (``check nodes'') and $m$ right nodes (``variable nodes'') and degree $k$ at each left node. Similarly, $\Psi^k_{\hat{m},\hat{n}}$ is the probability space whose elements are pairs $(\hat{A},\hat{b})$, $\hat{b}\in\{0,1\}^{\hat{n}}$, where $\hat{A}$ is either the incidence matrix of a $k$-uniform hypergraph $H$ with $\hat{m}$ nodes, $\hat{n}$ edges, and minimum degree 2 or the adjacency matrix $\hat{A}_H$ of a bipartite graph $H$ with $\hat{n}$ left nodes and $\hat{m}$ right nodes, with degree $k$ at each left node and minimum degree 2 at each right node. We use the same notation for both and let $\cGraphSetH$ be the set of all these graphs.\footnote{For simplicity we assume that for each left node a sequence of $k$ right nodes is chosen at random, allowing and ignoring repetitions. The difference from  $k$-uniform hypergraphs is negligible.}
The following lemma is central.

\begin{lemma}\label{lemma:Z2}
For any $\delta>0$ there exists $\ve = \ve(\delta)>0$ such that the 
following happens.
Let $H\in\cGraphSetH$ be uniformly random with
$\hat{n}<\hat{m}(1-\delta)$
and denote by $Z_{H}$ the number of solutions of the linear system
$\hat{A}_Hx = 0$ (over $\mathrm{GF}[2]$). Then
\begin{eqnarray}
\Pr[Z_{H} = 2^{\hat{m}-\hat{n}}]\ge \ve(\delta)>0 \, .
\end{eqnarray}
\end{lemma}
We note that a full proof of this lemma for the special case $k=3$, with $\Pr[Z_{H} = 2^{\hat{m}-\hat{n}}]=1-o(1)$, was given in~\cite{DubMan:2002}.
\begin{proof}[of Corollary \ref{corthresh} ({\normalfont assuming Lemma \ref{lemma:Z2}})]
 Consider the following two cases.
\begin{compactenum}[(i)]
\item $c^*_{k,2}<c<c_{k,2}$. Let a system $A_Gx=b$ be chosen at random from $\Phi^k_{m,n}$. Reducing $G$ to its 2-core $H$ leads to a system $\hat{A}_Hx=\hat{b}$ with $\hat{m}$ variables, $\hat{n}$ equations, and $\rank(\hat{A}_H)=\rank(A_G)-(m-\hat{m})$. The graph $H$ is random in $\cGraphSetH$.
By Propositions \ref{prop:two} and \ref{prop:three}, with high probability $\hat{n}\leq (1-\delta)\hat{m}$ for some $\delta=\delta(c)$, and $\hat{m}=\Theta(m)$. By Lemma \ref{lemma:Z2}, for $m$ large enough, we get $\Pr[Z_{H} = 2^{\hat{m}-\hat{n}}]\geq \ve(\delta)>0$.
This implies $\Pr[A_Gx=b\text{ is satisfiable}] \geq \Pr[A_G \text{ has full row rank}]=\Pr[Z_G=2^{m-n}]\geq \ve(\delta)$.

\item $c>c_{k,2}$. Let $A_Gx=b$ and its reduced version $\hat{A}_Hx=\hat{b}$ be as in (i).
By Propositions \ref{prop:two} and \ref{prop:three}, with high probability $\hat{n}\geq (1+\delta)\hat{m}$ for some $\delta=\delta(c)$, and $\hat{m}=\Theta(m)$. We have $\rank(\hat{A})\leq \hat{m}$, and by the randomness of $\hat{b}$ we have $\Pr[\hat{A}_Hx=\hat{b}\text{ is satisfiable}]\leq2^{\hat{m}-\hat{n}}\leq2^{-\delta\hat{m}}$.
\end{compactenum}
Combining parts (i) and (ii) with Friedgut's Theorem (Fact \ref{fact:Friedgut}) shows that $\lim_{m\to\infty} c_k(m)=c_{k,2}$, which implies Corollary \ref{corthresh}.
\end{proof}

We now move to the proof of Lemma \ref{lemma:Z2}, which focuses on the
2-core $H$ of the graph $G$, and we condition on its number of nodes. With
a slight abuse of notation we will drop the ``hat'' from our
notations. In other words, we now let $H$ be a uniformly random graph
from $\cGraphSet$ and let $\gamma = n / m$ (see Theorem \ref{thresh}).

It is convenient to introduce some additional notation. Given a
formal series $p(z)$, $\coeff[p(z),z^r]$ denotes the coefficient of $z^r$
in $p(z)$. We further introduce the notations 
\begin{align}
&q(z) = (e^{z}-1-z)\, ,\;\;\;\;\;\;\;\;\;\;\; Q(z) = \frac{z q'(z)}{q(z)}\, ,\\
&p_k(z) = \frac{1}{2}(1+z)^k+
\frac{1}{2}(1-z)^k\, ,\;\;\;\;\;\;\; P_k(z) = \frac{z p_k'(z)}{p_k(z)}\, .
\end{align}
It is easy to see that $z\mapsto Q(z)$ is a strictly increasing function
with $\lim_{z\to 0}Q(z)=2$, and $\lim_{z\to\infty}Q(z) = \infty$. 
Further $z\mapsto P_k(z)$ is strictly increasing with 
$\lim_{z\to 0}P_k(z)=0$, and $\lim_{z\to\infty}P_k(z) = k$ for $k$ even,
and $k-1$ otherwise.

Further we define the domain sets

\begin{eqnarray}
\domain&= &\big\{(w,l)\in\integers^2\,:\, 
0\le w\le m\, , 2w\le l\le kn-2(m-w)\, , l\mbox{ even }\big\}\, , \\
\domainGam(\ve)&= &\big\{(\omega,\lambda)\in\reals^2\,:\, 
\ve\le \omega\le 1-\ve\, , 
\tfrac{2\omega}{k\gamma}+\ve
\le \lambda\le 1-\tfrac{2(1-\omega)}{k\gamma}-\ve
\big\}\, ,\\
\domain(\ve)&= &\big\{(w,l)\in\domain\,:\, 
(\tfrac{w}{m},\tfrac{l}{kn})\in \domainGam(\ve)\big\}\, ,\\
\domainCompl(\ve)&= & \domain\setminus \domain(\ve)\, .
\end{eqnarray}

The assertion of Lemma \ref{lemma:Z2} now follows from the following
sequence of lemmas, to be proven in the subsections below.
\begin{lemma}\label{lemma:Comb}
Let $Z_{H}$ be the number of solutions of the linear system $A_Hx=0$.
Then
\begin{align}
\E[Z_H] & =  \frac{1}{N_0}\sum_{(w,l)\in\domain} N(w,l)\, ,
\end{align}
where we define
\begin{align}
N_0 & =  (kn)!\,\coeff[(e^z-1-z)^m,z^{kn}]\, ,\\
N(w,l) & =  \binom{m}{w}\, l!(kn-l)!\,
\coeff[(e^z-1-z)^w,z^l]\,\coeff[(e^z-1-z)^{m-w},z^{kn-l}]\,
\coeff[p_k(z)^n,z^l]\, .
\end{align}
\end{lemma}

\begin{lemma}\label{lemma:Boundary}
For any $\delta>0$ there exists $\ve>0$ such that, if $n\le m(1-\delta)$,
then
\begin{align}
\frac{1}{N_0}\cdot\sum_{(w,l)\in\domainCompl(\ve)}\! \! N(w,l)\le 2^{m\delta}\, .
\end{align}
\end{lemma}

\begin{lemma}\label{lemma:Exp}
For any $\delta>0, \ve>0$ there exists $C= C(\delta,\ve)$
such that, if $m\delta\le n\le m(1-\delta)$ and $(w,l)\in\domain(\ve)$,
then 
\begin{align}
\frac{N(w,l)}{N_0}\le \frac{C}{m}\, \exp\Big(m\,\psi\big(\tfrac{w}{m},\tfrac{l}{kn}\big)
\Big)\, ,
\end{align}
where, letting $h(z) = -z\log z-(1-z)\log(1-z)$,\footnote{$\log$ means logarithm to the base $e$} we define
\begin{align}
\psi(\omega,\lambda)  &=  h(\omega)-k\gamma\, h(\lambda)-
\log q(s)+ k\gamma\log s\label{eq:Phi}\\
&+\omega \log q(a)-k\gamma\lambda \log a+(1-\omega)\log q(b)-k\gamma(1-\lambda)
\log b\nonumber\\
& +\gamma\log p_k(c)-k\gamma\lambda \log c\, .\nonumber
\end{align}
Finally, $a=a(\omega,\lambda)$, $b = b(\omega,\lambda)$,
$c = c(\omega,\lambda)$, and $s$ are the unique non-negative solutions
of
\begin{eqnarray}
Q(s) = k\gamma\, , \;\;\;\;\; Q(a) = \frac{k\gamma\lambda}{\omega}\, ,
\;\;\;\;\;Q(b) = \frac{k\gamma(1-\lambda)}{(1-\omega)}\, ,
\label{Qeq}\\
P_k(c) = k\lambda\, .\label{Peq}
\end{eqnarray}
\end{lemma}

\begin{lemma}\label{lemma:Calculus}
For any $\gamma<1$, the function $\psi: \domainGam(0)\to\reals$ achieves its
unique global maximum at $(\omega,\lambda) =(1/2,1/2)$,
with $\psi(1/2,1/2) = (1-\gamma)\log 2$.

Further, there exists $\xi>0$ such that 
$-\Hess_{\psi}(1/2,1/2)\succeq\, \xi\, I_2$.\footnote{$\Hess_\psi$ denotes the Hessian matrix of $\psi$ and $I_2$ the $2\times 2$ unit matrix}
\end{lemma}

Finally, let us recall a well known fact about lattice sums
(see for instance \cite{BR}).
\begin{lemma}\label{lemma:Sum}
Let $D$ be an open domain in $\reals^d$, and $F: D\to\reals$
be continuously differentiable, achieving its unique maximum in
$z_*\in D$, with $\Hess_F(z_*)\succeq \xi I_{d}$ for some 
$\xi>0$.
Then there exists $C>0$ such that, for any $\delta\ge 0$
\begin{eqnarray}
\sum_{x\in \integers^d\, :\; x\,\delta \in D} \exp\Big(
\tfrac{1}{\delta}F(\delta x)\Big)
\le \frac{C}{\delta^{d/2}}\, \exp\Big(\tfrac{1}{\delta}F(z_*)\Big)\, .
\end{eqnarray}
\end{lemma}

\begin{proof}[of Lemma \ref{lemma:Z2}]
The proof is simply obtained by putting together
Lemmas \ref{lemma:Comb}, \ref{lemma:Boundary}, \ref{lemma:Exp},
\ref{lemma:Calculus}
and using Lemma \ref{lemma:Sum} (with $F(x_1,x_2)=
\psi(x_1,2x_2/(k\gamma))$, $d=2$ and $\delta = 1/n$)
to bound the sum.
\end{proof}
%
%

\subsection{Proof of Lemma \ref{lemma:Comb}}

Clearly $N_0$ is the number of graphs in 
$\cGraphSet$. Indeed it is the number of way of putting 
$nk$ distinct balls in $m$ bins in such a way that 
each bin contains at least $2$ balls.

The claim follows by proving that, for each 
$(w,l)\in\domain$, $N(w,l)$ is the number of couples 
$(H,x)$ where $H\in \cGraphSet$ and $x\in\{0,1\}^m$
with $A_Hx = 0 \mod 2$, such that $x$ has $w$ ones
and $H$ has $l$ edges incident on variable (right) nodes $i$
such that $x_i=1$. Indeed, $\binom{m}{w}$ gives the number
of ways of choosing the ones. Paint by red the $l$ edges 
incident on these nodes, and by blue the other $(kn-l)$ edges.
The coefficient factors give the number of ways of attributing
red/blue edges to nodes on the two sides. The factorials give the number 
of ways of matching edges of the same color on the two sides.
\qed
%
%
\subsection{Proof of Lemma \ref{lemma:Exp}}

Let us start by proving a lower bound on $N_0$. For any $s>0$,
we have
\begin{eqnarray}
N_0 = (kn)! \frac{q(s)^m}{s^{kn}}\, \Pr_{s}\Big[\sum_{i=1}^mX_i =kn\Big]
\, , 
\end{eqnarray}
where $X_1,\dots, X_m$ are i.i.d. Poisson random variable (with parameter $s$) conditioned to 
$X_i\ge 2$, i.e., for any $q\ge 2$,
\begin{eqnarray}
\Pr_s[X_i = q] = \frac{1}{e^s-1-s}\, \frac{s^q}{q!}\, .
\end{eqnarray}
By assumption $s$ is chosen such that
$\E_s[X_i]=Q(s)=k\gamma\in (k\delta,k(1-\delta))$. By the local central 
limit theorem  for lattice random variables of 
\cite[Corollary 22.3]{BR}, we have 
$\Pr_{s}\big[\sum_{i=1}^mX_i =kn\big] \ge C'/\sqrt{m}$ for sone constant 
$C'(\delta)$, whence, using Stirling's formula 
\begin{eqnarray}
N_0 \ge C_1(\delta) \left(\frac{kn}{e}\right)^{kn} 
\frac{q(s)^m}{s^{kn}}\, .\label{eq:BoundN0}
\end{eqnarray}

Consider now $N(w,l)$. By the central limit theorem for the sum of 
Bernoulli random variables, for any $m\delta\le w\le m(1-\delta)$, we have
\begin{eqnarray}
\binom{m}{w}\le \frac{C_2(\delta)}{\sqrt{m}}\, e^{mh(w/m)}\, .
\end{eqnarray}
Treating the coefficient terms as above,
and using Stirling's formula for $l,(kn-l)=\Theta(m)$, we get
\begin{eqnarray}
N(w,l) \le \frac{C_3(\delta)}{m}\,e^{mh(w/m)}\, \left(\frac{l}{e}\right)^l
 \left(\frac{knl}{e}\right)^{(kn-l)}\frac{q(a)^w}{a^l}\, 
\frac{q(b)^{m-w}}{b^{kn-l}}\, \frac{p_k(c)^n}{c^l}\, .\label{eq:BoundNwl}
\end{eqnarray}
The claim is proved by taking the ratio of the bounds 
(\ref{eq:BoundNwl}) and (\ref{eq:BoundN0}).
\qed
%
%
\subsection{Proof of Lemma \ref{lemma:Calculus}, outline}

We now present an outline of the proof of Lemma~\ref{lemma:Calculus}. Appendix~\ref{app:fullproof5} contains the additional details for a complete proof.

For $(\omega,\lambda)=(1/2,1/2)$, Eqs.~(\ref{Qeq}), (\ref{Peq}) 
admit the unique solution $a=b=s$ and $c=1$. A straightforward calculation
yields $\psi(1/2,1/2) = (1-\gamma)\log 2$.

Call $\Psi(\omega,\lambda;a,b,c)$ the right hand side of
Eq.~(\ref{eq:Phi}). Notice that the derivatives of $\Psi$
with respect to $a,b,c$ vanish by 
Eqs.~(\ref{Qeq}), (\ref{Peq}). Therefore it is easy to compute the 
partial derivatives
\begin{eqnarray}
\frac{\partial \psi}{\partial \omega} & = & \log \frac{1-\omega}{\omega}
+\log\frac{q(a)}{q(b)}\, ,\\
\frac{\partial \psi}{\partial \lambda} & = & -k\gamma
\log \frac{1-\lambda}{\lambda}-k\gamma\log\frac{a}{b}-k\gamma\log c\, .
\end{eqnarray}
Using the fact that $a=b=s$ and $c=1$ at $(\omega,\lambda)=(1/2,1/2)$,
we get that the gradient of $\psi$ vanishes at $(1/2,1/2)$,
and again, $\psi(1/2,1/2) = (1-\gamma)\log 2$. 

By a somewhat longer calculation, we obtain the following 
second derivatives
\begin{eqnarray}
\left.\frac{\partial^2 \psi}{\partial \omega^2}
\right|_{1/2,1/2} & = & -4\,\Big(1+\frac{(k\gamma)^2}{s^2C}\Big)\, ,\\
\left.
\frac{\partial^2 \psi}{\partial \lambda\partial\omega}\right|_{1/2,1/2} & = & 
4\, \frac{(k\gamma)^2}{s^2C}\, , \\
\left.\frac{\partial^2 \psi}{\partial \omega^2}
\right|_{1/2,1/2} & = & -4\, \frac{(k\gamma)^2}{s^2C}\, ,
\end{eqnarray}
with
\begin{eqnarray}
C= \frac{q''(s)}{q(s)}-\frac{q'(s)^2}{q(s)^2}+\frac{k\gamma}{s^2}>0\, .
\end{eqnarray}
It is easy to deduce that $-\Hess_\psi(1/2,1/2)$ is positive definite.

The function $\psi:\domainGam(0)\to\reals$ is continuous in   
$\domainGam(0)$ and differentiable in its interior. Further,
we have the following asymptotic behaviors (first two at fixed $\lambda$,
second two at fixed $\omega$):
\begin{align}
\lim_{\omega\to 0}&\frac{\partial\psi}{\partial\omega}=+ \infty\, ,
\;\;\;\;\;\;\;
\;\;\;\;\;\;\;
\;\;\;\;\;\;\;
\;\;\;\;\;\;
\lim_{\omega\to 1}\frac{\partial\psi}{\partial\omega}=- \infty\, ,\\
\lim_{\lambda\to 2\omega/(k\gamma)}
&\frac{\partial\psi}{\partial\lambda}=+ \infty\, ,
\;\;\;\;\;\;\;\;\;\;
\lim_{\lambda\to 1-2(1-\omega)/(k\gamma)}
\frac{\partial\psi}{\partial\lambda}=- \infty\, .
\end{align}
Therefore any global maximum of $\psi$ must be a stationary point in the 
interior of $\domainGam(0)$. We next will prove that 
$(1/2,1/2)$ is the only such point.

Notice that  $\Psi(\omega,\lambda;a,b,c)$ is convex with respect to
$a,b,c$. As a consequence 
\begin{eqnarray}
\psi(\omega,\lambda)=\min_{a,b,c}\Psi(\omega,\lambda;a,b,c)\, .
\label{eq:Variational}
\end{eqnarray}
We will construct an upper bound on $\psi$ by choosing $a,b,c$
appropriately. 
The first remark is that 
\begin{eqnarray}
\Psi(1-\omega,1-\lambda;b,a,1/c) = \Psi(\omega,\lambda;a,b,c)
-\gamma\, \log \frac{p_k(c)}{c^{k}p_k(1/c)} \, .
\end{eqnarray}
Since, for $c\in[0,1]$ (which is guaranteed by Eq.~(\ref{Peq})
for $\lambda\in [0,1/2]$) we have $p_k(c)\ge c^{k}p_k(1/c)$,
we can restrict without loss of generality to $\lambda\le 1/2$
(whence $c\in [0,1]$).

Next notice that, maximizing $\Psi$ over $\omega$, we get 
$\Psi(\omega,\lambda;a,b,c)\le \Psi_1(\lambda;a,b,c)$,
where
\begin{align}
\Psi_1(\lambda;a,b,c)  = & \log\big(q(a)+q(b)\big)-k\gamma\, h(\lambda)-
\log q(s)+ k\gamma\log s\\
&-k\gamma\lambda \log a-k\gamma(1-\lambda)
\log b
 +\gamma\log p_k(c)-k\gamma\lambda \log c\, .\nonumber
\end{align}

Next fix $c = c(\lambda) = b\lambda/(a-a\lambda)$. Since
this transformation is invertible, we can as well keep $c$
as a free parameter, and let $\lambda =ac/(ac+b)$.
If we let $\Psi_2(a,b,c) = \Psi_1(ac/(ac+b);a,b,c)$,
we get
\begin{eqnarray}
\Psi_2(a,b,c) =  & \log\big(q(a)+q(b)\big)-
\log q(s)+\gamma\log p_k(c)\\
&-k\gamma\log(ac+b)+k\gamma\lambda \log s\, .\nonumber
\end{eqnarray}
Also, without loss of generality, we can rescale $a$ by a factor $s$,
and set $b=s$, therefore defining $\Psi_3(a,c)=\Psi_2(sa,s,c)$.
If we introduce the notation
\begin{eqnarray}
\Lambda_s(x)=\frac{q(sx)}{q(s)} = \frac{e^{sx}-1-sx}{e^{s}-1-s}\,,
\end{eqnarray}
we get the expression
\begin{eqnarray}
\Psi_3(a,c) = -k\gamma \log(1+ac) 
+\log\big(1+\Lambda_s(a)\big)+\gamma\log p_k(c)\, .
\end{eqnarray}
By the above derivation we have the following
relation with $\psi(\omega,\lambda)$:
\begin{align}
&\psi(\omega,\lambda)\le \left.\Psi_4(c)\right|_{c=\lambda/a_*(c)(1-\lambda)}
\, ,\\
&\Psi_4(c) = \Psi_3(a_*(c),c)\, ,\;\;\;\;\;\;\;
a_*(c) = \arg\min_{a\ge 0}\Psi_3(a,c)\, .
\end{align}
A direct calculation shows that $a_*(1) = 1$
and $\Psi_3(1,1) = (1-\gamma)\log 2$. This point
corresponds to $(\omega,\lambda)=(1/2,1/2)$ through the above derivation.
We will show that $c=1$ is indeed the global maximum of $\Psi_4(c)$
for $c\in [0,1]$, which implies the assertion.

Maximizing $\Psi_3(a,c)$ with respect to $c$ implies $a_*(c)$ to be
the unique non-negative solution of the stationarity condition
\begin{eqnarray}
a = \frac{(1+c)^{k-1}-(1-c)^{k-1}}{(1+c)^{k-1}+(1-c)^{k-1}}\, .
\label{Beq}
\end{eqnarray}
On the other hand, the stationarity condition with respect
to $a$ yields
\begin{eqnarray}
c = \frac{\lambda_s(a)}{1+\Lambda_s(a)-a\lambda_s(a)}\, .
\label{Ceq}
\end{eqnarray}
where we used the fact that $\Lambda_s'(1) = k\gamma$
and defined $\lambda_s(x) = \Lambda_s'(x)/\Lambda_s'(1)$.

Equations (\ref{Beq}) and (\ref{Ceq}) admit the solutions
$a=c=0$ and $a=c=1$, and is easy to check that these are
both local maxima of $\Psi_4$. We will show that they admit only one
more solution with $c\in(0,1)$, that necessarily is a local minimum of
$\Psi_4$. Indeed, if we let $a=\tanh x$, $c= \tanh y$, Eq.~(\ref{Beq})
becomes
\begin{eqnarray}
x = (k-1) y\, .
\end{eqnarray}
Our claim is therefore implied by Lemma \ref{lemma:Elementary}
below.
\qed

\begin{lemma}\label{lemma:Elementary}
For $s> 0$, let
\begin{eqnarray}
\Lambda_s(t) = \frac{e^{st}-1-st}{e^{s}-1-s}\, ,
\;\;\;\;\;\;\;
\lambda_s(t) = \frac{e^{st}-1}{e^{s}-1}\, .
\end{eqnarray}
Define $F_s:\reals\to\reals$ by
\begin{eqnarray}
F_s(x) = \atanh\big(f(\tanh x)\big)\, ,\;\;\;\;\;\;\;\;\;
f_s(t)= \frac{\lambda_s(t)}{1+\Lambda_s(t)-t\lambda_s(t)}\, .
\end{eqnarray}
Then $F_s$ is convex on $[0,\infty)$.
\end{lemma}
\begin{proof}
This can be seen simply by graphing $F_s(x)$, or by some calculus which
we omit.
\end{proof}%

%
%
\subsection{Proof of Lemma \ref{lemma:Boundary}}

The proof is analogous to the one of Lemma \ref{lemma:Exp}.
We have just to be careful to the values of $w,l$ near the boundary of
the domain $\domain$. Luckily we only need a loose upper bound.
Equation (\ref{eq:BoundN0}) remains true in the present case (as it
only hinges on $n=\Theta(m)$). On the other hand 
using   $\binom{m}{w}\le \exp(mh(w/m))$, $\coeff[f(x)^k,x^l]\le
f(a)^k/a^l$ and $m!\le \sqrt{2\pi}\, (m/e)^{m+1/2}$, we get
\begin{eqnarray}
N(w,l) \le 2\pi\, e^{-kn-1} e^{mh(w/m)}\, l^{l+1/2}
 (kn-l)^{(kn-l+1/2)}\frac{q(a)^w}{a^l}\, 
\frac{q(b)^{m-w}}{b^{kn-l}}\, \frac{p_k(c)^n}{c^l}\, .
\end{eqnarray}
for any $a,b,c>0$. Taking the ratio, and bounding polynomial factors 
$\sqrt{l(kn-l)}\le C m$
we get 
\begin{eqnarray}
\frac{N(w,l)}{N_0}\le C\, m\, \exp\big(m\psi(w/m,\lambda/kn)\big)\, ,
\end{eqnarray}
whence 
\begin{eqnarray}
\frac{1}{N_0}\sum_{(w,l)\in\domainCompl(\ve)} N(w,l)\le
Cm^3\, \exp\big(m\sup\{ \psi(\omega,\lambda):\, (\omega,\lambda)\in
\domainGam(0)\setminus \domainGam(\ve)\} \big)
\end{eqnarray}
with $\psi(\omega,\lambda)$ defined as in Eq.~(\ref{eq:Phi}).
Notice that $\psi:\domainGam(\delta)\to \reals$ is a continuous function.
It is therefore sufficient to show that it is strictly smaller than 
$(1-\gamma)\log 2$ on the boundaries of its domain.
This indeed follows from Lemma \ref{lemma:Calculus}.
\qed

\fi

\input{./endlichgsst.tex}

\end{document}

%% file: figures/xGS_n=10E5_d=3_k=1.tex
\begingroup
  \makeatletter
  \providecommand\color[2][]{%
    \GenericError{(gnuplot) \space\space\space\@spaces}{%
      Package color not loaded in conjunction with
      terminal option `colourtext'%
    }{See the gnuplot documentation for explanation.%
    }{Either use 'blacktext' in gnuplot or load the package
      color.sty in LaTeX.}%
    \renewcommand\color[2][]{}%
  }%
  \providecommand\includegraphics[2][]{%
    \GenericError{(gnuplot) \space\space\space\@spaces}{%
      Package graphicx or graphics not loaded%
    }{See the gnuplot documentation for explanation.%
    }{The gnuplot epslatex terminal needs graphicx.sty or graphics.sty.}%
    \renewcommand\includegraphics[2][]{}%
  }%
  \providecommand\rotatebox[2]{#2}%
  \@ifundefined{ifGPcolor}{%
    \newif\ifGPcolor
    \GPcolorfalse
  }{}%
  \@ifundefined{ifGPblacktext}{%
    \newif\ifGPblacktext
    \GPblacktexttrue
  }{}%
  \let\gplgaddtomacro\g@addto@macro
  \gdef\gplbacktext{}%
  \gdef\gplfronttext{}%
  \makeatother
  \ifGPblacktext
    \def\colorrgb#1{}%
    \def\colorgray#1{}%
  \else
    \ifGPcolor
      \def\colorrgb#1{\color[rgb]{#1}}%
      \def\colorgray#1{\color[gray]{#1}}%
      \expandafter\def\csname LTw\endcsname{\color{white}}%
      \expandafter\def\csname LTb\endcsname{\color{black}}%
      \expandafter\def\csname LTa\endcsname{\color{black}}%
      \expandafter\def\csname LT0\endcsname{\color[rgb]{1,0,0}}%
      \expandafter\def\csname LT1\endcsname{\color[rgb]{0,1,0}}%
      \expandafter\def\csname LT2\endcsname{\color[rgb]{0,0,1}}%
      \expandafter\def\csname LT3\endcsname{\color[rgb]{1,0,1}}%
      \expandafter\def\csname LT4\endcsname{\color[rgb]{0,1,1}}%
      \expandafter\def\csname LT5\endcsname{\color[rgb]{1,1,0}}%
      \expandafter\def\csname LT6\endcsname{\color[rgb]{0,0,0}}%
      \expandafter\def\csname LT7\endcsname{\color[rgb]{1,0.3,0}}%
      \expandafter\def\csname LT8\endcsname{\color[rgb]{0.5,0.5,0.5}}%
    \else
      \def\colorrgb#1{\color{black}}%
      \def\colorgray#1{\color[gray]{#1}}%
      \expandafter\def\csname LTw\endcsname{\color{white}}%
      \expandafter\def\csname LTb\endcsname{\color{black}}%
      \expandafter\def\csname LTa\endcsname{\color{black}}%
      \expandafter\def\csname LT0\endcsname{\color{black}}%
      \expandafter\def\csname LT1\endcsname{\color{black}}%
      \expandafter\def\csname LT2\endcsname{\color{black}}%
      \expandafter\def\csname LT3\endcsname{\color{black}}%
      \expandafter\def\csname LT4\endcsname{\color{black}}%
      \expandafter\def\csname LT5\endcsname{\color{black}}%
      \expandafter\def\csname LT6\endcsname{\color{black}}%
      \expandafter\def\csname LT7\endcsname{\color{black}}%
      \expandafter\def\csname LT8\endcsname{\color{black}}%
    \fi
  \fi
  \setlength{\unitlength}{0.0500bp}%
  \begin{picture}(7200.00,5040.00)%
    \gplgaddtomacro\gplbacktext{%
      \csname LTb\endcsname%
      \put(990,723){\makebox(0,0)[r]{\strut{} 0}}%
      \put(990,1091){\makebox(0,0)[r]{\strut{} 0.1}}%
      \put(990,1460){\makebox(0,0)[r]{\strut{} 0.2}}%
      \put(990,1828){\makebox(0,0)[r]{\strut{} 0.3}}%
      \put(990,2197){\makebox(0,0)[r]{\strut{} 0.4}}%
      \put(990,2565){\makebox(0,0)[r]{\strut{} 0.5}}%
      \put(990,2934){\makebox(0,0)[r]{\strut{} 0.6}}%
      \put(990,3302){\makebox(0,0)[r]{\strut{} 0.7}}%
      \put(990,3671){\makebox(0,0)[r]{\strut{} 0.8}}%
      \put(990,4039){\makebox(0,0)[r]{\strut{} 0.9}}%
      \put(990,4408){\makebox(0,0)[r]{\strut{} 1}}%
      \put(1256,440){\makebox(0,0){\strut{} 0.914}}%
      \put(1961,440){\makebox(0,0){\strut{} 0.915}}%
      \put(2666,440){\makebox(0,0){\strut{} 0.916}}%
      \put(3371,440){\makebox(0,0){\strut{} 0.917}}%
      \put(4076,440){\makebox(0,0){\strut{} 0.918}}%
      \put(4781,440){\makebox(0,0){\strut{} 0.919}}%
      \put(5486,440){\makebox(0,0){\strut{} 0.92}}%
      \put(6191,440){\makebox(0,0){\strut{} 0.921}}%
      \put(220,2749){\rotatebox{90}{\makebox(0,0){\strut{}failure rate among $100$ attempts}}}%
      \put(4005,110){\makebox(0,0){\strut{}$\beta$}}%
      \put(4041,926){\makebox(0,0)[l]{\strut{}\labIncr$a=0.91785$}}%
      \put(4041,1331){\makebox(0,0)[l]{\strut{}\labIncr$\sum_{res}=0.050985$}}%
    }%
    \gplgaddtomacro\gplfronttext{%
      \csname LTb\endcsname%
      \put(2172,4603){\makebox(0,0)[l]{\strut{}measured data}}%
      \csname LTb\endcsname%
      \put(2172,4383){\makebox(0,0)[l]{\strut{}$(1+e^{-(\beta-a)/b)})^{-1}$}}%
    }%
    \gplbacktext
    \put(0,0){\includegraphics{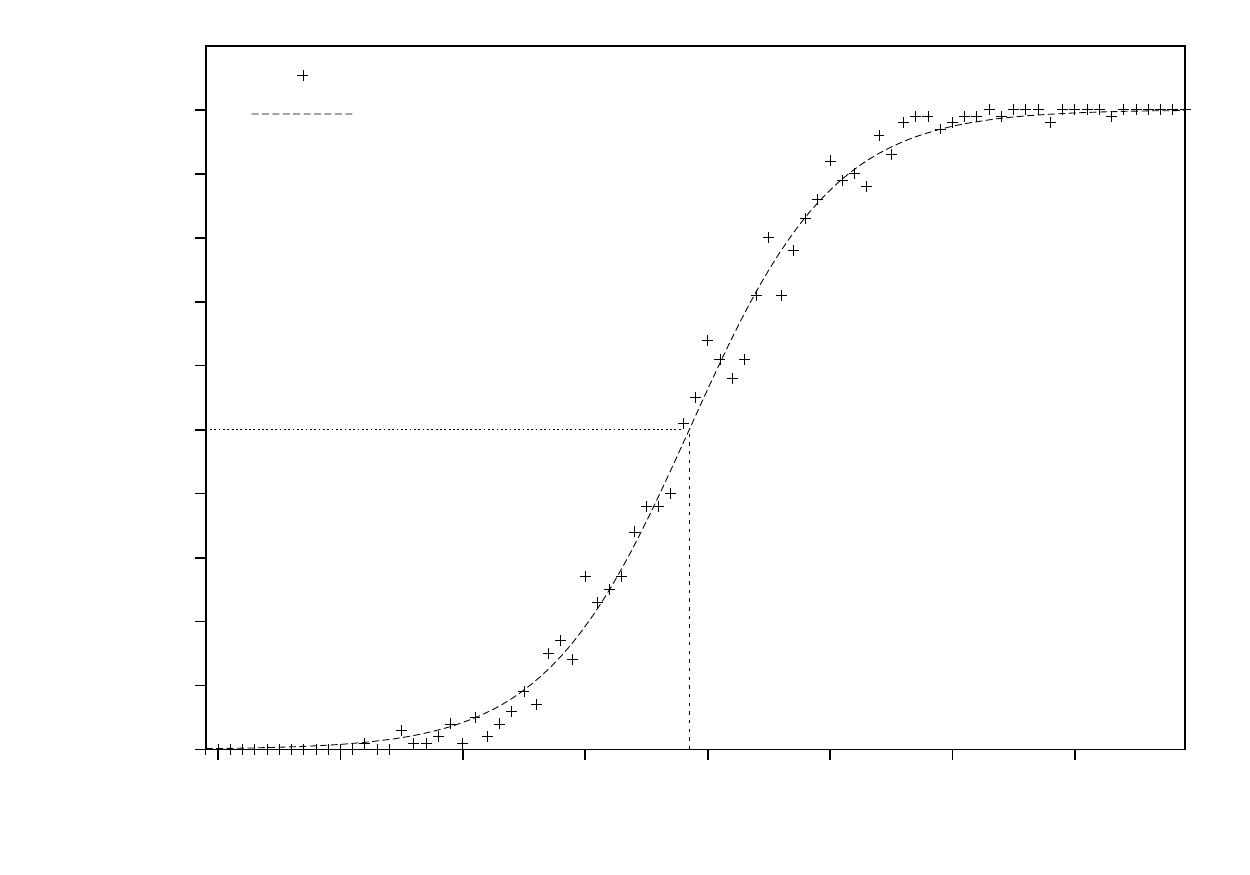}}%
    \gplfronttext
  \end{picture}%
\endgroup

%% file: figures/xGS_n=10E6_d=3_k=1.tex
\begingroup
  \makeatletter
  \providecommand\color[2][]{%
    \GenericError{(gnuplot) \space\space\space\@spaces}{%
      Package color not loaded in conjunction with
      terminal option `colourtext'%
    }{See the gnuplot documentation for explanation.%
    }{Either use 'blacktext' in gnuplot or load the package
      color.sty in LaTeX.}%
    \renewcommand\color[2][]{}%
  }%
  \providecommand\includegraphics[2][]{%
    \GenericError{(gnuplot) \space\space\space\@spaces}{%
      Package graphicx or graphics not loaded%
    }{See the gnuplot documentation for explanation.%
    }{The gnuplot epslatex terminal needs graphicx.sty or graphics.sty.}%
    \renewcommand\includegraphics[2][]{}%
  }%
  \providecommand\rotatebox[2]{#2}%
  \@ifundefined{ifGPcolor}{%
    \newif\ifGPcolor
    \GPcolorfalse
  }{}%
  \@ifundefined{ifGPblacktext}{%
    \newif\ifGPblacktext
    \GPblacktexttrue
  }{}%
  \let\gplgaddtomacro\g@addto@macro
  \gdef\gplbacktext{}%
  \gdef\gplfronttext{}%
  \makeatother
  \ifGPblacktext
    \def\colorrgb#1{}%
    \def\colorgray#1{}%
  \else
    \ifGPcolor
      \def\colorrgb#1{\color[rgb]{#1}}%
      \def\colorgray#1{\color[gray]{#1}}%
      \expandafter\def\csname LTw\endcsname{\color{white}}%
      \expandafter\def\csname LTb\endcsname{\color{black}}%
      \expandafter\def\csname LTa\endcsname{\color{black}}%
      \expandafter\def\csname LT0\endcsname{\color[rgb]{1,0,0}}%
      \expandafter\def\csname LT1\endcsname{\color[rgb]{0,1,0}}%
      \expandafter\def\csname LT2\endcsname{\color[rgb]{0,0,1}}%
      \expandafter\def\csname LT3\endcsname{\color[rgb]{1,0,1}}%
      \expandafter\def\csname LT4\endcsname{\color[rgb]{0,1,1}}%
      \expandafter\def\csname LT5\endcsname{\color[rgb]{1,1,0}}%
      \expandafter\def\csname LT6\endcsname{\color[rgb]{0,0,0}}%
      \expandafter\def\csname LT7\endcsname{\color[rgb]{1,0.3,0}}%
      \expandafter\def\csname LT8\endcsname{\color[rgb]{0.5,0.5,0.5}}%
    \else
      \def\colorrgb#1{\color{black}}%
      \def\colorgray#1{\color[gray]{#1}}%
      \expandafter\def\csname LTw\endcsname{\color{white}}%
      \expandafter\def\csname LTb\endcsname{\color{black}}%
      \expandafter\def\csname LTa\endcsname{\color{black}}%
      \expandafter\def\csname LT0\endcsname{\color{black}}%
      \expandafter\def\csname LT1\endcsname{\color{black}}%
      \expandafter\def\csname LT2\endcsname{\color{black}}%
      \expandafter\def\csname LT3\endcsname{\color{black}}%
      \expandafter\def\csname LT4\endcsname{\color{black}}%
      \expandafter\def\csname LT5\endcsname{\color{black}}%
      \expandafter\def\csname LT6\endcsname{\color{black}}%
      \expandafter\def\csname LT7\endcsname{\color{black}}%
      \expandafter\def\csname LT8\endcsname{\color{black}}%
    \fi
  \fi
  \setlength{\unitlength}{0.0500bp}%
  \begin{picture}(7200.00,5040.00)%
    \gplgaddtomacro\gplbacktext{%
      \csname LTb\endcsname%
      \put(990,723){\makebox(0,0)[r]{\strut{} 0}}%
      \put(990,1091){\makebox(0,0)[r]{\strut{} 0.1}}%
      \put(990,1460){\makebox(0,0)[r]{\strut{} 0.2}}%
      \put(990,1828){\makebox(0,0)[r]{\strut{} 0.3}}%
      \put(990,2197){\makebox(0,0)[r]{\strut{} 0.4}}%
      \put(990,2565){\makebox(0,0)[r]{\strut{} 0.5}}%
      \put(990,2934){\makebox(0,0)[r]{\strut{} 0.6}}%
      \put(990,3302){\makebox(0,0)[r]{\strut{} 0.7}}%
      \put(990,3671){\makebox(0,0)[r]{\strut{} 0.8}}%
      \put(990,4039){\makebox(0,0)[r]{\strut{} 0.9}}%
      \put(990,4408){\makebox(0,0)[r]{\strut{} 1}}%
      \put(1256,440){\makebox(0,0){\strut{} 0.914}}%
      \put(1961,440){\makebox(0,0){\strut{} 0.915}}%
      \put(2666,440){\makebox(0,0){\strut{} 0.916}}%
      \put(3371,440){\makebox(0,0){\strut{} 0.917}}%
      \put(4076,440){\makebox(0,0){\strut{} 0.918}}%
      \put(4781,440){\makebox(0,0){\strut{} 0.919}}%
      \put(5486,440){\makebox(0,0){\strut{} 0.92}}%
      \put(6191,440){\makebox(0,0){\strut{} 0.921}}%
      \put(220,2749){\rotatebox{90}{\makebox(0,0){\strut{}failure rate among $100$ attempts}}}%
      \put(4005,110){\makebox(0,0){\strut{}$\beta$}}%
      \put(4092,926){\makebox(0,0)[l]{\strut{}\labIncr$a=0.917923$}}%
      \put(4092,1331){\makebox(0,0)[l]{\strut{}\labIncr$\sum_{res}=0.0147659$}}%
    }%
    \gplgaddtomacro\gplfronttext{%
      \csname LTb\endcsname%
      \put(2172,4603){\makebox(0,0)[l]{\strut{}measured data}}%
      \csname LTb\endcsname%
      \put(2172,4383){\makebox(0,0)[l]{\strut{}$(1+e^{-(\beta-a)/b)})^{-1}$}}%
    }%
    \gplbacktext
    \put(0,0){\includegraphics{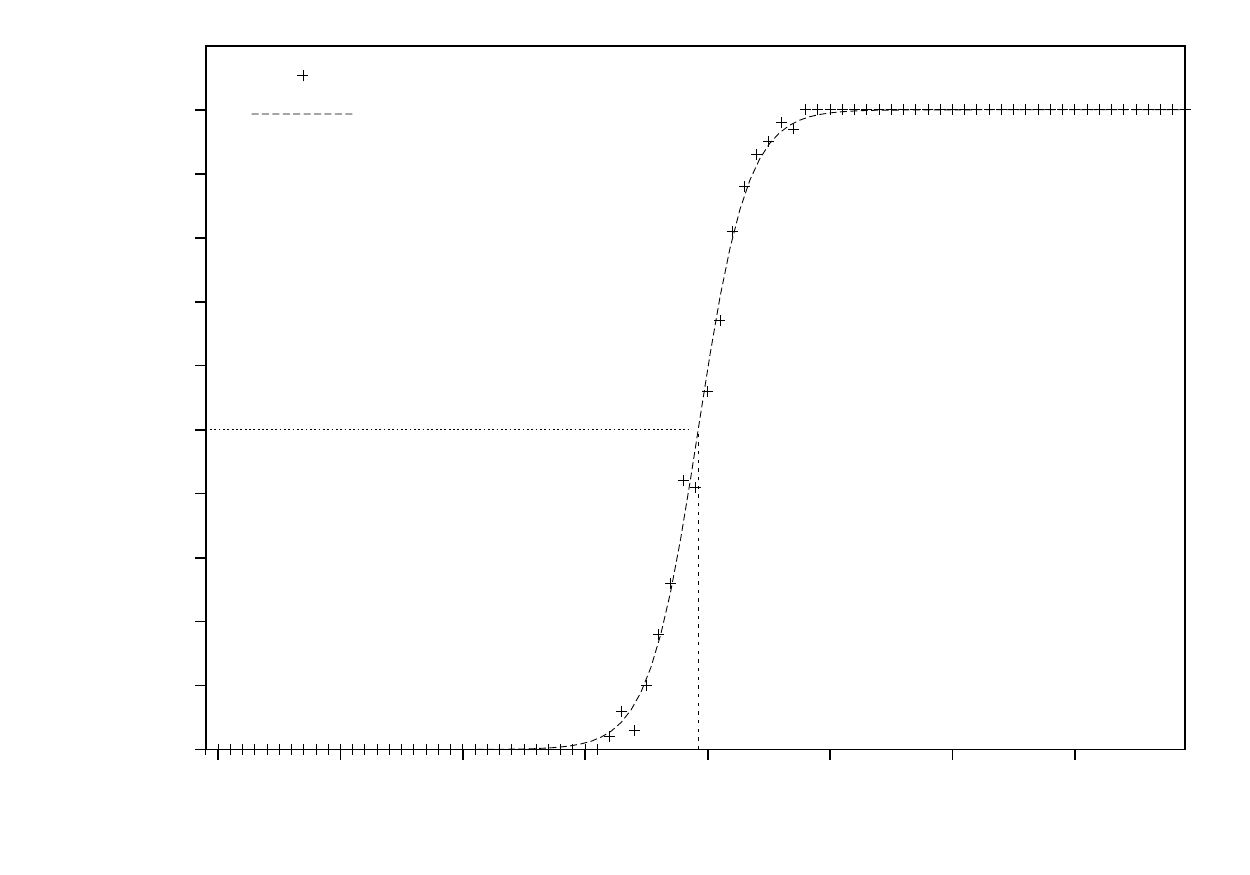}}%
    \gplfronttext
  \end{picture}%
\endgroup

%% file: figures/xkthulhu_MTW_n=10E5_d=4_k=1.tex
\begingroup
  \makeatletter
  \providecommand\color[2][]{%
    \GenericError{(gnuplot) \space\space\space\@spaces}{%
      Package color not loaded in conjunction with
      terminal option `colourtext'%
    }{See the gnuplot documentation for explanation.%
    }{Either use 'blacktext' in gnuplot or load the package
      color.sty in LaTeX.}%
    \renewcommand\color[2][]{}%
  }%
  \providecommand\includegraphics[2][]{%
    \GenericError{(gnuplot) \space\space\space\@spaces}{%
      Package graphicx or graphics not loaded%
    }{See the gnuplot documentation for explanation.%
    }{The gnuplot epslatex terminal needs graphicx.sty or graphics.sty.}%
    \renewcommand\includegraphics[2][]{}%
  }%
  \providecommand\rotatebox[2]{#2}%
  \@ifundefined{ifGPcolor}{%
    \newif\ifGPcolor
    \GPcolorfalse
  }{}%
  \@ifundefined{ifGPblacktext}{%
    \newif\ifGPblacktext
    \GPblacktexttrue
  }{}%
  \let\gplgaddtomacro\g@addto@macro
  \gdef\gplbacktext{}%
  \gdef\gplfronttext{}%
  \makeatother
  \ifGPblacktext
    \def\colorrgb#1{}%
    \def\colorgray#1{}%
  \else
    \ifGPcolor
      \def\colorrgb#1{\color[rgb]{#1}}%
      \def\colorgray#1{\color[gray]{#1}}%
      \expandafter\def\csname LTw\endcsname{\color{white}}%
      \expandafter\def\csname LTb\endcsname{\color{black}}%
      \expandafter\def\csname LTa\endcsname{\color{black}}%
      \expandafter\def\csname LT0\endcsname{\color[rgb]{1,0,0}}%
      \expandafter\def\csname LT1\endcsname{\color[rgb]{0,1,0}}%
      \expandafter\def\csname LT2\endcsname{\color[rgb]{0,0,1}}%
      \expandafter\def\csname LT3\endcsname{\color[rgb]{1,0,1}}%
      \expandafter\def\csname LT4\endcsname{\color[rgb]{0,1,1}}%
      \expandafter\def\csname LT5\endcsname{\color[rgb]{1,1,0}}%
      \expandafter\def\csname LT6\endcsname{\color[rgb]{0,0,0}}%
      \expandafter\def\csname LT7\endcsname{\color[rgb]{1,0.3,0}}%
      \expandafter\def\csname LT8\endcsname{\color[rgb]{0.5,0.5,0.5}}%
    \else
      \def\colorrgb#1{\color{black}}%
      \def\colorgray#1{\color[gray]{#1}}%
      \expandafter\def\csname LTw\endcsname{\color{white}}%
      \expandafter\def\csname LTb\endcsname{\color{black}}%
      \expandafter\def\csname LTa\endcsname{\color{black}}%
      \expandafter\def\csname LT0\endcsname{\color{black}}%
      \expandafter\def\csname LT1\endcsname{\color{black}}%
      \expandafter\def\csname LT2\endcsname{\color{black}}%
      \expandafter\def\csname LT3\endcsname{\color{black}}%
      \expandafter\def\csname LT4\endcsname{\color{black}}%
      \expandafter\def\csname LT5\endcsname{\color{black}}%
      \expandafter\def\csname LT6\endcsname{\color{black}}%
      \expandafter\def\csname LT7\endcsname{\color{black}}%
      \expandafter\def\csname LT8\endcsname{\color{black}}%
    \fi
  \fi
  \setlength{\unitlength}{0.0500bp}%
  \begin{picture}(7200.00,5040.00)%
    \gplgaddtomacro\gplbacktext{%
      \csname LTb\endcsname%
      \put(990,723){\makebox(0,0)[r]{\strut{} 0}}%
      \put(990,1091){\makebox(0,0)[r]{\strut{} 0.1}}%
      \put(990,1460){\makebox(0,0)[r]{\strut{} 0.2}}%
      \put(990,1828){\makebox(0,0)[r]{\strut{} 0.3}}%
      \put(990,2197){\makebox(0,0)[r]{\strut{} 0.4}}%
      \put(990,2565){\makebox(0,0)[r]{\strut{} 0.5}}%
      \put(990,2934){\makebox(0,0)[r]{\strut{} 0.6}}%
      \put(990,3302){\makebox(0,0)[r]{\strut{} 0.7}}%
      \put(990,3671){\makebox(0,0)[r]{\strut{} 0.8}}%
      \put(990,4039){\makebox(0,0)[r]{\strut{} 0.9}}%
      \put(990,4408){\makebox(0,0)[r]{\strut{} 1}}%
      \put(1326,440){\makebox(0,0){\strut{} 0.973}}%
      \put(2031,440){\makebox(0,0){\strut{} 0.974}}%
      \put(2736,440){\makebox(0,0){\strut{} 0.975}}%
      \put(3441,440){\makebox(0,0){\strut{} 0.976}}%
      \put(4147,440){\makebox(0,0){\strut{} 0.977}}%
      \put(4852,440){\makebox(0,0){\strut{} 0.978}}%
      \put(5557,440){\makebox(0,0){\strut{} 0.979}}%
      \put(6262,440){\makebox(0,0){\strut{} 0.98}}%
      \put(220,2749){\rotatebox{90}{\makebox(0,0){\strut{}failure rate among $100$ attempts}}}%
      \put(4005,110){\makebox(0,0){\strut{}$\beta$}}%
      \put(4028,926){\makebox(0,0)[l]{\strut{}\labIncr$a=0.976732$}}%
      \put(4028,1331){\makebox(0,0)[l]{\strut{}\labIncr$\sum_{res}=0.0242203$}}%
    }%
    \gplgaddtomacro\gplfronttext{%
      \csname LTb\endcsname%
      \put(2172,4603){\makebox(0,0)[l]{\strut{}measured data}}%
      \csname LTb\endcsname%
      \put(2172,4383){\makebox(0,0)[l]{\strut{}$(1+e^{-(\beta-a)/b)})^{-1}$}}%
    }%
    \gplbacktext
    \put(0,0){\includegraphics{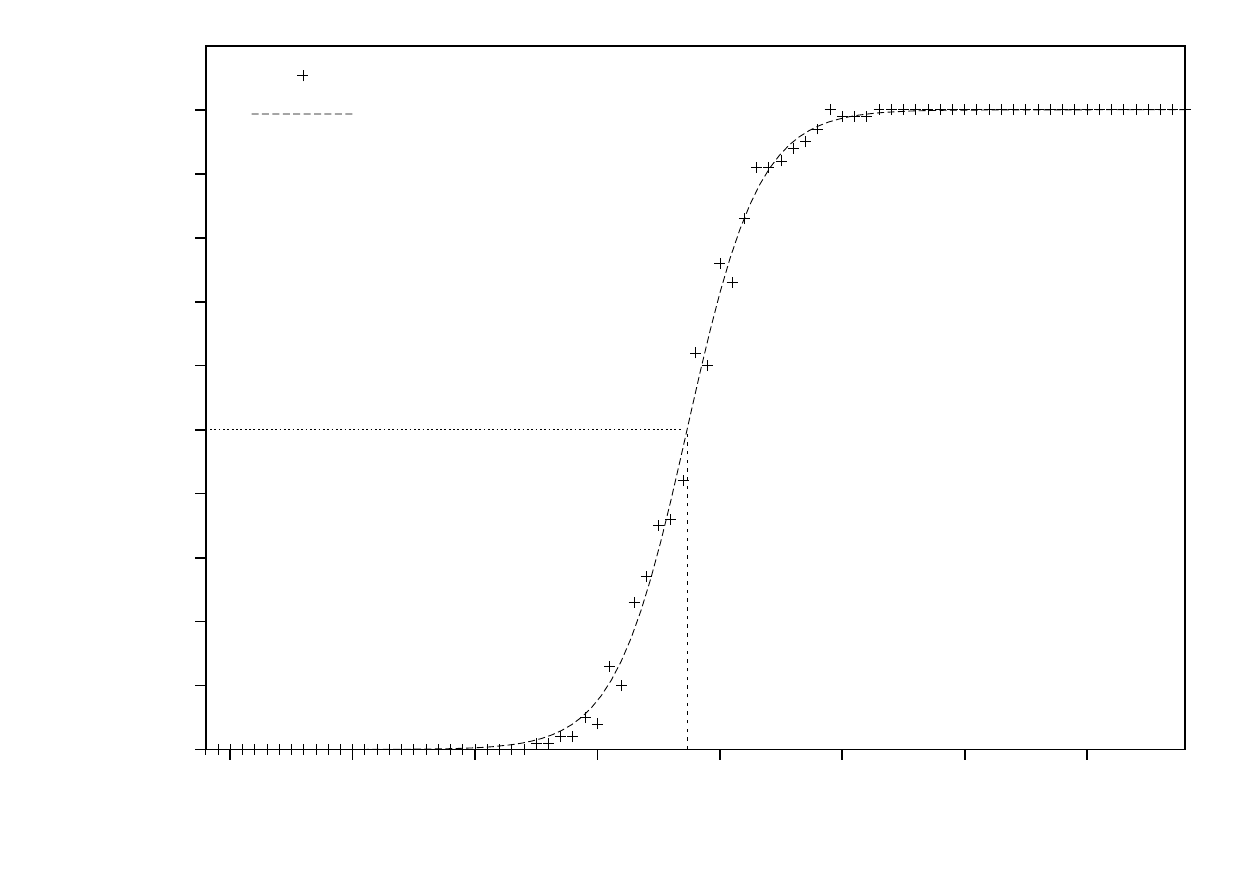}}%
    \gplfronttext
  \end{picture}%
\endgroup

%% file: figures/xkthulhu_MTW_n=10E6_d=4_k=1.tex
\begingroup
  \makeatletter
  \providecommand\color[2][]{%
    \GenericError{(gnuplot) \space\space\space\@spaces}{%
      Package color not loaded in conjunction with
      terminal option `colourtext'%
    }{See the gnuplot documentation for explanation.%
    }{Either use 'blacktext' in gnuplot or load the package
      color.sty in LaTeX.}%
    \renewcommand\color[2][]{}%
  }%
  \providecommand\includegraphics[2][]{%
    \GenericError{(gnuplot) \space\space\space\@spaces}{%
      Package graphicx or graphics not loaded%
    }{See the gnuplot documentation for explanation.%
    }{The gnuplot epslatex terminal needs graphicx.sty or graphics.sty.}%
    \renewcommand\includegraphics[2][]{}%
  }%
  \providecommand\rotatebox[2]{#2}%
  \@ifundefined{ifGPcolor}{%
    \newif\ifGPcolor
    \GPcolorfalse
  }{}%
  \@ifundefined{ifGPblacktext}{%
    \newif\ifGPblacktext
    \GPblacktexttrue
  }{}%
  \let\gplgaddtomacro\g@addto@macro
  \gdef\gplbacktext{}%
  \gdef\gplfronttext{}%
  \makeatother
  \ifGPblacktext
    \def\colorrgb#1{}%
    \def\colorgray#1{}%
  \else
    \ifGPcolor
      \def\colorrgb#1{\color[rgb]{#1}}%
      \def\colorgray#1{\color[gray]{#1}}%
      \expandafter\def\csname LTw\endcsname{\color{white}}%
      \expandafter\def\csname LTb\endcsname{\color{black}}%
      \expandafter\def\csname LTa\endcsname{\color{black}}%
      \expandafter\def\csname LT0\endcsname{\color[rgb]{1,0,0}}%
      \expandafter\def\csname LT1\endcsname{\color[rgb]{0,1,0}}%
      \expandafter\def\csname LT2\endcsname{\color[rgb]{0,0,1}}%
      \expandafter\def\csname LT3\endcsname{\color[rgb]{1,0,1}}%
      \expandafter\def\csname LT4\endcsname{\color[rgb]{0,1,1}}%
      \expandafter\def\csname LT5\endcsname{\color[rgb]{1,1,0}}%
      \expandafter\def\csname LT6\endcsname{\color[rgb]{0,0,0}}%
      \expandafter\def\csname LT7\endcsname{\color[rgb]{1,0.3,0}}%
      \expandafter\def\csname LT8\endcsname{\color[rgb]{0.5,0.5,0.5}}%
    \else
      \def\colorrgb#1{\color{black}}%
      \def\colorgray#1{\color[gray]{#1}}%
      \expandafter\def\csname LTw\endcsname{\color{white}}%
      \expandafter\def\csname LTb\endcsname{\color{black}}%
      \expandafter\def\csname LTa\endcsname{\color{black}}%
      \expandafter\def\csname LT0\endcsname{\color{black}}%
      \expandafter\def\csname LT1\endcsname{\color{black}}%
      \expandafter\def\csname LT2\endcsname{\color{black}}%
      \expandafter\def\csname LT3\endcsname{\color{black}}%
      \expandafter\def\csname LT4\endcsname{\color{black}}%
      \expandafter\def\csname LT5\endcsname{\color{black}}%
      \expandafter\def\csname LT6\endcsname{\color{black}}%
      \expandafter\def\csname LT7\endcsname{\color{black}}%
      \expandafter\def\csname LT8\endcsname{\color{black}}%
    \fi
  \fi
  \setlength{\unitlength}{0.0500bp}%
  \begin{picture}(7200.00,5040.00)%
    \gplgaddtomacro\gplbacktext{%
      \csname LTb\endcsname%
      \put(990,723){\makebox(0,0)[r]{\strut{} 0}}%
      \put(990,1091){\makebox(0,0)[r]{\strut{} 0.1}}%
      \put(990,1460){\makebox(0,0)[r]{\strut{} 0.2}}%
      \put(990,1828){\makebox(0,0)[r]{\strut{} 0.3}}%
      \put(990,2197){\makebox(0,0)[r]{\strut{} 0.4}}%
      \put(990,2565){\makebox(0,0)[r]{\strut{} 0.5}}%
      \put(990,2934){\makebox(0,0)[r]{\strut{} 0.6}}%
      \put(990,3302){\makebox(0,0)[r]{\strut{} 0.7}}%
      \put(990,3671){\makebox(0,0)[r]{\strut{} 0.8}}%
      \put(990,4039){\makebox(0,0)[r]{\strut{} 0.9}}%
      \put(990,4408){\makebox(0,0)[r]{\strut{} 1}}%
      \put(1326,440){\makebox(0,0){\strut{} 0.973}}%
      \put(2031,440){\makebox(0,0){\strut{} 0.974}}%
      \put(2736,440){\makebox(0,0){\strut{} 0.975}}%
      \put(3441,440){\makebox(0,0){\strut{} 0.976}}%
      \put(4147,440){\makebox(0,0){\strut{} 0.977}}%
      \put(4852,440){\makebox(0,0){\strut{} 0.978}}%
      \put(5557,440){\makebox(0,0){\strut{} 0.979}}%
      \put(6262,440){\makebox(0,0){\strut{} 0.98}}%
      \put(220,2749){\rotatebox{90}{\makebox(0,0){\strut{}failure rate among $100$ attempts}}}%
      \put(4005,110){\makebox(0,0){\strut{}$\beta$}}%
      \put(4045,926){\makebox(0,0)[l]{\strut{}\labIncr$a=0.976756$}}%
      \put(4045,1331){\makebox(0,0)[l]{\strut{}\labIncr$\sum_{res}=0.0035227$}}%
    }%
    \gplgaddtomacro\gplfronttext{%
      \csname LTb\endcsname%
      \put(2172,4603){\makebox(0,0)[l]{\strut{}measured data}}%
      \csname LTb\endcsname%
      \put(2172,4383){\makebox(0,0)[l]{\strut{}$(1+e^{-(\beta-a)/b)})^{-1}$}}%
    }%
    \gplbacktext
    \put(0,0){\includegraphics{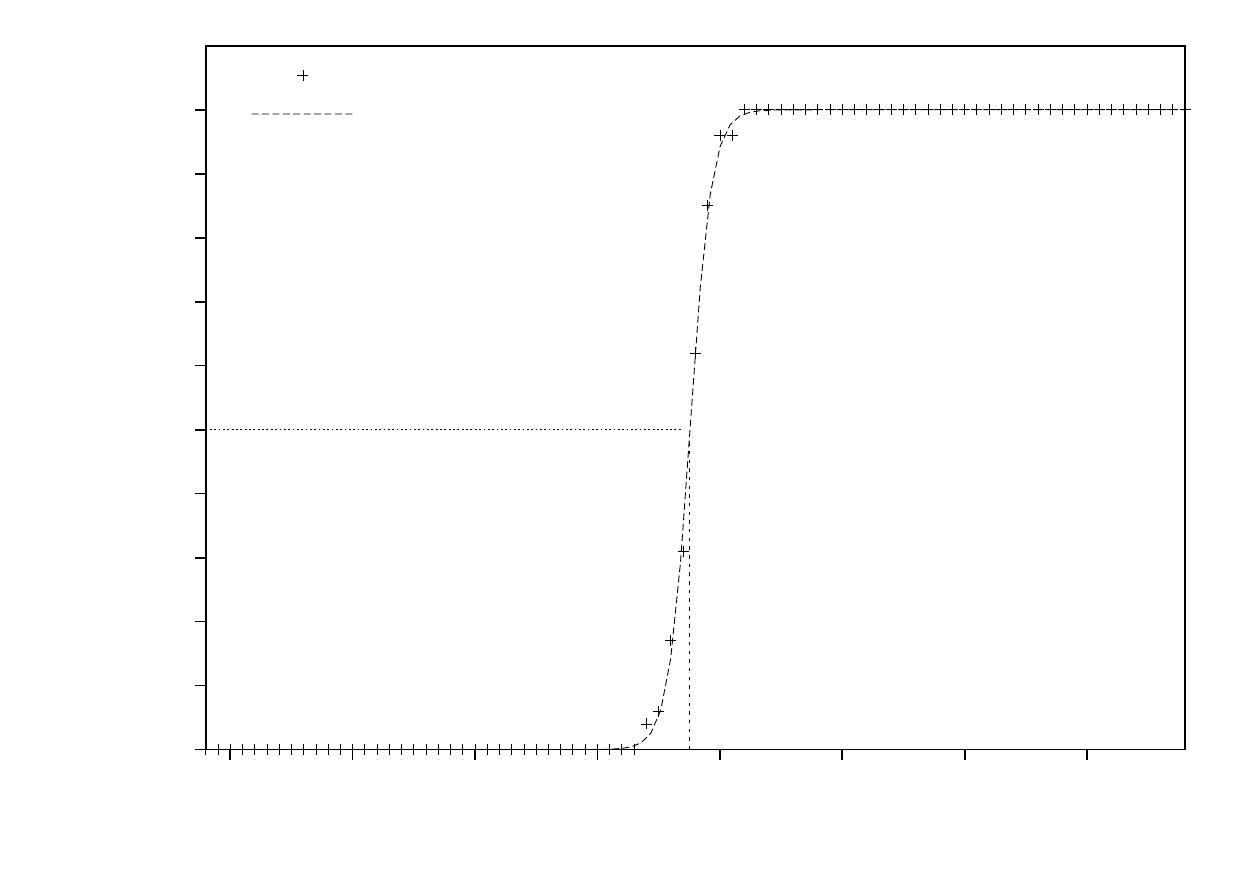}}%
    \gplfronttext
  \end{picture}%
\endgroup

%% file: figures/xkthulhu_MTW_n=10E5_d=5_k=1.tex
\begingroup
  \makeatletter
  \providecommand\color[2][]{%
    \GenericError{(gnuplot) \space\space\space\@spaces}{%
      Package color not loaded in conjunction with
      terminal option `colourtext'%
    }{See the gnuplot documentation for explanation.%
    }{Either use 'blacktext' in gnuplot or load the package
      color.sty in LaTeX.}%
    \renewcommand\color[2][]{}%
  }%
  \providecommand\includegraphics[2][]{%
    \GenericError{(gnuplot) \space\space\space\@spaces}{%
      Package graphicx or graphics not loaded%
    }{See the gnuplot documentation for explanation.%
    }{The gnuplot epslatex terminal needs graphicx.sty or graphics.sty.}%
    \renewcommand\includegraphics[2][]{}%
  }%
  \providecommand\rotatebox[2]{#2}%
  \@ifundefined{ifGPcolor}{%
    \newif\ifGPcolor
    \GPcolorfalse
  }{}%
  \@ifundefined{ifGPblacktext}{%
    \newif\ifGPblacktext
    \GPblacktexttrue
  }{}%
  \let\gplgaddtomacro\g@addto@macro
  \gdef\gplbacktext{}%
  \gdef\gplfronttext{}%
  \makeatother
  \ifGPblacktext
    \def\colorrgb#1{}%
    \def\colorgray#1{}%
  \else
    \ifGPcolor
      \def\colorrgb#1{\color[rgb]{#1}}%
      \def\colorgray#1{\color[gray]{#1}}%
      \expandafter\def\csname LTw\endcsname{\color{white}}%
      \expandafter\def\csname LTb\endcsname{\color{black}}%
      \expandafter\def\csname LTa\endcsname{\color{black}}%
      \expandafter\def\csname LT0\endcsname{\color[rgb]{1,0,0}}%
      \expandafter\def\csname LT1\endcsname{\color[rgb]{0,1,0}}%
      \expandafter\def\csname LT2\endcsname{\color[rgb]{0,0,1}}%
      \expandafter\def\csname LT3\endcsname{\color[rgb]{1,0,1}}%
      \expandafter\def\csname LT4\endcsname{\color[rgb]{0,1,1}}%
      \expandafter\def\csname LT5\endcsname{\color[rgb]{1,1,0}}%
      \expandafter\def\csname LT6\endcsname{\color[rgb]{0,0,0}}%
      \expandafter\def\csname LT7\endcsname{\color[rgb]{1,0.3,0}}%
      \expandafter\def\csname LT8\endcsname{\color[rgb]{0.5,0.5,0.5}}%
    \else
      \def\colorrgb#1{\color{black}}%
      \def\colorgray#1{\color[gray]{#1}}%
      \expandafter\def\csname LTw\endcsname{\color{white}}%
      \expandafter\def\csname LTb\endcsname{\color{black}}%
      \expandafter\def\csname LTa\endcsname{\color{black}}%
      \expandafter\def\csname LT0\endcsname{\color{black}}%
      \expandafter\def\csname LT1\endcsname{\color{black}}%
      \expandafter\def\csname LT2\endcsname{\color{black}}%
      \expandafter\def\csname LT3\endcsname{\color{black}}%
      \expandafter\def\csname LT4\endcsname{\color{black}}%
      \expandafter\def\csname LT5\endcsname{\color{black}}%
      \expandafter\def\csname LT6\endcsname{\color{black}}%
      \expandafter\def\csname LT7\endcsname{\color{black}}%
      \expandafter\def\csname LT8\endcsname{\color{black}}%
    \fi
  \fi
  \setlength{\unitlength}{0.0500bp}%
  \begin{picture}(7200.00,5040.00)%
    \gplgaddtomacro\gplbacktext{%
      \csname LTb\endcsname%
      \put(990,723){\makebox(0,0)[r]{\strut{} 0}}%
      \put(990,1091){\makebox(0,0)[r]{\strut{} 0.1}}%
      \put(990,1460){\makebox(0,0)[r]{\strut{} 0.2}}%
      \put(990,1828){\makebox(0,0)[r]{\strut{} 0.3}}%
      \put(990,2197){\makebox(0,0)[r]{\strut{} 0.4}}%
      \put(990,2565){\makebox(0,0)[r]{\strut{} 0.5}}%
      \put(990,2934){\makebox(0,0)[r]{\strut{} 0.6}}%
      \put(990,3302){\makebox(0,0)[r]{\strut{} 0.7}}%
      \put(990,3671){\makebox(0,0)[r]{\strut{} 0.8}}%
      \put(990,4039){\makebox(0,0)[r]{\strut{} 0.9}}%
      \put(990,4408){\makebox(0,0)[r]{\strut{} 1}}%
      \put(1608,440){\makebox(0,0){\strut{} 0.989}}%
      \put(2313,440){\makebox(0,0){\strut{} 0.99}}%
      \put(3018,440){\makebox(0,0){\strut{} 0.991}}%
      \put(3723,440){\makebox(0,0){\strut{} 0.992}}%
      \put(4429,440){\makebox(0,0){\strut{} 0.993}}%
      \put(5134,440){\makebox(0,0){\strut{} 0.994}}%
      \put(5839,440){\makebox(0,0){\strut{} 0.995}}%
      \put(6544,440){\makebox(0,0){\strut{} 0.996}}%
      \put(220,2749){\rotatebox{90}{\makebox(0,0){\strut{}failure rate among $100$ attempts}}}%
      \put(4005,110){\makebox(0,0){\strut{}$\beta$}}%
      \put(4092,926){\makebox(0,0)[l]{\strut{}\labIncr$a=0.992423$}}%
      \put(4092,1331){\makebox(0,0)[l]{\strut{}\labIncr$\sum_{res}=0.0115793$}}%
    }%
    \gplgaddtomacro\gplfronttext{%
      \csname LTb\endcsname%
      \put(2172,4603){\makebox(0,0)[l]{\strut{}measured data}}%
      \csname LTb\endcsname%
      \put(2172,4383){\makebox(0,0)[l]{\strut{}$(1+e^{-(\beta-a)/b)})^{-1}$}}%
    }%
    \gplbacktext
    \put(0,0){\includegraphics{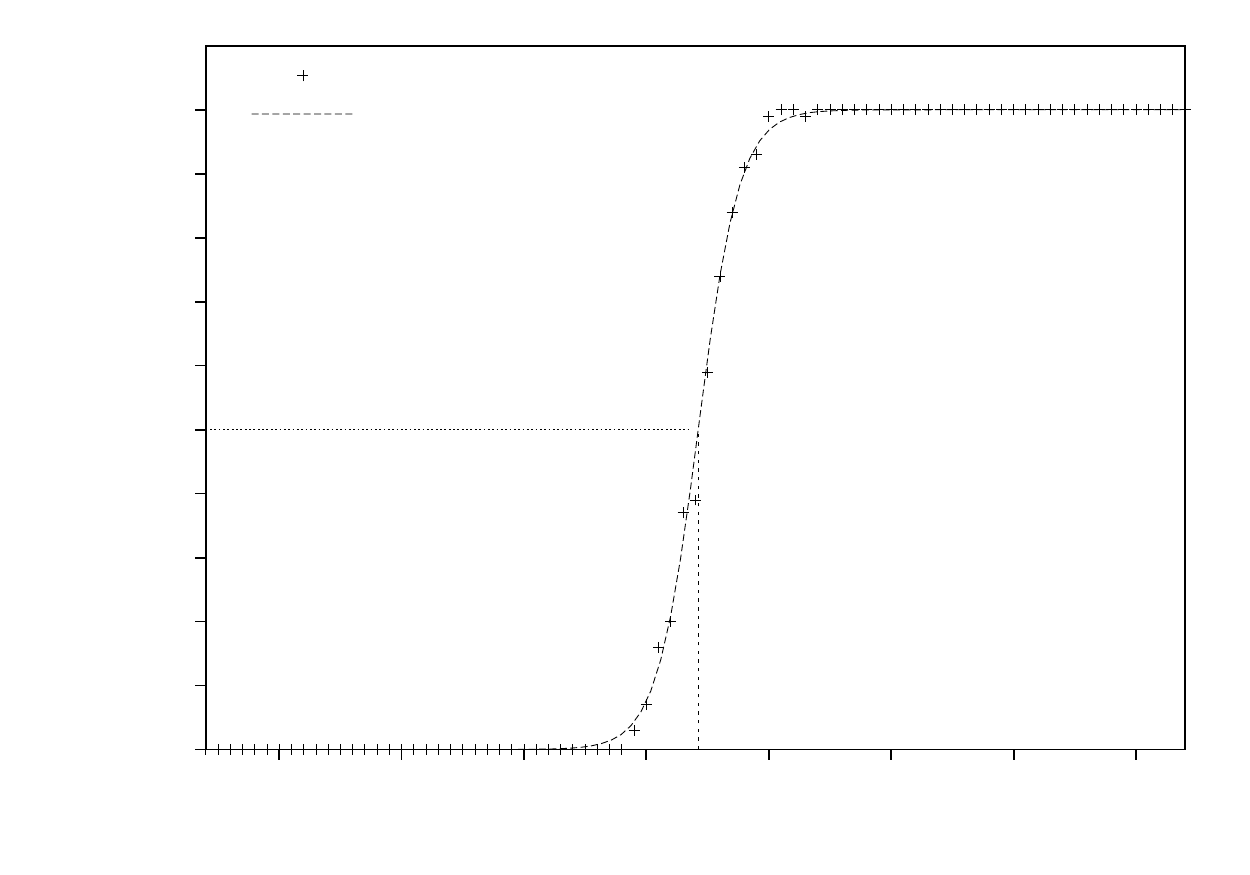}}%
    \gplfronttext
  \end{picture}%
\endgroup

%% file: figures/xkthulhu_MTW_n=10E6_d=5_k=1.tex
\begingroup
  \makeatletter
  \providecommand\color[2][]{%
    \GenericError{(gnuplot) \space\space\space\@spaces}{%
      Package color not loaded in conjunction with
      terminal option `colourtext'%
    }{See the gnuplot documentation for explanation.%
    }{Either use 'blacktext' in gnuplot or load the package
      color.sty in LaTeX.}%
    \renewcommand\color[2][]{}%
  }%
  \providecommand\includegraphics[2][]{%
    \GenericError{(gnuplot) \space\space\space\@spaces}{%
      Package graphicx or graphics not loaded%
    }{See the gnuplot documentation for explanation.%
    }{The gnuplot epslatex terminal needs graphicx.sty or graphics.sty.}%
    \renewcommand\includegraphics[2][]{}%
  }%
  \providecommand\rotatebox[2]{#2}%
  \@ifundefined{ifGPcolor}{%
    \newif\ifGPcolor
    \GPcolorfalse
  }{}%
  \@ifundefined{ifGPblacktext}{%
    \newif\ifGPblacktext
    \GPblacktexttrue
  }{}%
  \let\gplgaddtomacro\g@addto@macro
  \gdef\gplbacktext{}%
  \gdef\gplfronttext{}%
  \makeatother
  \ifGPblacktext
    \def\colorrgb#1{}%
    \def\colorgray#1{}%
  \else
    \ifGPcolor
      \def\colorrgb#1{\color[rgb]{#1}}%
      \def\colorgray#1{\color[gray]{#1}}%
      \expandafter\def\csname LTw\endcsname{\color{white}}%
      \expandafter\def\csname LTb\endcsname{\color{black}}%
      \expandafter\def\csname LTa\endcsname{\color{black}}%
      \expandafter\def\csname LT0\endcsname{\color[rgb]{1,0,0}}%
      \expandafter\def\csname LT1\endcsname{\color[rgb]{0,1,0}}%
      \expandafter\def\csname LT2\endcsname{\color[rgb]{0,0,1}}%
      \expandafter\def\csname LT3\endcsname{\color[rgb]{1,0,1}}%
      \expandafter\def\csname LT4\endcsname{\color[rgb]{0,1,1}}%
      \expandafter\def\csname LT5\endcsname{\color[rgb]{1,1,0}}%
      \expandafter\def\csname LT6\endcsname{\color[rgb]{0,0,0}}%
      \expandafter\def\csname LT7\endcsname{\color[rgb]{1,0.3,0}}%
      \expandafter\def\csname LT8\endcsname{\color[rgb]{0.5,0.5,0.5}}%
    \else
      \def\colorrgb#1{\color{black}}%
      \def\colorgray#1{\color[gray]{#1}}%
      \expandafter\def\csname LTw\endcsname{\color{white}}%
      \expandafter\def\csname LTb\endcsname{\color{black}}%
      \expandafter\def\csname LTa\endcsname{\color{black}}%
      \expandafter\def\csname LT0\endcsname{\color{black}}%
      \expandafter\def\csname LT1\endcsname{\color{black}}%
      \expandafter\def\csname LT2\endcsname{\color{black}}%
      \expandafter\def\csname LT3\endcsname{\color{black}}%
      \expandafter\def\csname LT4\endcsname{\color{black}}%
      \expandafter\def\csname LT5\endcsname{\color{black}}%
      \expandafter\def\csname LT6\endcsname{\color{black}}%
      \expandafter\def\csname LT7\endcsname{\color{black}}%
      \expandafter\def\csname LT8\endcsname{\color{black}}%
    \fi
  \fi
  \setlength{\unitlength}{0.0500bp}%
  \begin{picture}(7200.00,5040.00)%
    \gplgaddtomacro\gplbacktext{%
      \csname LTb\endcsname%
      \put(990,723){\makebox(0,0)[r]{\strut{} 0}}%
      \put(990,1091){\makebox(0,0)[r]{\strut{} 0.1}}%
      \put(990,1460){\makebox(0,0)[r]{\strut{} 0.2}}%
      \put(990,1828){\makebox(0,0)[r]{\strut{} 0.3}}%
      \put(990,2197){\makebox(0,0)[r]{\strut{} 0.4}}%
      \put(990,2565){\makebox(0,0)[r]{\strut{} 0.5}}%
      \put(990,2934){\makebox(0,0)[r]{\strut{} 0.6}}%
      \put(990,3302){\makebox(0,0)[r]{\strut{} 0.7}}%
      \put(990,3671){\makebox(0,0)[r]{\strut{} 0.8}}%
      \put(990,4039){\makebox(0,0)[r]{\strut{} 0.9}}%
      \put(990,4408){\makebox(0,0)[r]{\strut{} 1}}%
      \put(1608,440){\makebox(0,0){\strut{} 0.989}}%
      \put(2313,440){\makebox(0,0){\strut{} 0.99}}%
      \put(3018,440){\makebox(0,0){\strut{} 0.991}}%
      \put(3723,440){\makebox(0,0){\strut{} 0.992}}%
      \put(4429,440){\makebox(0,0){\strut{} 0.993}}%
      \put(5134,440){\makebox(0,0){\strut{} 0.994}}%
      \put(5839,440){\makebox(0,0){\strut{} 0.995}}%
      \put(6544,440){\makebox(0,0){\strut{} 0.996}}%
      \put(220,2749){\rotatebox{90}{\makebox(0,0){\strut{}failure rate among $100$ attempts}}}%
      \put(4005,110){\makebox(0,0){\strut{}$\beta$}}%
      \put(4105,926){\makebox(0,0)[l]{\strut{}\labIncr$a=0.992441$}}%
      \put(4105,1331){\makebox(0,0)[l]{\strut{}\labIncr$\sum_{res}=0.0019705$}}%
    }%
    \gplgaddtomacro\gplfronttext{%
      \csname LTb\endcsname%
      \put(2172,4603){\makebox(0,0)[l]{\strut{}measured data}}%
      \csname LTb\endcsname%
      \put(2172,4383){\makebox(0,0)[l]{\strut{}$(1+e^{-(\beta-a)/b)})^{-1}$}}%
    }%
    \gplbacktext
    \put(0,0){\includegraphics{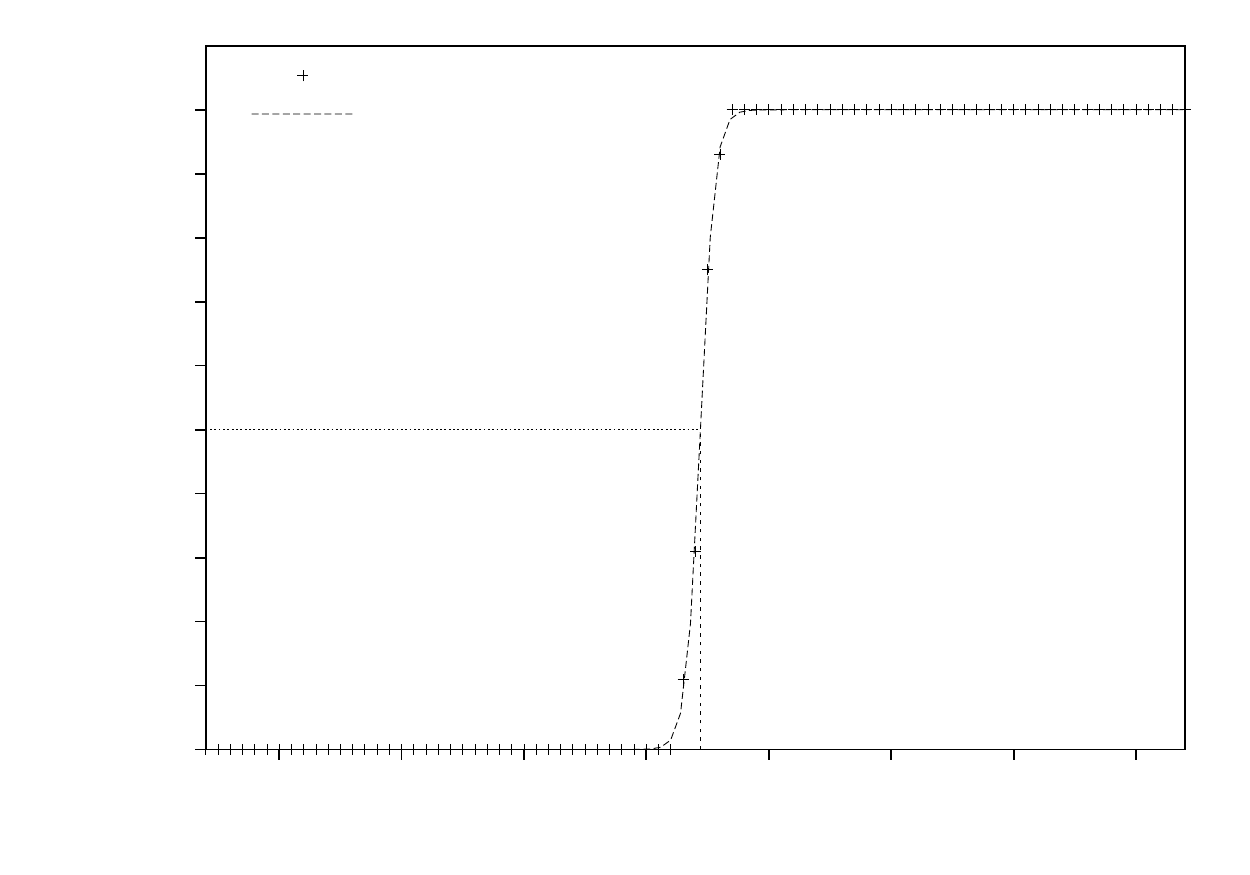}}%
    \gplfronttext
  \end{picture}%
\endgroup

%% file: figures/xGSvsHC_n=10E5_d=3_k=1.tex
\begingroup
  \makeatletter
  \providecommand\color[2][]{%
    \GenericError{(gnuplot) \space\space\space\@spaces}{%
      Package color not loaded in conjunction with
      terminal option `colourtext'%
    }{See the gnuplot documentation for explanation.%
    }{Either use 'blacktext' in gnuplot or load the package
      color.sty in LaTeX.}%
    \renewcommand\color[2][]{}%
  }%
  \providecommand\includegraphics[2][]{%
    \GenericError{(gnuplot) \space\space\space\@spaces}{%
      Package graphicx or graphics not loaded%
    }{See the gnuplot documentation for explanation.%
    }{The gnuplot epslatex terminal needs graphicx.sty or graphics.sty.}%
    \renewcommand\includegraphics[2][]{}%
  }%
  \providecommand\rotatebox[2]{#2}%
  \@ifundefined{ifGPcolor}{%
    \newif\ifGPcolor
    \GPcolorfalse
  }{}%
  \@ifundefined{ifGPblacktext}{%
    \newif\ifGPblacktext
    \GPblacktexttrue
  }{}%
  \let\gplgaddtomacro\g@addto@macro
  \gdef\gplbacktext{}%
  \gdef\gplfronttext{}%
  \makeatother
  \ifGPblacktext
    \def\colorrgb#1{}%
    \def\colorgray#1{}%
  \else
    \ifGPcolor
      \def\colorrgb#1{\color[rgb]{#1}}%
      \def\colorgray#1{\color[gray]{#1}}%
      \expandafter\def\csname LTw\endcsname{\color{white}}%
      \expandafter\def\csname LTb\endcsname{\color{black}}%
      \expandafter\def\csname LTa\endcsname{\color{black}}%
      \expandafter\def\csname LT0\endcsname{\color[rgb]{1,0,0}}%
      \expandafter\def\csname LT1\endcsname{\color[rgb]{0,1,0}}%
      \expandafter\def\csname LT2\endcsname{\color[rgb]{0,0,1}}%
      \expandafter\def\csname LT3\endcsname{\color[rgb]{1,0,1}}%
      \expandafter\def\csname LT4\endcsname{\color[rgb]{0,1,1}}%
      \expandafter\def\csname LT5\endcsname{\color[rgb]{1,1,0}}%
      \expandafter\def\csname LT6\endcsname{\color[rgb]{0,0,0}}%
      \expandafter\def\csname LT7\endcsname{\color[rgb]{1,0.3,0}}%
      \expandafter\def\csname LT8\endcsname{\color[rgb]{0.5,0.5,0.5}}%
    \else
      \def\colorrgb#1{\color{black}}%
      \def\colorgray#1{\color[gray]{#1}}%
      \expandafter\def\csname LTw\endcsname{\color{white}}%
      \expandafter\def\csname LTb\endcsname{\color{black}}%
      \expandafter\def\csname LTa\endcsname{\color{black}}%
      \expandafter\def\csname LT0\endcsname{\color{black}}%
      \expandafter\def\csname LT1\endcsname{\color{black}}%
      \expandafter\def\csname LT2\endcsname{\color{black}}%
      \expandafter\def\csname LT3\endcsname{\color{black}}%
      \expandafter\def\csname LT4\endcsname{\color{black}}%
      \expandafter\def\csname LT5\endcsname{\color{black}}%
      \expandafter\def\csname LT6\endcsname{\color{black}}%
      \expandafter\def\csname LT7\endcsname{\color{black}}%
      \expandafter\def\csname LT8\endcsname{\color{black}}%
    \fi
  \fi
  \setlength{\unitlength}{0.0500bp}%
  \begin{picture}(7200.00,5040.00)%
    \gplgaddtomacro\gplbacktext{%
      \csname LTb\endcsname%
      \put(990,723){\makebox(0,0)[r]{\strut{} 0}}%
      \put(990,1091){\makebox(0,0)[r]{\strut{} 0.1}}%
      \put(990,1460){\makebox(0,0)[r]{\strut{} 0.2}}%
      \put(990,1828){\makebox(0,0)[r]{\strut{} 0.3}}%
      \put(990,2197){\makebox(0,0)[r]{\strut{} 0.4}}%
      \put(990,2565){\makebox(0,0)[r]{\strut{} 0.5}}%
      \put(990,2934){\makebox(0,0)[r]{\strut{} 0.6}}%
      \put(990,3302){\makebox(0,0)[r]{\strut{} 0.7}}%
      \put(990,3671){\makebox(0,0)[r]{\strut{} 0.8}}%
      \put(990,4039){\makebox(0,0)[r]{\strut{} 0.9}}%
      \put(990,4408){\makebox(0,0)[r]{\strut{} 1}}%
      \put(1326,440){\makebox(0,0){\strut{} 0.916}}%
      \put(2031,440){\makebox(0,0){\strut{} 0.9165}}%
      \put(2736,440){\makebox(0,0){\strut{} 0.917}}%
      \put(3441,440){\makebox(0,0){\strut{} 0.9175}}%
      \put(4147,440){\makebox(0,0){\strut{} 0.918}}%
      \put(4852,440){\makebox(0,0){\strut{} 0.9185}}%
      \put(5557,440){\makebox(0,0){\strut{} 0.919}}%
      \put(6262,440){\makebox(0,0){\strut{} 0.9195}}%
      \put(220,2749){\rotatebox{90}{\makebox(0,0){\strut{}failure rate among $100$ attempts}}}%
      \put(4005,110){\makebox(0,0){\strut{}$\beta$}}%
      \put(4640,926){\makebox(0,0)[l]{\strut{}\labIncr$a_{\rm pm} = 0.917919$}}%
      \put(4640,1331){\makebox(0,0)[l]{\strut{}\labIncr$a_{\rm gs} = 0.91785$}}%
    }%
    \gplgaddtomacro\gplfronttext{%
      \csname LTb\endcsname%
      \put(2172,4603){\makebox(0,0)[l]{\strut{}generalized selfless (gs)}}%
      \csname LTb\endcsname%
      \put(2172,4383){\makebox(0,0)[l]{\strut{}$ \sigma(\beta;a_{\rm gs}, b_{\rm gs} )$}}%
      \csname LTb\endcsname%
      \put(2172,4163){\makebox(0,0)[l]{\strut{}perfect matching (pm)}}%
      \csname LTb\endcsname%
      \put(2172,3943){\makebox(0,0)[l]{\strut{}$ \sigma(\beta;a_{\rm pm}, b_{\rm pm} )$}}%
    }%
    \gplbacktext
    \put(0,0){\includegraphics{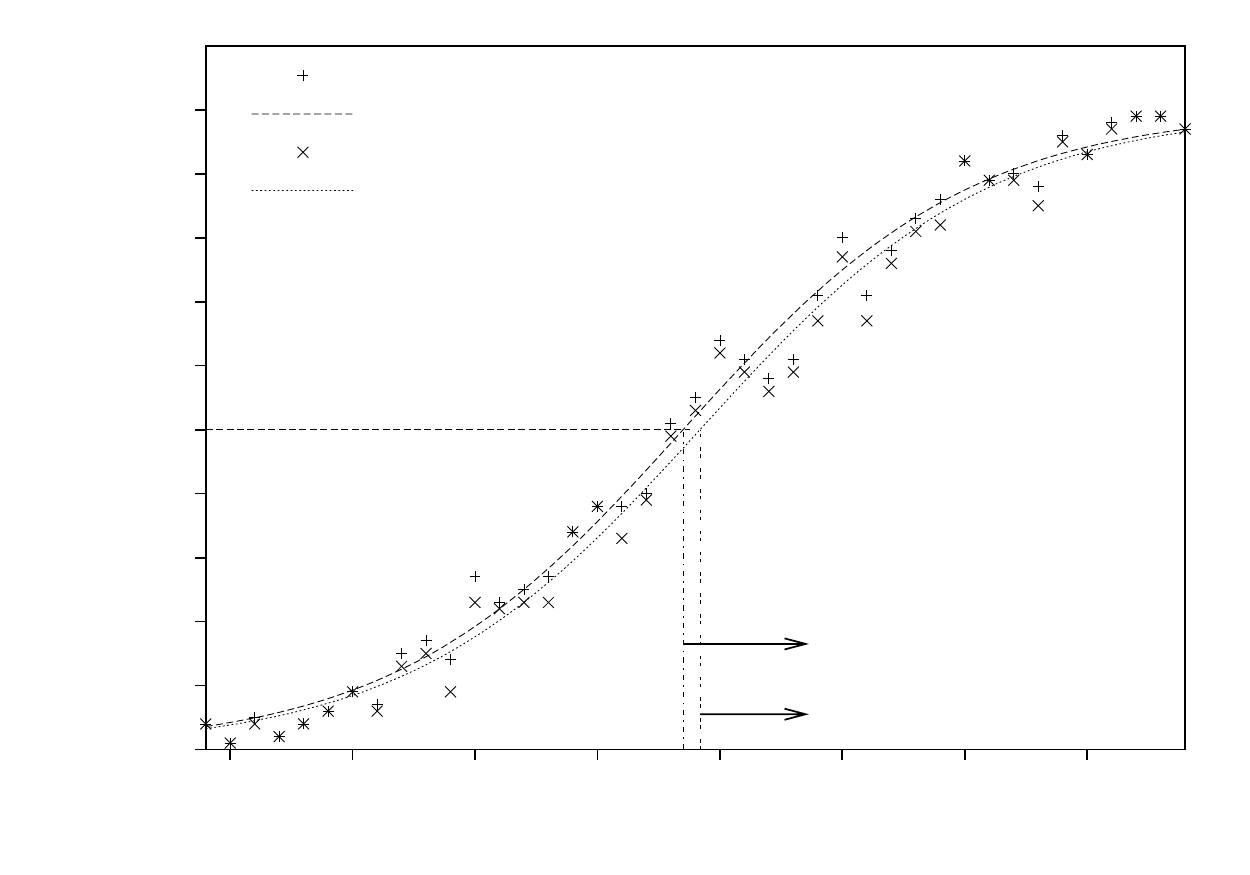}}%
    \gplfronttext
  \end{picture}%
\endgroup

%% file: figures/xGSvsHC_n=10E6_d=3_k=1.tex
\begingroup
  \makeatletter
  \providecommand\color[2][]{%
    \GenericError{(gnuplot) \space\space\space\@spaces}{%
      Package color not loaded in conjunction with
      terminal option `colourtext'%
    }{See the gnuplot documentation for explanation.%
    }{Either use 'blacktext' in gnuplot or load the package
      color.sty in LaTeX.}%
    \renewcommand\color[2][]{}%
  }%
  \providecommand\includegraphics[2][]{%
    \GenericError{(gnuplot) \space\space\space\@spaces}{%
      Package graphicx or graphics not loaded%
    }{See the gnuplot documentation for explanation.%
    }{The gnuplot epslatex terminal needs graphicx.sty or graphics.sty.}%
    \renewcommand\includegraphics[2][]{}%
  }%
  \providecommand\rotatebox[2]{#2}%
  \@ifundefined{ifGPcolor}{%
    \newif\ifGPcolor
    \GPcolorfalse
  }{}%
  \@ifundefined{ifGPblacktext}{%
    \newif\ifGPblacktext
    \GPblacktexttrue
  }{}%
  \let\gplgaddtomacro\g@addto@macro
  \gdef\gplbacktext{}%
  \gdef\gplfronttext{}%
  \makeatother
  \ifGPblacktext
    \def\colorrgb#1{}%
    \def\colorgray#1{}%
  \else
    \ifGPcolor
      \def\colorrgb#1{\color[rgb]{#1}}%
      \def\colorgray#1{\color[gray]{#1}}%
      \expandafter\def\csname LTw\endcsname{\color{white}}%
      \expandafter\def\csname LTb\endcsname{\color{black}}%
      \expandafter\def\csname LTa\endcsname{\color{black}}%
      \expandafter\def\csname LT0\endcsname{\color[rgb]{1,0,0}}%
      \expandafter\def\csname LT1\endcsname{\color[rgb]{0,1,0}}%
      \expandafter\def\csname LT2\endcsname{\color[rgb]{0,0,1}}%
      \expandafter\def\csname LT3\endcsname{\color[rgb]{1,0,1}}%
      \expandafter\def\csname LT4\endcsname{\color[rgb]{0,1,1}}%
      \expandafter\def\csname LT5\endcsname{\color[rgb]{1,1,0}}%
      \expandafter\def\csname LT6\endcsname{\color[rgb]{0,0,0}}%
      \expandafter\def\csname LT7\endcsname{\color[rgb]{1,0.3,0}}%
      \expandafter\def\csname LT8\endcsname{\color[rgb]{0.5,0.5,0.5}}%
    \else
      \def\colorrgb#1{\color{black}}%
      \def\colorgray#1{\color[gray]{#1}}%
      \expandafter\def\csname LTw\endcsname{\color{white}}%
      \expandafter\def\csname LTb\endcsname{\color{black}}%
      \expandafter\def\csname LTa\endcsname{\color{black}}%
      \expandafter\def\csname LT0\endcsname{\color{black}}%
      \expandafter\def\csname LT1\endcsname{\color{black}}%
      \expandafter\def\csname LT2\endcsname{\color{black}}%
      \expandafter\def\csname LT3\endcsname{\color{black}}%
      \expandafter\def\csname LT4\endcsname{\color{black}}%
      \expandafter\def\csname LT5\endcsname{\color{black}}%
      \expandafter\def\csname LT6\endcsname{\color{black}}%
      \expandafter\def\csname LT7\endcsname{\color{black}}%
      \expandafter\def\csname LT8\endcsname{\color{black}}%
    \fi
  \fi
  \setlength{\unitlength}{0.0500bp}%
  \begin{picture}(7200.00,5040.00)%
    \gplgaddtomacro\gplbacktext{%
      \csname LTb\endcsname%
      \put(990,723){\makebox(0,0)[r]{\strut{} 0}}%
      \put(990,1091){\makebox(0,0)[r]{\strut{} 0.1}}%
      \put(990,1460){\makebox(0,0)[r]{\strut{} 0.2}}%
      \put(990,1828){\makebox(0,0)[r]{\strut{} 0.3}}%
      \put(990,2197){\makebox(0,0)[r]{\strut{} 0.4}}%
      \put(990,2565){\makebox(0,0)[r]{\strut{} 0.5}}%
      \put(990,2934){\makebox(0,0)[r]{\strut{} 0.6}}%
      \put(990,3302){\makebox(0,0)[r]{\strut{} 0.7}}%
      \put(990,3671){\makebox(0,0)[r]{\strut{} 0.8}}%
      \put(990,4039){\makebox(0,0)[r]{\strut{} 0.9}}%
      \put(990,4408){\makebox(0,0)[r]{\strut{} 1}}%
      \put(1326,440){\makebox(0,0){\strut{} 0.916}}%
      \put(2031,440){\makebox(0,0){\strut{} 0.9165}}%
      \put(2736,440){\makebox(0,0){\strut{} 0.917}}%
      \put(3441,440){\makebox(0,0){\strut{} 0.9175}}%
      \put(4147,440){\makebox(0,0){\strut{} 0.918}}%
      \put(4852,440){\makebox(0,0){\strut{} 0.9185}}%
      \put(5557,440){\makebox(0,0){\strut{} 0.919}}%
      \put(6262,440){\makebox(0,0){\strut{} 0.9195}}%
      \put(220,2749){\rotatebox{90}{\makebox(0,0){\strut{}failure rate among $100$ attempts}}}%
      \put(4005,110){\makebox(0,0){\strut{}$\beta$}}%
      \put(4743,926){\makebox(0,0)[l]{\strut{}\labIncr$a_{\rm pm} = 0.917929$}}%
      \put(4743,1331){\makebox(0,0)[l]{\strut{}\labIncr$a_{\rm gs} = 0.917923$}}%
    }%
    \gplgaddtomacro\gplfronttext{%
      \csname LTb\endcsname%
      \put(2172,4603){\makebox(0,0)[l]{\strut{}generalized selfless (gs)}}%
      \csname LTb\endcsname%
      \put(2172,4383){\makebox(0,0)[l]{\strut{}$ \sigma(\beta;a_{\rm gs}, b_{\rm gs} )$}}%
      \csname LTb\endcsname%
      \put(2172,4163){\makebox(0,0)[l]{\strut{}perfect matching (pm)}}%
      \csname LTb\endcsname%
      \put(2172,3943){\makebox(0,0)[l]{\strut{}$ \sigma(\beta;a_{\rm pm}, b_{\rm pm} )$}}%
    }%
    \gplbacktext
    \put(0,0){\includegraphics{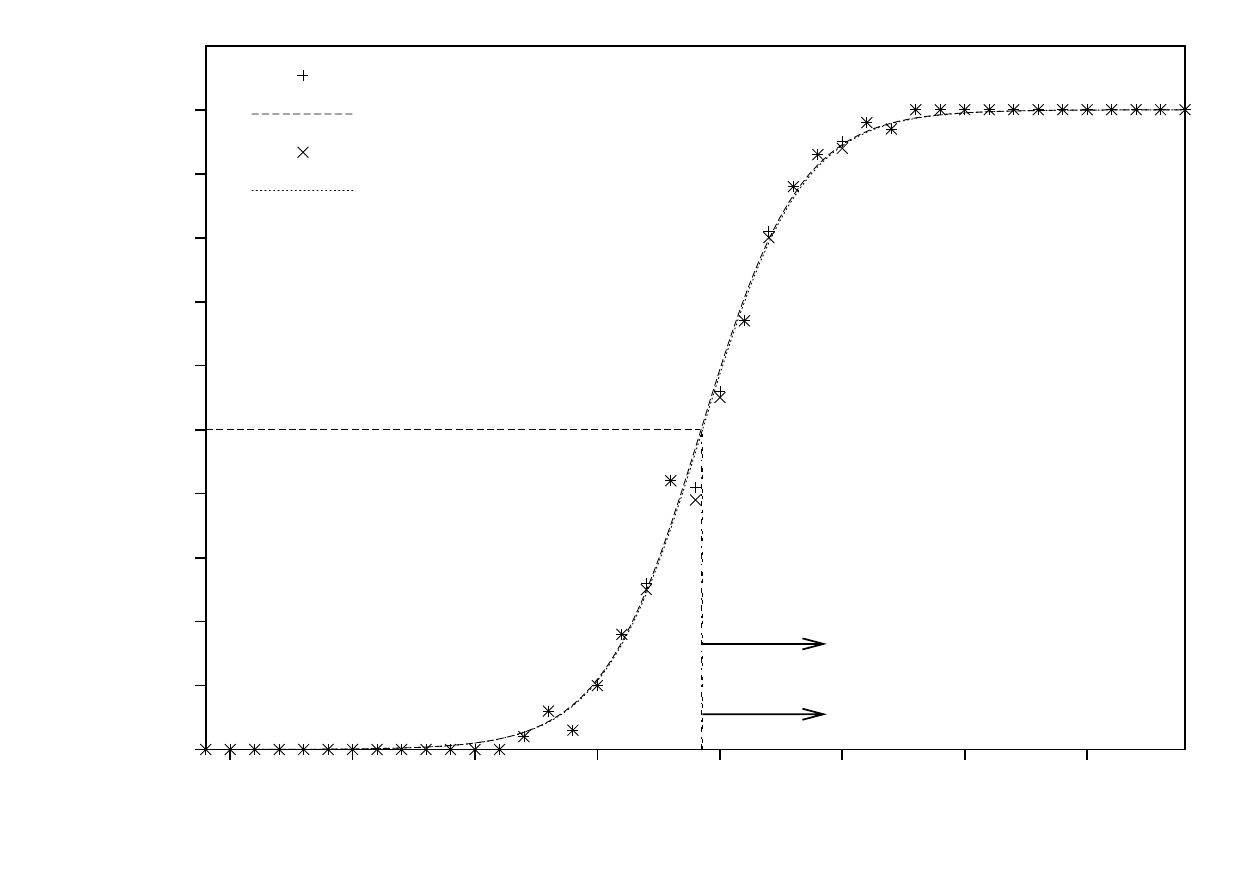}}%
    \gplfronttext
  \end{picture}%
\endgroup

%% file: figures/xkthulhu_MTW_n=10E5_d=3_k=2.tex
\begingroup
  \makeatletter
  \providecommand\color[2][]{%
    \GenericError{(gnuplot) \space\space\space\@spaces}{%
      Package color not loaded in conjunction with
      terminal option `colourtext'%
    }{See the gnuplot documentation for explanation.%
    }{Either use 'blacktext' in gnuplot or load the package
      color.sty in LaTeX.}%
    \renewcommand\color[2][]{}%
  }%
  \providecommand\includegraphics[2][]{%
    \GenericError{(gnuplot) \space\space\space\@spaces}{%
      Package graphicx or graphics not loaded%
    }{See the gnuplot documentation for explanation.%
    }{The gnuplot epslatex terminal needs graphicx.sty or graphics.sty.}%
    \renewcommand\includegraphics[2][]{}%
  }%
  \providecommand\rotatebox[2]{#2}%
  \@ifundefined{ifGPcolor}{%
    \newif\ifGPcolor
    \GPcolorfalse
  }{}%
  \@ifundefined{ifGPblacktext}{%
    \newif\ifGPblacktext
    \GPblacktexttrue
  }{}%
  \let\gplgaddtomacro\g@addto@macro
  \gdef\gplbacktext{}%
  \gdef\gplfronttext{}%
  \makeatother
  \ifGPblacktext
    \def\colorrgb#1{}%
    \def\colorgray#1{}%
  \else
    \ifGPcolor
      \def\colorrgb#1{\color[rgb]{#1}}%
      \def\colorgray#1{\color[gray]{#1}}%
      \expandafter\def\csname LTw\endcsname{\color{white}}%
      \expandafter\def\csname LTb\endcsname{\color{black}}%
      \expandafter\def\csname LTa\endcsname{\color{black}}%
      \expandafter\def\csname LT0\endcsname{\color[rgb]{1,0,0}}%
      \expandafter\def\csname LT1\endcsname{\color[rgb]{0,1,0}}%
      \expandafter\def\csname LT2\endcsname{\color[rgb]{0,0,1}}%
      \expandafter\def\csname LT3\endcsname{\color[rgb]{1,0,1}}%
      \expandafter\def\csname LT4\endcsname{\color[rgb]{0,1,1}}%
      \expandafter\def\csname LT5\endcsname{\color[rgb]{1,1,0}}%
      \expandafter\def\csname LT6\endcsname{\color[rgb]{0,0,0}}%
      \expandafter\def\csname LT7\endcsname{\color[rgb]{1,0.3,0}}%
      \expandafter\def\csname LT8\endcsname{\color[rgb]{0.5,0.5,0.5}}%
    \else
      \def\colorrgb#1{\color{black}}%
      \def\colorgray#1{\color[gray]{#1}}%
      \expandafter\def\csname LTw\endcsname{\color{white}}%
      \expandafter\def\csname LTb\endcsname{\color{black}}%
      \expandafter\def\csname LTa\endcsname{\color{black}}%
      \expandafter\def\csname LT0\endcsname{\color{black}}%
      \expandafter\def\csname LT1\endcsname{\color{black}}%
      \expandafter\def\csname LT2\endcsname{\color{black}}%
      \expandafter\def\csname LT3\endcsname{\color{black}}%
      \expandafter\def\csname LT4\endcsname{\color{black}}%
      \expandafter\def\csname LT5\endcsname{\color{black}}%
      \expandafter\def\csname LT6\endcsname{\color{black}}%
      \expandafter\def\csname LT7\endcsname{\color{black}}%
      \expandafter\def\csname LT8\endcsname{\color{black}}%
    \fi
  \fi
  \setlength{\unitlength}{0.0500bp}%
  \begin{picture}(7200.00,5040.00)%
    \gplgaddtomacro\gplbacktext{%
      \csname LTb\endcsname%
      \put(990,723){\makebox(0,0)[r]{\strut{} 0}}%
      \put(990,1091){\makebox(0,0)[r]{\strut{} 0.1}}%
      \put(990,1460){\makebox(0,0)[r]{\strut{} 0.2}}%
      \put(990,1828){\makebox(0,0)[r]{\strut{} 0.3}}%
      \put(990,2197){\makebox(0,0)[r]{\strut{} 0.4}}%
      \put(990,2565){\makebox(0,0)[r]{\strut{} 0.5}}%
      \put(990,2934){\makebox(0,0)[r]{\strut{} 0.6}}%
      \put(990,3302){\makebox(0,0)[r]{\strut{} 0.7}}%
      \put(990,3671){\makebox(0,0)[r]{\strut{} 0.8}}%
      \put(990,4039){\makebox(0,0)[r]{\strut{} 0.9}}%
      \put(990,4408){\makebox(0,0)[r]{\strut{} 1}}%
      \put(1608,440){\makebox(0,0){\strut{} 1.973}}%
      \put(2313,440){\makebox(0,0){\strut{} 1.974}}%
      \put(3018,440){\makebox(0,0){\strut{} 1.975}}%
      \put(3723,440){\makebox(0,0){\strut{} 1.976}}%
      \put(4429,440){\makebox(0,0){\strut{} 1.977}}%
      \put(5134,440){\makebox(0,0){\strut{} 1.978}}%
      \put(5839,440){\makebox(0,0){\strut{} 1.979}}%
      \put(6544,440){\makebox(0,0){\strut{} 1.98}}%
      \put(220,2749){\rotatebox{90}{\makebox(0,0){\strut{}failure rate among $100$ attempts}}}%
      \put(4005,110){\makebox(0,0){\strut{}$\beta$}}%
      \put(4062,926){\makebox(0,0)[l]{\strut{}\labIncr$a=1.97638$}}%
      \put(4062,1331){\makebox(0,0)[l]{\strut{}\labIncr$\sum_{res}=0.0423689$}}%
    }%
    \gplgaddtomacro\gplfronttext{%
      \csname LTb\endcsname%
      \put(2172,4603){\makebox(0,0)[l]{\strut{}measured data}}%
      \csname LTb\endcsname%
      \put(2172,4383){\makebox(0,0)[l]{\strut{}$(1+e^{-(\beta-a)/b)})^{-1}$}}%
    }%
    \gplbacktext
    \put(0,0){\includegraphics{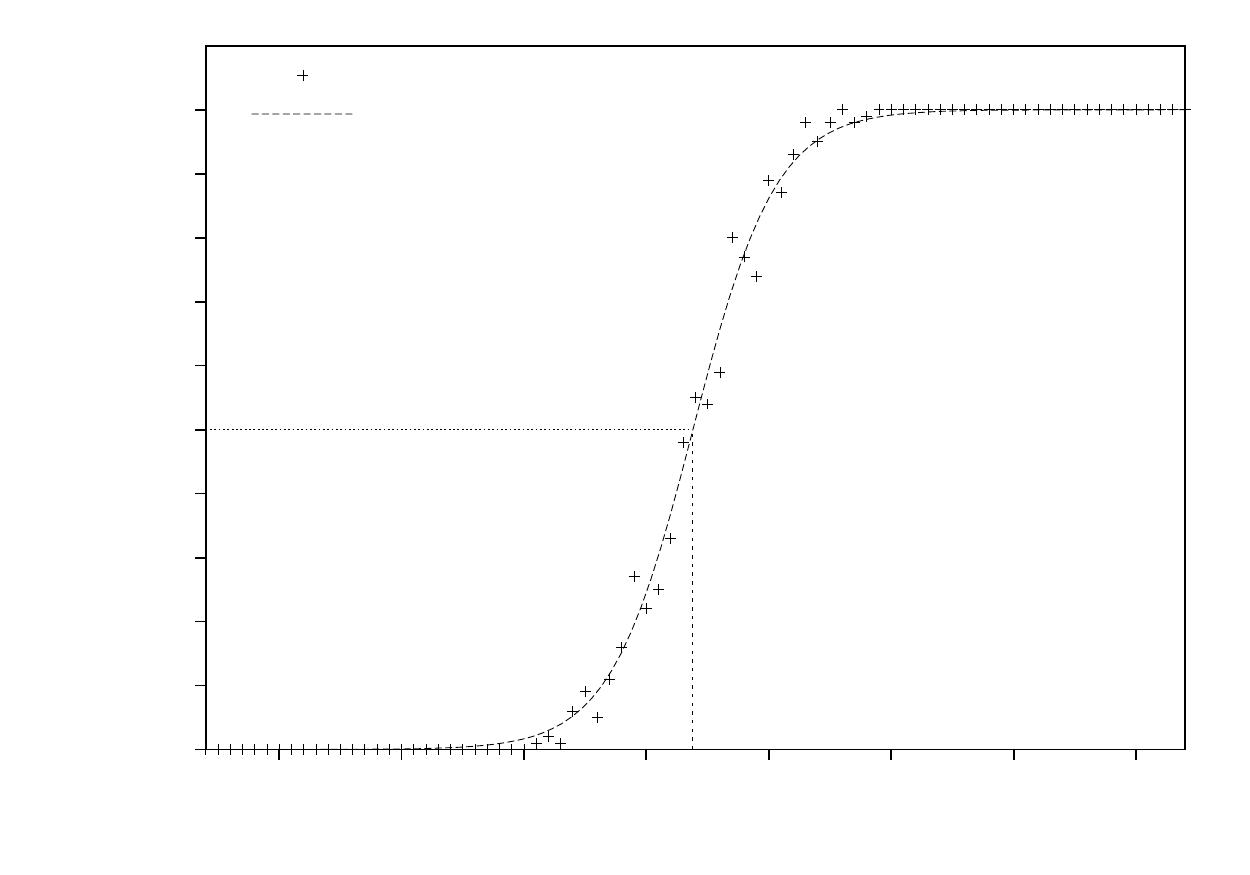}}%
    \gplfronttext
  \end{picture}%
\endgroup

%% file: figures/xkthulhu_MTW_n=10E6_d=3_k=2.tex
\begingroup
  \makeatletter
  \providecommand\color[2][]{%
    \GenericError{(gnuplot) \space\space\space\@spaces}{%
      Package color not loaded in conjunction with
      terminal option `colourtext'%
    }{See the gnuplot documentation for explanation.%
    }{Either use 'blacktext' in gnuplot or load the package
      color.sty in LaTeX.}%
    \renewcommand\color[2][]{}%
  }%
  \providecommand\includegraphics[2][]{%
    \GenericError{(gnuplot) \space\space\space\@spaces}{%
      Package graphicx or graphics not loaded%
    }{See the gnuplot documentation for explanation.%
    }{The gnuplot epslatex terminal needs graphicx.sty or graphics.sty.}%
    \renewcommand\includegraphics[2][]{}%
  }%
  \providecommand\rotatebox[2]{#2}%
  \@ifundefined{ifGPcolor}{%
    \newif\ifGPcolor
    \GPcolorfalse
  }{}%
  \@ifundefined{ifGPblacktext}{%
    \newif\ifGPblacktext
    \GPblacktexttrue
  }{}%
  \let\gplgaddtomacro\g@addto@macro
  \gdef\gplbacktext{}%
  \gdef\gplfronttext{}%
  \makeatother
  \ifGPblacktext
    \def\colorrgb#1{}%
    \def\colorgray#1{}%
  \else
    \ifGPcolor
      \def\colorrgb#1{\color[rgb]{#1}}%
      \def\colorgray#1{\color[gray]{#1}}%
      \expandafter\def\csname LTw\endcsname{\color{white}}%
      \expandafter\def\csname LTb\endcsname{\color{black}}%
      \expandafter\def\csname LTa\endcsname{\color{black}}%
      \expandafter\def\csname LT0\endcsname{\color[rgb]{1,0,0}}%
      \expandafter\def\csname LT1\endcsname{\color[rgb]{0,1,0}}%
      \expandafter\def\csname LT2\endcsname{\color[rgb]{0,0,1}}%
      \expandafter\def\csname LT3\endcsname{\color[rgb]{1,0,1}}%
      \expandafter\def\csname LT4\endcsname{\color[rgb]{0,1,1}}%
      \expandafter\def\csname LT5\endcsname{\color[rgb]{1,1,0}}%
      \expandafter\def\csname LT6\endcsname{\color[rgb]{0,0,0}}%
      \expandafter\def\csname LT7\endcsname{\color[rgb]{1,0.3,0}}%
      \expandafter\def\csname LT8\endcsname{\color[rgb]{0.5,0.5,0.5}}%
    \else
      \def\colorrgb#1{\color{black}}%
      \def\colorgray#1{\color[gray]{#1}}%
      \expandafter\def\csname LTw\endcsname{\color{white}}%
      \expandafter\def\csname LTb\endcsname{\color{black}}%
      \expandafter\def\csname LTa\endcsname{\color{black}}%
      \expandafter\def\csname LT0\endcsname{\color{black}}%
      \expandafter\def\csname LT1\endcsname{\color{black}}%
      \expandafter\def\csname LT2\endcsname{\color{black}}%
      \expandafter\def\csname LT3\endcsname{\color{black}}%
      \expandafter\def\csname LT4\endcsname{\color{black}}%
      \expandafter\def\csname LT5\endcsname{\color{black}}%
      \expandafter\def\csname LT6\endcsname{\color{black}}%
      \expandafter\def\csname LT7\endcsname{\color{black}}%
      \expandafter\def\csname LT8\endcsname{\color{black}}%
    \fi
  \fi
  \setlength{\unitlength}{0.0500bp}%
  \begin{picture}(7200.00,5040.00)%
    \gplgaddtomacro\gplbacktext{%
      \csname LTb\endcsname%
      \put(990,723){\makebox(0,0)[r]{\strut{} 0}}%
      \put(990,1091){\makebox(0,0)[r]{\strut{} 0.1}}%
      \put(990,1460){\makebox(0,0)[r]{\strut{} 0.2}}%
      \put(990,1828){\makebox(0,0)[r]{\strut{} 0.3}}%
      \put(990,2197){\makebox(0,0)[r]{\strut{} 0.4}}%
      \put(990,2565){\makebox(0,0)[r]{\strut{} 0.5}}%
      \put(990,2934){\makebox(0,0)[r]{\strut{} 0.6}}%
      \put(990,3302){\makebox(0,0)[r]{\strut{} 0.7}}%
      \put(990,3671){\makebox(0,0)[r]{\strut{} 0.8}}%
      \put(990,4039){\makebox(0,0)[r]{\strut{} 0.9}}%
      \put(990,4408){\makebox(0,0)[r]{\strut{} 1}}%
      \put(1608,440){\makebox(0,0){\strut{} 1.973}}%
      \put(2313,440){\makebox(0,0){\strut{} 1.974}}%
      \put(3018,440){\makebox(0,0){\strut{} 1.975}}%
      \put(3723,440){\makebox(0,0){\strut{} 1.976}}%
      \put(4429,440){\makebox(0,0){\strut{} 1.977}}%
      \put(5134,440){\makebox(0,0){\strut{} 1.978}}%
      \put(5839,440){\makebox(0,0){\strut{} 1.979}}%
      \put(6544,440){\makebox(0,0){\strut{} 1.98}}%
      \put(220,2749){\rotatebox{90}{\makebox(0,0){\strut{}failure rate among $100$ attempts}}}%
      \put(4005,110){\makebox(0,0){\strut{}$\beta$}}%
      \put(4069,926){\makebox(0,0)[l]{\strut{}\labIncr$a=1.97639$}}%
      \put(4069,1331){\makebox(0,0)[l]{\strut{}\labIncr$\sum_{res}=0.00146084$}}%
    }%
    \gplgaddtomacro\gplfronttext{%
      \csname LTb\endcsname%
      \put(2172,4603){\makebox(0,0)[l]{\strut{}measured data}}%
      \csname LTb\endcsname%
      \put(2172,4383){\makebox(0,0)[l]{\strut{}$(1+e^{-(\beta-a)/b)})^{-1}$}}%
    }%
    \gplbacktext
    \put(0,0){\includegraphics{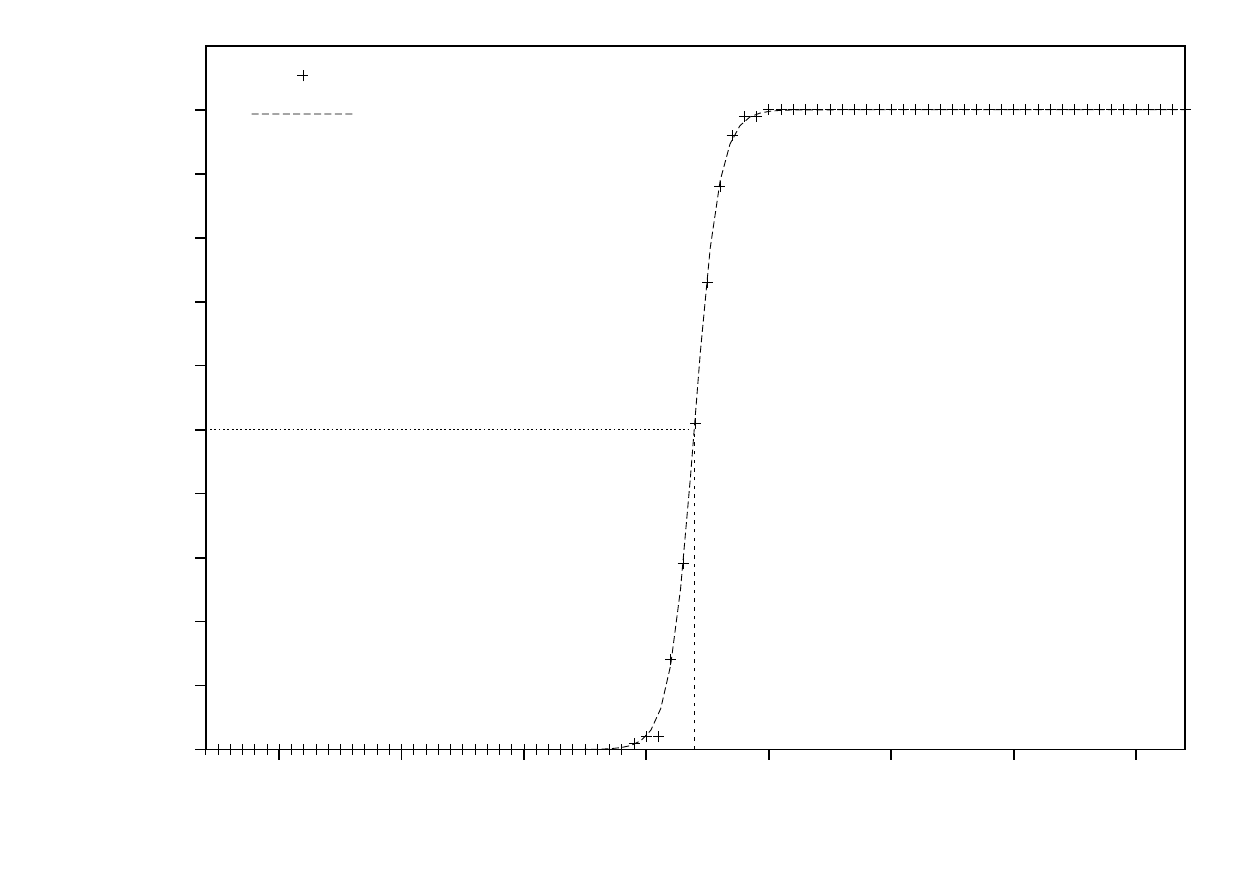}}%
    \gplfronttext
  \end{picture}%
\endgroup

%% file: endlichgsst.tex
\section{Appendix D: Proof of Lemma 5}\label{app:fullproof5}

This appendix contains the full proof of Lemma 5 of Appendix C.

\subsection{Relation to Appendix C}

We use the following notation:
\begin{eqnarray}
q(x) := \exp(x) - x -1 \, \, \,  \nonumber \\
 Q(x) := \frac{x \cdot q'(x)}{q(x)} \,= \, \frac{x(\exp(x)-1)}{\exp(x)-x-1} \nonumber\\
p(x) = \frac{1}{2} \cdot \mbox{PPLUS}(x, k) = \sum_{j \, \, \mbox{ even } } {k \choose j} \cdot x^j, 
\nonumber \\
P(x) := \frac{x p'(x)}{p(x)} \nonumber \\
\mbox{PPLUS}(x, y) \, = \, (1+x)^y \, \, + (1-x)^y , x  \mbox{ arbitrary  if  } \, \, y \ge  0\, \,  \mbox{ integer ,} 1 \ge x \ge 0  
\mbox{ if  } y > 0  \nonumber \\
\mbox{PMINUS}(x,  y) \, = \, (1+x)^y \, \, - (1-x)^y , x  \mbox{ arbitrary  if  }\, \, y \ge 0\, \, \mbox{ integer, } 1 \ge x \ge 0  
\mbox{ if  } y > 0. \nonumber 
\end{eqnarray}
Observe that $ Q(x) $ is the expectation of the integer random variable with probability of $i\ge 2$ 
being  $ =x^i / (i!q(x)). $ Similarly for $P(x).$ 

We have 
\begin{eqnarray}
 Q(x) \mbox{ is strictly monotonously increasing  for } x >  0. \nonumber  \\
 \lim_{x \to \infty } Q(x)- x = 0 \, \mbox{ and } \,
Q(x) >x  \label{baQ1} \\
\lim_{x\to 0} Q(x) =  2 \nonumber \\
P(x) \mbox{ is strictly monotonously increasing from } 0 \mbox{ to } \, \, k \, \, \mbox{ if  } k \,\mbox{ even, }\, \, \,
 k-1\, \mbox{ if } \, k \, \mbox{ odd .} \label{baP1}
\end{eqnarray}
 
\noindent
{\it Proof that $Q(x)$ is increasing.} Simple calculus: 
\begin{eqnarray}
\frac{1}{Q(x)} \,= \, \frac{1}{x} \,- \, \frac{1}{\exp(x)-1}  \nonumber \\
\frac{d}{dx} \frac{1}{Q(x)} \,= \, - \frac{1}{x^2} \, + \, \frac{\exp(x)}{(\exp(x)-1)^2} \, < \,0 
\Longleftrightarrow x^2\,+2 \,< \,\exp(x)+ \frac{1}{\exp(x)}  \label{baQ11} \\
\mbox{ With several differentiations, using } \, \, \frac{d}{dx} \frac{1}{\exp(x)} \,= \, - \frac{1}{\exp(x)} \nonumber \\ \mbox{ the right-hand-side inequality of (\ref{baQ11})  follows  from }\,
 2 \le \exp(x)+\frac{1}{\exp(x)} \nonumber \\
\mbox{ which follows from } (\exp(x)-1)^2 \ge 0. \nonumber 
\end{eqnarray}

\begin{eqnarray} 
n\, = \,\sharp \mbox{variables} \, \, \, , \, \, \, m\,= \, \sharp \mbox{  equations, }\, \, \gamma:= m/n ,\, \, \gamma \,<\, 1 \, ,\, k \, =\, \sharp \mbox{  variables per equation .} \,  \nonumber \\ \mbox{ Note: We have exchanged the meaning of }\, n \,\mbox{ and }\, m \,\mbox{ when  compared to Appendix C.} \nonumber \\
\, \,
k \gamma= Q\left(s(k, \gamma) \right) \mbox{ defines } \,s(k, \gamma)\,\ge 2 \, \, \mbox{ (cf. (\ref{baQ1}.))} \label{baS}
\end{eqnarray}
As $Q(2.0) \,= 2.911 \dots $  (and $Q(2.2) = 3.03\dots $) and $k\ge 3$ 
the assumption $s(k, \gamma)\,\ge 2$ can be made without loss of generality.

\begin{eqnarray}
\mbox{ For }\, \,\omega, \lambda \,\,  \mbox{ we assume throughout, that there  exists an } \, \varepsilon > 0\, \mbox{ such that } \nonumber \\
\varepsilon \le \omega \le 1-\varepsilon \,\, \mbox{ and } \,\, 
\frac{2\omega}{k \gamma}+ \varepsilon \le \lambda \le \frac{2(1- \omega)}{k\gamma}- \varepsilon.\label{baOMLA}
\end{eqnarray}
Condition (\ref{baOMLA})  ensures that we stay away from the boundary of the domain allowed for 
$ \omega, \lambda $ (cf. (15) of Appendix C.)

\begin{definition} \label{defPsi}
For $a, b, c, s> 0$  
\begin{eqnarray}
\Psi_1(\omega,  a, b, s) \, \, 
 \,:= \, \left(\frac{q(a)}{q(s)\omega}\right)^\omega \left(\frac{q(b)}{(1- \omega)\cdot q(s)}\right)^{1 - \omega} \label{defPsi1} \nonumber \\ 
\Psi_2(\lambda, a, b, c, s) \,\,:= 
    \left(\left( \frac{\lambda \cdot s}{a \cdot c}\right)^\lambda \cdot \left(\frac{(1 - \lambda)\cdot s}{b}\right)^{1-\lambda}\right)^k\cdot p(c)  \nonumber  \\
\Psi(\omega, \lambda, a, b, c, s) := \Psi_1(\omega,  a, b, s) \cdot \big( \Psi_2(\lambda, a, b, c, s)\big)^\gamma.  \nonumber 
\end{eqnarray}
\end{definition}

\begin{definition} \label{defpsi}
\begin{eqnarray}
\psi(\omega, \lambda) := \Psi \big(\omega ,\lambda, a, b, c, s)  \big),\, \, \mbox{ with }\,\,s=s(k, \gamma)  \nonumber \\
\mbox{ and } \, \, a, b, c \mbox{ given by } Q(a) = \frac{\lambda k \gamma}{\omega} , \,\, \, Q(b) = \frac{(1-\lambda)k \gamma }{1-\omega}, \,\, \,  \, P(c) \,= \, \lambda \cdot k  \, \, \mbox{ (Recall (\ref{baQ1}), (\ref{baP1}) )} \label{baABC} 
\end{eqnarray}
\end{definition}
Oberserve that (\ref{baOMLA}) ensures that $a, b, c>0$ in Definition \ref{defpsi}. 

\begin{lemma}[Formula (40) from Appendix C]\label{baMIN}
\begin{eqnarray}
\psi(\omega, \lambda) = 
\min_{a, b, c>0} \Psi \big(\omega ,\lambda, a, b, c, s) \nonumber  
\end{eqnarray}
\end{lemma}

The point left  unfinished in Appendix C is a full proof of the following theorem. 
\begin{theorem}[Lemma 5 in Appendix C] \label{main}
For $\gamma<1\, \, \, \, \psi( \omega, \lambda )$ achieves its unique global maximum over $\omega, \lambda$ saitsfying (\ref{baOMLA})  for  $(\omega, \lambda)=(1/2, 1/2).$ And $\psi(1/2, 1/2) \,= \, 2^{1-\gamma}.$
\end{theorem}
We have
 $\, \psi(1/2, 1/2)\,=  \Psi(1/2, 1/2, s, s, 1, s)\, =\,2^{1-\gamma}. $ 

As already observed in Appendix C  we can restrict attention to  $\lambda \le 1/2.$ 
\begin{lemma} \label{halb}
For $\lambda > 1/2$ we have $\psi(\omega, \lambda) \le \psi (1 - \omega, \, \,\, 1 - \lambda).$ 
\end{lemma} 
\begin{proof} 
\begin{eqnarray}
\mbox{ Let } \hat{\lambda} = 1 - \lambda \le 1/2 \mbox{ and }   \hat{\omega}= 1- \omega. \mbox{ Let } a, b, c \mbox{ be such that } \psi ( \hat{\omega}, \hat{\lambda} ) = \Psi(\hat{\omega}, \hat{\lambda},a, b, c, s)\nonumber \\
\mbox{ Then } c < 1 \mbox{ as } P(1)= \frac{1}{2}k \mbox{ (cf. (\ref{baABC})}) \nonumber \\
\Psi_1(\omega,  b, a, s) \,= \,\Psi_1(\hat{\omega}, a, b, s) .\nonumber \\
\Psi_2(\hat{\lambda}, a, b, c, s)  \, \, = \, \, \left(\left( \frac{\hat{\lambda} \cdot s}{a \cdot c}\right)^{\hat{\lambda}} \cdot \left(\frac{\lambda\cdot s}{b}\right)^\lambda\right)^k\cdot p(c) \nonumber  \\
\Psi_2( \lambda , b, a, 1/c, s) = \left(\left( \frac{\lambda \cdot s}{b \cdot(1/c) } \right)^\lambda \cdot \left(\frac{\hat {\lambda}\cdot s}{a}\right)^{\hat{\lambda}}\right)^k \cdot p(1/c) \, \,\ \nonumber  \\
= \, \Psi_2( \hat{\lambda}, a, b, c, s) \cdot \left(\frac{1}{1/c}\right)^k \cdot \frac{p(1/c)}{p(c)} \le \Psi_2( \hat{\lambda}, a, b, c, s) \nonumber \\
\mbox{ as for  } 0 \le  c \le 1 \, \,  \, \,
 c^k \cdot \frac{\mbox{PPLUS}(1/c, k)}{\mbox{PPLUS}(c, k) }\,= \frac{(c+1)^k + (c-1)^k}{(1+c)^k+ (1-c)^k} \le 1.\nonumber 
\end{eqnarray}
Now, Lemma \ref{baMIN} implies the claim. 
\end{proof}
\hspace*{1cm} 
\begin{definition} \label{defGA}
For $a \ge 0, s> 0 $ and $ 0\le c \le 1$ we define
\begin{eqnarray}
\Gamma_1(a, s)= 1 + \frac{q(as)}{q(s)}   \nonumber \\
\Gamma_2(a, c, s)= \left(\frac{1}{1+ac}\right)^{Q(s)} \cdot \mbox{ \em PPLUS }(c, Q(s) ) \nonumber \\
\Gamma (a, c, s) = \Gamma_1(a, s) \cdot \Gamma_2(a, c, s) \nonumber 
\end{eqnarray}
\end{definition}

Next the key proposition. 
\begin{proposition} \label{schluessel}
 \begin{eqnarray}
\mbox{ Let  }\, \, s \ge 2. \, \,  \mbox{ For } \, \, 0 \le P \le 1\, \, \mbox{ there exist } 0 \le a, c \le 1 \, \,\, \mbox{ with } \, \,  P=ac , \nonumber \\ \mbox{ such that }  \Gamma(a, c, s) \le 2, 
\mbox{ equality only for }\,P=0\, \mbox{ or } \, P=1 .\, \nonumber 
\end{eqnarray}
\end{proposition}

\noindent
{\it Proof of Theorem \ref{main} from Proposition \ref{schluessel}.} 
\begin{eqnarray}
\mbox{ With } s =s(k, \gamma) \mbox{ for } b \mbox{ and } as \mbox{ with } a > 0 \mbox{ for } a  \mbox{ in Definition \ref{defPsi} } \nonumber \\ \mbox{ we have } \psi(\omega, \lambda) \le \Psi(\omega, \, \lambda, as,\, s,\,  c,\, s)  \, \mbox{( Lemma \ref{baMIN} .)}  \nonumber \\
\Psi_1(\omega, as, s, s )\, =\,\left(\frac{q(as)}{q(s)\omega}\right)^\omega \left(\frac{q(s)}{(1- \omega)\cdot q(s)}\right)^{1 - \omega}  \, \le \, 1 \,+ \, \frac{q(as)}{q(s)}\, = \, \Gamma_1(a, s) \, \, \mbox{ (AGM inequality ) }\nonumber \\
P:=\frac{\lambda}{1-\lambda}.  \mbox{  Then } \lambda\,= \, \frac{P}{1+P}\, \, , \, \,1- \lambda= \frac{1}{1+P}. \mbox{ Let  } ac=P .\nonumber \\
 \Psi_2(\lambda, as, s, c, s) \,= \, \left( \left( \frac{\lambda s}{asc}\right)^\lambda \cdot 
\left( \frac{(1-\lambda)s}{s}\right)^{1-\lambda}\right)^k \cdot p(c)  \, \, \nonumber \\
= \, \, \left(\left(\frac{1}{1+ac}\right)^\lambda \cdot \left (\frac{1}{1+ac} \right)^{1-\lambda}\right)^k \cdot p(c)
=\, \, \left(\frac{1}{1+ac} \right)^k \cdot p(c). \, \, \,  \label{schlmain}  
\end{eqnarray}

The definition of $\Psi$ requires $\Psi_2^\gamma$ therefore: $p(c)\,= \,$
\begin{eqnarray}
= \frac{1}{2}  \,\left((1+c)^{k\gamma/\gamma}\,+ (1-c)^{k\gamma/\gamma}\right) \, \, \le  \frac{1}{2} \,\left((1+c)^{k\gamma}\,+ (1-c)^{k\gamma} \right)^{1/\gamma} \,= \, \frac{1}{2} \left(\mbox{PPLUS}(c, k\gamma)\right)^{1/\gamma}\label{schlmainn} \\
\mbox{ Note } x_1^y + x_2^y \, \le \, (x_1 +x_2)^y  \mbox{ for } y \ge 1, x_1, x_2 \ge 0,  
\mbox{ equality only for } y=1 \mbox{ or one of the }  x_i=0. \nonumber \\
\mbox{ By } Q(s)=k\gamma \, \, \mbox{ (\ref{schlmain}) and (\ref{schlmainn}) implies }\, \, \nonumber \\ \,\left(\Psi_2(\lambda,\, as, s, c, s)\right)^\gamma \, \,
\le \, \left(\frac{1}{1+ac}\right)^{Q(s)}(p(c))^\gamma \,\le \,\nonumber \\
 \le \,\left(\frac{1}{1+ac}\right)^{Q(s)}\left(\frac{1}{2}\right)^\gamma \mbox{PPLUS}(c, Q(s)) \,= \,\left(\frac{1}{2}\right)^\gamma\Gamma_2(a, c, s). \nonumber 
 \end{eqnarray}
For $0 \le\lambda\le 1/2$ we have $0 \le P\le 1$ and Proposition \ref{schluessel} applies. With  $a, c$ which satisfy  Proposition \ref{schluessel}  we have Theorem \ref{main} (using Lemma \ref{halb}.)
Note that the bound $2^{1-\gamma}$ is only reached for $\omega=1/2$ and $P=1$ that is $\lambda=1/2.$ 
(The case $P=0$ need not be considered for  Theorem \ref{main}.)
\hspace*{1cm} \hfill \qed. 

\subsection{Proof of Proposition \ref{schluessel}}

\begin{remark}
We prove Proposition \ref{schluessel} only for $s\ge 6.$ We have $Q(6)=6.09\dots$ and the proof covers
all $k\gamma \ge 6.09 \dots.$ Reading the details it should be clear that we can also find a proof 
for $6 \ge s \ge 2 . $
Some additional, purely technical effort seems unavoidable for this.
\end{remark}

We usually write $Q$ instead of $Q(s)$ (cf. (\ref{baS}.))

We need to consider  $\Gamma(a, c, s)$ for  $0 \le a, c\, \le 1.$ It has the following properties. 
\begin{lemma} \label{lebagamma}
\begin{itemize}
\item[(a)]
$\Gamma(0, 0, s) = 2, \, \, \,   \Gamma(0, 1, s) = 2^{Q(s)},\,  $ 
$ \Gamma(1, 0, s) = 2 \cdot 2 = 4, \, \, \, \Gamma(1, 1, s) = 2 $ \vspace{1mm}
 \item[(b)] For each $0 \le c \le 1 $ $\Gamma(a, c)$ has only one extremum in $a.$ It is a minimum and $0 \le a_{\mbox{min}} \le 1.$ If $c=0$ then $a_{\mbox{min}} =0$ if $c=1$ then $a_{\mbox{min}}=1.$ \vspace{1mm}
\item[(c)] For each $0 \le a\le 1 $ $\Gamma(a, c)$ has only one extremum in $c.$ It is a minimum and $0 \le c_{\mbox{min}} \le 1.$ If $a=0$ then $c_{\mbox{min}} =0$ if $a=1$ then $c_{\mbox{min}}=1.$
\end{itemize} 
\end{lemma}
\begin{proof}
(b) \begin{eqnarray}
\frac{d}{da} ( \ln \Gamma( a, c, s) )\,= \, \frac{\frac{s (\exp(sa)-1)}{q(s)}}{\Gamma_(a, s)} - 
Q \cdot \frac{c}{1+ac} > = < 0 \nonumber \\ 
  \Longleftrightarrow \frac{\frac {\exp(sa)-1 }{\exp(s)-1}}{\Gamma_1(a, s)}\,- \,\frac{c}{1+ac} > = < 0 \, \, \mbox{ (Division with Q(s)) }     \nonumber \\
  \Longleftrightarrow \frac{K}{1+ L-aK} > = < c \, \mbox{ with } \, \, K=\frac{\exp(sa)-1}{\exp(s)-1}, \, L= 
  \frac{\exp(sa)-sa-1}{\exp(s)-s-1} \nonumber \\
 \frac{K}{1+ L-aK}  \mbox{ is strictly increasing in } 0 < a < 1 \nonumber \\
 \mbox{ Moreover } \frac{K}{1+ L-aK}_{|a=0}\,=\,0,\,\frac{K}{1+    L-aK}_{|a=1}\,= \, 1 \nonumber  
 \end{eqnarray} 
  
(c)\begin{eqnarray}
\frac{d}{dc} ( \ln \Gamma( a, c, s)) \,= \, -Q \cdot \frac{a}{1+ac} + Q \cdot 
\frac{\mbox{PMINUS}(c, Q-1)}{\mbox{PPLUS}(c, Q) } \,\, \, >=< \, \,\, \,0 \nonumber \\
\Longleftrightarrow \, \, \frac{\mbox{PMINUS}(c, Q-1)}{\mbox{PPLUS}(c, Q) } >=<   \frac{a}{1+ac} \nonumber \\
\Longleftrightarrow  \mbox{PMINUS}(c, Q-1)\cdot (1+ ac) >=< a \cdot \mbox{PPLUS}(c, Q) \nonumber \\
\Longleftrightarrow \mbox{PMINUS}(c, Q-1) >=< a \cdot ( \mbox{PPLUS}(c, Q)\,-\,c\cdot \mbox{PMINUS}(c, Q-1)) \nonumber  \\
\Longleftrightarrow \frac{\mbox{PMINUS}(c, Q-1)} {\mbox{PPLUS}(c, Q-1)} >=< a \label{baPPPM1} \\
\mbox{ We calculate below } \frac{\mbox{PMINUS}(c, Q-1)} {\mbox{PPLUS}(c, Q-1)} \mbox{ is strictly increasing in } 0<c<1.  \label{baPPPM2} \\
\mbox{ Moreover } \frac{\mbox{PMINUS}(c, Q-1)} {\mbox{PPLUS}(c, Q-1)}_{|c=0}\, = \,0, \, \,  \frac{\mbox{PMINUS}(c, Q-1)} {\mbox{PPLUS}(c, Q-1)}_{|c=1}\,= 1\, \nonumber 
\end{eqnarray}

 \noindent
 {\it Proof of (\ref{baPPPM2})} 
\begin{eqnarray}
 \frac{d}{dc} \frac { \mbox{ PMINUS}(c, Q ) }{\mbox{PPLUS}(c, Q)}  \,= \, \frac{4(1-c^2)^{Q-1}}{(\mbox{PPLUS}(c, Q))^2} \label{baPPPM3}
\end{eqnarray}
\end{proof}

Some experimentation reveals that $\Gamma(a, c, s)<2$ only for  $a, c$ as follows:
\begin{itemize}
\item[-] An area like 
$1 > a \ge 1- \varepsilon $  and all $1 > c \ge \varepsilon.$ $\varepsilon$ decreasing in $s.$
The strip becomes narrower when $c\rightarrow 1.$
\item[-] An area like 
$0 < c \le  \varepsilon$ and all $a < 1- \varepsilon.$   
The strip becomes narrower when $c \rightarrow 0.$
\item[-] Observe that for each $0 \le P \le 1$ we can find an $a, c$ with $P=ac$ in the area described. 
\end{itemize}

Proposition \ref{schluessel} follows from the following three lemmas. 
First, we first single out $4$ argument pairs  $(a, c)$  for which we can bound 
$\Gamma(a, c, s)<2.$
\begin{lemma}  \label{kons} There is a constant $B<2$ such that $ \Gamma(a, c, s) < B $ for:
\begin{itemize}
\item[(a)] $ a=1-1/Q, \, \, c= 1/Q $ and  $s \ge 4.2.$ \vspace{1mm}
\item[(b)] $a=1-1/Q, \, \, c = 11/20$  and $ s\ge 5 .$ \vspace{1mm}
\item[(c)] $a= 1/2,\, \, c=1/Q  $ and $s \ge 4.8.$  \vspace{1mm}
\item[(d)] $a= 1/2,\, \, c=1/(2Q)  $ and $s \ge 3.0.$ 
\end{itemize}
\end{lemma} 

The next lemma deals with pairs $(a, c)$ with $ac$ in the neighbourhood of $0$
\begin{lemma} \label{cbei0} 
Let  $s\ge 4.2$ and $A= A(c, s) = c \cdot Q.$  Then 
\begin{eqnarray} \Gamma(A, c, s) \le 2  \mbox{ for } 0 \le c \le 1/(2Q)  , \mbox {with  equality only for } c =  0. \nonumber 
\end{eqnarray}
\end{lemma}

The next lemma treats $ac$ in the neighbourhood of $1.$
\begin{lemma} \label{cbei1} 
Let  $s \ge 6$ and $A= A(c, s) = 85/(100Q)  \cdot c + 1- 85/(100Q ). $  Then 
\begin{eqnarray} \Gamma(A, c, s) \le 2  \mbox{ for } 2/5 \le c \le 1   , \mbox {with  equality only for } c =  1. \nonumber 
\end{eqnarray}
\end{lemma}

\hspace*{1cm} \\
\noindent
{\it Proof of Proposition \ref{schluessel} for $s \ge 6$ from the preceding three lemmas.} 
The lemmas capture all $P=ac:$ 
\begin{itemize}
\item Lemma \ref{cbei0} captures $ 0 \le P \le  (1/2Q)Q(1/2Q)= 1/(4Q).$ \vspace{1mm}
\item Lemma \ref{kons} (d) and (c) with Lemma \ref{lebagamma} (b) capture 
$1/(4Q) \le P \le (1/2)(1/Q)$ \vspace{1mm}
\item Lemma \ref{kons} (c) and (a) with Lemma \ref{lebagamma} (c) capture 
$(1/2)((1/Q)\le P \le (1-1/Q)(1/Q)$ \vspace{1mm}
\item Lemma \ref{kons} (a) and (b) capture $(1-1/Q)(1/Q)\le P \le (1-1/Q)(11/20).$ \vspace{1mm}
\item Lemma \ref{cbei1} captures $(85/(100Q))(2/5)+1-85/(100Q))(2/5)  \le P \le 1.$
\end{itemize}
As  $(-(3/5)( 85/(100Q))+ 1)2/5\, \le \, 11/20-11/(20Q)  \Longleftrightarrow Q \ge 346/150=2.3\dots$ all $P$ are captured.

\subsection{ Proof of Lemma  \ref{kons} }

\noindent
{\bf Lemma \ref{kons} (repeated) } {\em 
 There is a constant $B<2$ such that $ \Gamma(a, c, s) < B $ for:
\begin{itemize}
\item[(a)] $ a=1-1/Q, \, \, c= 1/Q $ and  $s \ge 4.2.$ \vspace{1mm}
\item[(b)] $a=1-1/Q, \, \, c = 11/20$  and $ s\ge 5 .$ \vspace{1mm}
\item[(c)] $a= 1/2, \, \, c=1/Q  $ and $s \ge 4.8.$ \vspace{1mm}
\item[(d)] $a= 1/2, \, \, c= 1/(2Q)  $ and $s \ge 3.0\dots $ 
\end{itemize} }
\begin{proof} 
(a)  We have 
\begin{eqnarray}
\Gamma_1(1-1/Q \, ,\, s)\,= \, 1+ \frac{\exp((1-1/Q)s) - (1-1/Q)s-1}{\exp(s)-s-1}  \nonumber \\
\Gamma_2(1-1/Q, 1/Q, s) \,= \, \, \left(\frac{Q^2+Q } {Q^2 + Q-1 }\right)^Q\,+ \,\left(\frac{Q^2-Q } {Q^2 + Q-1 }\right)^Q. \nonumber
\end{eqnarray}

Further below we show: 
\begin{eqnarray}
\mbox{For } 0 \le a \le 1 \, \, \frac{\exp(as)-as-1}{\exp(s)-s-1}\, \le \,\frac{\exp(as)-1}{\exp(s)-1} \,\le \, \frac{\exp(as)}{\exp(s)} , \,
\mbox{equality only if } a=1 \mbox{ or } a=0 \label{baLK1}
\end{eqnarray}
Therefore 
\begin{eqnarray}
\Gamma_1(1-1/Q,\, \, s) \,<  \,1+ \frac{\exp((1-1/Q)s)}{\exp(s)}\,= 1 + \exp(-s/Q) . \label{bagamma11}
\end{eqnarray}
Concerning  $\Gamma_2(1-1/Q, \, 1/Q, \, s)$ we observe:
\begin{eqnarray}
\left(\frac{Q^2-Q } {Q^2 + Q-1 }\right)^Q \, = \,\left(1\,+ \, \frac{-2Q+1}{Q^2+Q-1}\right)^Q \, \le \exp\left(-\frac{(2Q-1)Q}{Q^2+Q-1} \right). \nonumber
\end{eqnarray}
Altogether we have
\begin{eqnarray}
\Gamma(1-1/Q, 1/Q, s)_{|s=4.2} 
\,< \,\left( 1+ \exp(-s/Q) \right) \left(\left( \frac{Q^2+Q } {Q^2 + Q-1}\right)^Q \, + \, \exp\left(-\frac{2Q^2-Q}{Q^2+Q-1}\right)\right)_{|s=4.2}\, \nonumber \\
=\, 1.9829 \dots .\label{1-1/q1/q1}\\
 \lim_{s\to \infty} \Gamma(1-1/Q, 1/Q, s)\,= \,  \lim_{s \to \infty} \,\left( 1+ \exp(-s/Q) \right) \left(\left( \frac{Q^2+Q } {Q^2 + Q-1}\right)^Q \, + \, \exp\left(-\frac{2Q^2-Q}{Q^2+Q-1}\right)\right)\,= \, \nonumber \\
\,=\,(1+\exp(-1))(1+\exp(-2)) \approx 1.55. \nonumber 
\end{eqnarray}
We show that  the function in (\ref{1-1/q1/q1}) is decreasing in $s\ge 4.2.$ 

We consider the factors of (\ref{1-1/q1/q1}) separately. 
\begin{eqnarray}
  1 + \exp(-s/Q) \,= \,1+ \exp \left(  - \frac{ \exp(s)-s-1} {\exp(s)-1}\right) \nonumber \\
\frac{\exp(s)-s-1}{\exp(s)-1}\,= \, 1 \,- \, \frac{s}{\exp(s)-1} , \,\mbox{ and }  \frac{s}{\exp(s)-1} \mbox{ is decreasing }. \nonumber \\
\mbox{ Therefore the leftmost factor in (\ref{1-1/q1/q1}) }, 1+\exp(-s/Q), \mbox{ is decreasing. }  \label{bagamma12}
\end{eqnarray}

We come to the second factor of (\ref{1-1/q1/q1}.)
Because of (\ref{baQ1}) we can consider $Q$ as an independent argument.  
\begin{eqnarray}
\frac{d}{dQ}\, \ln \left(\frac{Q^2+Q } {Q^2 + Q-1 }\right)^Q  \nonumber \\
= \,\ln(Q^2+Q) - \ln(Q^2+ Q -1) + \frac{Q(2Q+1)}{Q^2+Q}- \frac{Q(2Q+1)}{Q^2+Q-1} \nonumber \\ \,= \, \frac{1}{x}+ \frac{Q(2Q+1)}{Q^2+Q}- \frac{Q(2Q+1)}{Q^2+Q-1} \nonumber \\
\mbox{ for an } Q^2+Q-1 < x < Q^2+Q \mbox{ by the Mean Value Theorem.} \nonumber 
\end{eqnarray}
The preceding term is $<0$ by direct calculation with  $x=Q^2+Q-1.$  

\begin{eqnarray}
\frac{2Q^2-Q}{Q^2+Q-1} \mbox{  is increasing by  for } \,Q>1 \, \mbox{ by simple differentiation. }\nonumber   \\
\mbox{ Therefore }  \exp\left(-\frac{2Q^2-Q}{Q^2+Q-1}\right) \mbox{ is decreasing.} \nonumber 
\end{eqnarray}

\hspace*{2cm} \\
\noindent
{\it Proof of (\ref{baLK1}.)}
For $a=0$ and $a=1$ the first two terms are equal. 
\begin{eqnarray}
(\exp(as)-as-1)(\exp(s)-1)\, < \,(\exp(as)-1)(\exp(s)-s-1) \nonumber \\
\Longleftrightarrow \, -as(\exp(s)-1) < -s (\exp(as)-1) \nonumber \\ 
\Longleftrightarrow \, a(\exp(s)-1) >  \exp(as)-1  \, \, \mbox{ which holds by convexity for } \, \, 0<a<1.\nonumber
\end{eqnarray}
The first inequality is shown. 

For $a=1$ the terms of the second inequality are equal. Using $ a <1$  a simple computation shows the 
required inequality.   \\

 \hspace*{3cm} \\

(b)  
 \begin{eqnarray} 
 \mbox{We have }\,\,\Gamma_1(1-1/Q, s) \, < \, 1 \,+ \, \exp( -s/Q ) \mbox{ (by  (\ref{bagamma11})) } \nonumber \\
 \Gamma_2(1-1/Q, \,11/20, \, s) \,= \, \left(\frac{31}{31 - 11/Q} \right)^Q +  \left( \frac{ 9 }{31-11/Q} \right)^Q \nonumber \\
\Gamma(1-1/Q, 11/20, s)_{|s=5}\,<\, (1 \,+ \, \exp(-s/Q)) \Gamma_2(1-1/Q,\, 11/20,\, s)_{|s=5} \,= \, 1.9971 \dots  \label{1-1/q11/201} \\
\mbox{ And  } \lim_{s \to \infty}\,(1 \,+ \, \exp(-s/Q)) \Gamma_2(1-1/Q,\, 11/20,\, s)=  (1+\exp(-1))\exp(11/31) \,= \, 1.9505 \dots \nonumber 
 \end{eqnarray}
We show that the function in (\ref{1-1/q11/201}) is decreasing in $s.$ 
The first
factor is decreasing as seen in (\ref{bagamma12}.)

The two additive terms of $\Gamma_2(1-1/Q, 11/20, s)$ are considered separately. 
\begin{eqnarray}
\left(\frac{31}{ 31- 11/Q} \right)^Q  \,= \,  \frac{1}{(1-11/(31Q))^Q }  \nonumber \\
\ln (1-11/(31Q))^Q \,= \, Q(-\frac{11}{31Q}\,- \, \frac{1}{2} \left(\frac{11}{31Q}\right)^2 \,- \,\frac{1}{3}\left(\frac{11}{31/Q}\right)^3\,- \, \cdots ) \mbox{ (Logarithm series)} \nonumber \\
\mbox{This implies by termwise differentiation } \frac{d}{dQ}\,\ln (1-11/(31Q))^Q\,>\,0 \, \mbox{ for } Q >1. \nonumber \\
\mbox{ Therefore }\,  \frac{1}{(1-11/(31Q))^Q }   \mbox{ is decreasing .} \label{1-1/q11/202}
\end{eqnarray}

The last term, 
\begin{eqnarray}
\left(\frac{9}{(31-11Q)}\right)^Q\,= \, \left(\frac{1}{(1-11/(31Q))\cdot 31/9}\right)^Q \nonumber 
\end{eqnarray} 
is the product of two decreasing terms by (\ref{1-1/q11/202}.) \\

(c) We show below  the following statements (\ref{bagamma13}) and (\ref{1/21/q1}.) 
\begin{eqnarray}
\Gamma_1(a, s) \, \mbox{ is decreasing in } s \, \mbox{ for } \, 0<a<1. \label{bagamma13} \\
\Gamma_2(1/2, 1/Q, s) \,= \,\left( \frac{2Q+2}{2Q+ 1} \right)^Q \,+ \, \left( \frac{2Q-2}{2Q+ 1} \right)^Q  \nonumber \\
\,= \, \left(\frac{1+1/Q}{1+1/(2Q)}\right)^Q\,+ \, \left(\frac{1-1/Q}{1+1/(2Q)}\right)^Q 
\mbox{is increasing in } s. \label{1/21/q1}
\end{eqnarray}

We have 
\begin{eqnarray}
\lim_{s \to \infty} \, \Gamma_2(1/Q,\,1/2,\,s) \,=\, \exp(1/2)\,+ \, \exp(-3/2)\,= \, 1.8718 \dots \nonumber \\
\mbox{ With (\ref{1/21/q1}) we get } 
 \Gamma(1/2,\, 1/Q,\,s)_{|s=4.8} \,= \, \Gamma_1(1/2, s)\Gamma_2(1/2,\, 1/Q,\, s)_{|s=4.8} \,  \nonumber \\
 < \,  \Gamma_1(1/2, s)( \exp(1/2)\,+ \, \exp(-3/2) )_{|s=4.8} \,= \, 1.9951 \dots \nonumber 
\end{eqnarray}
With (\ref{bagamma13}) 
the preceding inequality implies the claim.

We prove (\ref{1/21/q1}.) 
The additive terms of $\Gamma_2(1/Q, 1/2, s)$ are treated independently. 
\begin{eqnarray}
\frac{d}{dQ} \ln \left( \frac{2Q+2}{2Q+1 } \right)^Q  \,
\,= \,\ln(2Q+2)  - \ln(2Q+1) + Q\frac{2}{2Q+2} - Q\frac{2}{2Q+1}  \nonumber \\
\,=\, \ln(x+1)  - \ln(x) + \frac{x-1}{x+1} - \frac{x-1}{x} \mbox{ with } x=2Q+1 . \nonumber \\
\ln(x+1)  - \ln(x) + \frac{x-1}{x+1} - \frac{x-1}{x}\,=\,  \frac{1}{y}  +  \frac{x-1}{x+1} - \frac{x-1}{x} \nonumber  \\
\mbox{ for an  } x <y < x+1 \mbox{ with the Mean Value Theorem.} \nonumber 
\end{eqnarray}
The very last expression for $y=x+1$ is $> 0$   and $((2Q+2)/(2Q+1))^Q$ is increasing. 

 \begin{eqnarray}
\ln \left(\frac{1-1/Q}{1+1/(2Q)}\right)^Q \,= \, Q\left(-\frac{1}{Q}\,- \, \frac{1}{2}\left(\frac{1}{Q}\right)^2\,- \, \frac{1}{3}\left(\frac{1}{Q}\right)^3-\cdots \right)\,- \, \nonumber \\
-Q\left(\frac{1}{2Q}\,- \, \frac{1}{2}\left(\frac{1}{2Q}\right)^2\, \,+ \, \frac{1}{3}\left(\frac{1}{3Q}\right)^3-\cdots \right) \mbox{(Logarithm series.) }\nonumber \\
\mbox{ Termwise differentiation shows that }\,\frac{d}{dQ} \ln \left(\frac{1-1/Q}{1+1/(2Q)}\right)^Q >0. \nonumber
\end{eqnarray}
And $((2Q-2)/(2Q+1))^Q$ is increasing, (\ref{1/21/q1}) is shown.  \\

\noindent
{\it Proof of (\ref{bagamma13}.)} 
\begin{eqnarray}
\frac{d}{ds} \ln\left( \frac{\exp(sa)-sa-1}{\exp(s)-s-1} \right) \,= \, \frac{a(\exp(sa)-1)}{\exp(sa)-sa-1}
- \frac{\exp(s)-1}{\exp(s)-s-1}\, <\,0 \Longleftrightarrow \nonumber \\
a(\exp(s)-s-1)\, \,  \,  < \, (\exp(s)-1)\frac{ \exp(sa)-sa-1}{\exp(sa)-1}\,\,= \, 
(\exp(s)-1)\left(1\, - \, \frac{sa}{\exp(sa)-1} \right)\label{1/21/q2}
\end{eqnarray}
For $a=0$ both sides of (\ref{1/21/q2}) are $0,$ for $a=1$ both sides are $\exp(s)-s-1.$ 
Given $s>0,$ the  right-hand-side is concave in $0<a<1$ and (\ref{1/21/q2}) holds for all $0<a<1.$ \\

(d) We show below (\ref{1/21/2q1})
\begin{eqnarray}
\Gamma_2(1/2, \, 1/(2Q),\,  s) \,= \,\left( \frac{4Q+2}{4Q+ 1} \right)^Q \,+ \, \left( \frac{4Q-2}{4Q+ 1} \right)^Q  \nonumber \\
\,= \, \left(\frac{1+1/(2Q)}{1+1/(4Q)}\right)^Q\,+ \, \left(\frac{1-1/(2Q)}{1+1/(4Q)}\right)^Q 
\mbox{is increasing in } s. \label{1/21/2q1}
\end{eqnarray}

We have 
\begin{eqnarray}
\lim_{s \to \infty} \, \Gamma_2(1/2,\,1/(2Q),\,s) \,=\, \exp(1/4)\,+ \, \exp(-3/4)\,= \, 1.7563 \dots \nonumber \\
\mbox{ With (\ref{1/21/2q1}) we get } 
 \Gamma(1/2,\, 1/(2Q),\,s)_{|s=3.0} \,= \, \Gamma_1(1/2, s)\Gamma_2(1/2,\, 1/Q,\, s)_{|s=3.0} \,  \nonumber \\
 < \,  \Gamma_1(1/2, s)( \exp(1/4)\,+ \, \exp(-3/4) )_{|s=3.0} \,= \, 1.972 \dots \nonumber 
\end{eqnarray}
With (\ref{bagamma13}) 
the preceding inequality implies the claim.  

To prove (\ref{1/21/2q1}) we proceed as in the proof of (\ref{1/21/q1}).
The $(4Q+2)/(4Q+ 1)$-term is treated as $(2Q+2)/(2Q+1)$
only with $x=4Q+1.$ And the $(1-1/(2Q))/(1+1/4Q)$-term is treated with the logarithm series. 
\hspace*{10cm} \end{proof}

\subsection{Proof of Lemma \ref{cbei0}}

\noindent
{\bf Lemma \ref{cbei0} (repeated) }{ \em
Let  $s\ge 4.2$ and $A= A(c, s) = c \cdot Q.$  Then 
\begin{eqnarray} \Gamma(A, c, s) \le 2  \mbox{ for } 0 \le c \le 1/(2Q)  , \mbox {with  equality only for } c =  0. \nonumber 
\end{eqnarray}}

\begin{proof}
We  consider $c$ as a function of $a:$ Let $C=C(a, s) = a/Q.$   Then  $0 \le C(a, s) \le 1/(2Q)$ iff
$0 \le a  \le 1/2. $  The claim  of the lemma is equivalent to:
\begin{eqnarray}
\mbox{ For } \, \,\, s\ge 4.2\,\, \, \, \, \Gamma(a, C, s) \le 2 \mbox{ for all }  0 \le a \le 1/2 ,\,   \mbox {with  equality only for } a =  0. \label{zielcbei0} 
\end{eqnarray}
We show  that $ \Gamma(a, C, s) $  is strictly decreasing in $0<a\le 1/2.$ 
As $C(0, s)=0$ and $\Gamma(0, 0, s)=2$ by Lemma \ref{lebagamma}(a) we get 
(\ref{zielcbei0}.) 

\begin{eqnarray}
\frac{d}{da} \ln \Gamma(a, C, s) \,= \,\frac {\frac{s (\exp(sa)-1)}{\exp(s)-s-1}}{\Gamma_1(a, s)}\, - \,Q\cdot \frac{2a/Q}{1+ aC}\,
+ \,Q \cdot \frac{1}{Q} \frac{\mbox{PMINUS}(C, Q-1)}{\mbox{PPLUS}(C, Q)} >=< 0 \nonumber  \\
\Longleftrightarrow 
\frac{ \frac{\exp(sa)-1}{\exp(s)-1}}{\Gamma_1(a, s)} \,- \, \frac{2a/Q}{1+ aC}\,
+ \,\frac{1}{Q} \frac{\mbox{PMINUS}(C, Q-1)}{\mbox{PPLUS}(C, Q)} >=< 0  \mbox{ (Division with  } Q, \, C=a/Q \mbox{)} \label{cbei0ablsum}. \\
\mbox{With } \,\,  a=0 \, \, \mbox{ derivative is }  0. \nonumber 
\end{eqnarray}

We split the left-hand-side of (\ref{cbei0ablsum}) into two additive terms.  
The following two inequalities directly imply that
 $\Gamma(a, C, s)$ is  decreasing. 
\begin{eqnarray} 
\frac{ \frac{\exp(sa)-1}{\exp(s)-1}}{ \Gamma_1(a, s) } \,- \, \frac{a/Q}{1+ aC}\,< \,0\,  \label{cbei0abl1} \\
- \, \frac{a/Q}{1+ aC}\,
+ \,\frac{1}{Q} \frac{\mbox{PMINUS}(C, Q-1)}{\mbox{PPLUS}(C, Q)} < 0 \label{cbei0abl2} \,  
\end{eqnarray} 

\noindent
{\em Proof of (\ref{cbei0abl2}) for  $0 < a <1 $\, and $ s \ge 2. $ }
With the calculation leading to (\ref{baPPPM1}) we have:
\begin{eqnarray}
- \, \frac{a/Q}{1+ aC}\,
+ \,\frac{1}{Q} \frac{\mbox{PMINUS}(C, Q-1)}{\mbox{PPLUS}(C, Q)} >=< 0 \nonumber   \\ 
\Longleftrightarrow 
\mbox{PMINUS}(C, Q-1) \, >=< \, 
\, \, a \cdot \mbox{PPLUS} (C, Q-1)  \nonumber \\ 
\Longleftrightarrow
\left( \frac{1+C}{1-C} \right)^{Q-1} \,= \, \left( \frac{1+a/Q}{1-a/Q} \right)^{Q-1}       \,>=< \, \frac{1+a}{1-a} \nonumber \\
\Longleftrightarrow 
(Q-1)\left(\frac{a}{Q} + \frac{1}{3}\left(\frac{a}{Q}\right)^3\,+ \, \frac{1}{5}\left(\frac{a}{Q}\right)^5\,+ \, \cdots  \right)\,>=< \, a + \frac{a^3}{3}+ \frac{a^5}{5} + \cdots \nonumber 
\\ \mbox{ (Logarithm series , } a < 1, Q \ge Q(2) > 2 \mbox{)}  \nonumber
\end{eqnarray}
Here, the  $<$-case of the last inequality  holds because $(Q-1)(1/k)(a/Q)^k< a^k/k$ as $Q\ge Q(2)> 2. $ Inequality (\ref{cbei0abl2}) is proved. \\

\noindent
{\em Proof of (\ref{cbei0abl1}) for $0<a\le 1/2$ and $s \ge 4.2.$}  
We prove further below: 
\begin{eqnarray} 
 a\frac{\exp(sa)-1}{\exp(s)-1}\, \le \, \frac{\exp(sa)-sa-1}{\exp(s)-s-1} \mbox{ for } 0 \le a \le 1, \mbox{ equality only for }\, a=1,  a=0. \label{baLK2}
\end{eqnarray}

\begin{eqnarray}
\mbox{ Abbreviating } K=K(a, s) = \frac{\exp(sa)-1}{\exp(s)-1} \mbox{ and } L=L(a, s) \,= \, \frac{\exp(sa)-sa-1}{\exp(s)-s-1} \, \,\nonumber \\  \mbox{ we have } \Gamma_1(a, s)= 1+L.  \mbox{ Then (\ref{cbei0abl1}) becomes } \nonumber \\
\frac{K}{1+L} \,< \, \frac{a/Q}{1+ aC}  \nonumber \\
\Longleftrightarrow K \cdot\left(1+ a \cdot \frac{a}{Q}\right) \, <\, \frac{a}{Q}\cdot( 1 + L) \, \, \, 
\mbox{ ( } C=a/Q \mbox{)} \nonumber  \\
\Longleftrightarrow K \,< \, \frac{a}{Q} \cdot(1+ L - a\cdot  K) \label{cbei0abl11}
\end{eqnarray} 
With (\ref{baLK2}) inequality (\ref{cbei0abl11}) follows from 
\begin{eqnarray}
K< \frac{a}{Q}  
\Longleftrightarrow \frac{s(\exp(sa)-1)}{\exp(s)-s-1}< a.  \label{cbei0abl12} 
\end{eqnarray}

To show the right-hand-side of (\ref{cbei0abl12}) for $0 < a \le 1/2$   
it is sufficient to show it for $a=1/2$. This as
both sides of  the inequality are $0$ for $a=0$ and its  left-hand-side is convex in $a.$  
We fix $a=1/2$ from now on.  The right-hand-side of  (\ref{cbei0abl12}) is equivalent to 
\begin{eqnarray}
s \exp(sa)-a \exp(s)< s-a(s+1) = (1-a)s-a   \label{cbei0abl13} .\\
\mbox{ With }\,  s=4.2 \,\mbox{(and }a=1/2 \mbox{) we get }\, 0.95\dots < 1.6. \nonumber
\end{eqnarray}
To get (\ref{cbei0abl13}) for $ s \ge 4.2$ we observe that  its right-hand-side  is increasing in $s.$ 
We show that the left-hand-side is decreasing in $s.$ 
\begin{eqnarray}
\frac{d}{ds}\left( s \exp(sa)-a \exp(s) \right) \,= \,sa \exp(sa) + \exp(sa)- a\exp(s)<0 \nonumber \\
\Longleftrightarrow  1+sa < a \exp(s(1-a)) \label{cbei0abl14}   \\
 \mbox{ With } s=4.2 \mbox{ we get }    3.1<4.8. \nonumber
 \end{eqnarray}
As the derivative with respect to $s$ of the right-hand-side of (\ref{cbei0abl14}) is  $>1/2$ for $s= 4.2$  and increasing, inequality (\ref{cbei0abl14})  holds for all $s \ge 4.2$ finishing the argument.   \\

\noindent
{\it Proof of (\ref{baLK2}).}
For $a=0, a=1$ the claim holds. For $0<a<1$ it is equivalent to 
\begin{eqnarray}
a < \frac{\exp(s)-1}{\exp(s)-s-1}\left(1 \,- \, \frac{sa}{\exp(sa)-1}\right) \nonumber 
\end{eqnarray}
This inequality holds because its right-hand-side is concave in $a.$  \\
\hfill \hspace*{3cm}          \end{proof}

\subsection{Proof of Lemma \ref{cbei1}}

\noindent
{\bf Lemma \ref{cbei1} (repeated) }{ \em
Let  $s \ge 6$ and $A= A(c, s) = 85/(100Q)  \cdot c + 1- 85/(100Q ). $  Then 
\begin{eqnarray} \Gamma(A, c, s) \le 2  \mbox{ for } 2/5 \le c \le 1   , \mbox {with  equality only for } c =  1. \nonumber 
\end{eqnarray}}

\begin{proof}  We show that 
 $\Gamma(A, c ,s ) $  is strictly increasing for  $ 2/5  \le c <1.$ 
As $A(1, s)=1$ and $\Gamma(1, 1, s)=2$ by \ref{lebagamma}(a) this implies the  lemma.  

\begin{eqnarray} 
\mbox{ We abbreviate  } A' = \frac{d}{dc} A(c, s) \,= \, \frac{85}{100Q}. \nonumber \\
\frac{d}{dc} \ln \Gamma(A, c, s)= \frac{\frac{s(\exp(A\cdot s)-1)}{\exp(s)-s-1}\cdot A'}{ \Gamma_1(A, s) } - Q \cdot \frac{A'c+A}{1+Ac}  + 
 Q \cdot \frac{ \mbox{PMINUS}(c, Q-1 )}{ \mbox{ PPLUS}(c, Q) } \, > = <0  \, \nonumber \\
\Longleftrightarrow 
\frac{\frac{\exp(As)-1}{\exp(s)-1}\cdot A'}{\Gamma_1(A, s)}\, - \,  \frac{A'c+A}{1+Ac}  \,+ \,\frac{ \mbox{PMINUS}(c, Q-1 )}{ \mbox{ PPLUS}(c, Q) } \, > = <0 \,(\mbox{Division with } Q. ) \label{cbei1abl} \\
\mbox{Observe that  } c=1 \mbox{ in  (\ref{cbei1abl})  gives } \frac{A'}{2}- \frac{A'+1}{2} + \frac{1}{2}\,= \,0. \nonumber
\end{eqnarray}

The following inequalites   imply that $\Gamma(A, c, s)$ is increasing. 
\begin{eqnarray}
\frac{ \frac{\exp(sA)-1}{\exp(s)-1}\cdot A'}{ \Gamma_1(A, s)} -  \frac{cA'}{1+Ac}  > 0 \label{cbei1abl1} 
\\ 
- \frac{A}{1+Ac}  +   \frac{ \mbox{ PMINUS }(c, Q-1 )} {  \mbox{ PPLUS}(c, Q) }>  0 . \label{cbei1abl2}.
\end{eqnarray}

\noindent 
{\it Proof of (\ref{cbei1abl2}) for $\,s\ge 4\,$
 and  $\,2/5 \le c <1.$}  
We show further below:
\begin{eqnarray}
\mbox{ Given } \, s\ge 2 \,\, \, \, \,\frac { \mbox{ PMINUS}(c, Q-1 ) }{\mbox{PPLUS}(c, Q-1)}  
\mbox{ is concave for } 0<c<1 \label{baPPPM4}
\end{eqnarray}

By (\ref{baPPPM1}) inequality (\ref{cbei1abl2}) is equivalent to:  
\begin{eqnarray}
\frac { \mbox{ PMINUS}(c, Q-1 ) }{\mbox{PPLUS}(c, Q-1)} >  A(c, s). \label{cbei1abl21} \\
\mbox{ For } c=1 \mbox{   both sides of  (\ref{cbei1abl21}) are } \, 1 \label{cbei1abl22}.
\end{eqnarray} 
We show that (\ref{cbei1abl21}) holds for $c=2/5$ and $s \ge 4.0.$  Then, by (\ref{baPPPM4}), it holds for all $2/5\le c<1$ and we have  (\ref{cbei1abl2}.)   
\begin{eqnarray} 
 \frac{\mbox{ PMINUS } (2/5, Q-1 ) } {\mbox{ PPLUS }(2/5, Q-1)} \, =  \,\frac{7^{Q-1}-3^{Q-1}}{7^{Q-1}+3^{Q-1}}\,= \,1- 2 \frac{3^{Q-1}}{7^{Q-1}+3^{Q-1}} \mbox{ and }
 A(2/5, s)\,= \,1-\frac{51}{100Q}. \nonumber \\
\mbox{ Then }\frac{\mbox{ PMINUS } (2/5, Q-1 ) } {\mbox{ PPLUS }(2/5, Q-1)} \,> \,(A/2/5, s) \, \Longleftrightarrow\,
 \frac{100Q}{51} \,< \,\frac{1}{2}\left(\left(\frac{7}{3} \right)^{Q-1}+1\right) \label{cbei1abl23}.\\
  \mbox{With } Q=4.3 \mbox{ the right-hand-side of  (\ref{cbei1abl23}) becomes } \, \, 8.43 \dots \,< \,8.69 \dots \nonumber.
\end{eqnarray}
Considering the derivative shows that (\ref{cbei1abl23}) holds for all $Q\ge 4.3.$ 
$Q=Q(s)$ is increasing in $s$ (cf. (\ref{baQ1}).)  We have $Q(4) \,= \, 4.32$  and inequality (\ref{cbei1abl23}) holds for $s\ge 4.$\\
 
\noindent
{\it Proof of (\ref{baPPPM4}.)} The derivative of
the fraction is calculated in (\ref{baPPPM3}). Considering the numerator and denominator 
separately one sees that it is decreasing in $c.$ \\

\noindent
{\it Proof of (\ref{cbei1abl1}) for $\, s \ge 6\,$ and $ \, , \,  0 < c <1.$ }
We abbreviate 
\begin{eqnarray} 
K=K(A, s) = \frac{\exp(As)-1)}{\exp(s)-1} \, \, \,\mbox{ and } L=L(A, s) =\frac { \exp(As)-As-1}{\exp(s)-s-1}   \nonumber 
\end{eqnarray}
We have $\Gamma_1(A, s)= 1 +L.$ We divide with $A'$  and  (\ref{cbei1abl1}) is equivalent to
\begin{eqnarray}
\frac{ K }{ \Gamma_1(A, s) } \,= \, \frac{K}{1 + L} > \frac{c}{1+A \cdot c} \Longleftrightarrow K(1+Ac)-Lc > c. \label{cbei1abl11}
\end{eqnarray} 
As $A<1$ we have  $K > L $ by (\ref{baLK1}) and  
(\ref{cbei1abl11})  follows from 
\begin{eqnarray}
 K(1+Ac)- Kc \,= \,K(1+Ac-c)\, > \, c \label{cbei1abl12}.
\end{eqnarray} 
As $( A - 1 )c \ge  A-1 $ ( by $  A  \le 1, c \le 1$)  inequality (\ref{cbei1abl12}) follows from 
\begin{eqnarray}
   K \cdot ( 1+ A-1 ) \,= \, K \cdot A \ge c    \label{cbei1abl13}.
\end{eqnarray} 

We need to show (\ref{cbei1abl13}) for $0 < c < 1.$ This becomes easier when we consider
$c$ as a function of $a.$ 
The inverse function of $A(c, s)$  is
\begin{eqnarray}   
C = C(a, s) = 100Q/85  \cdot a + 1 - 100Q/85. \nonumber \\ 
\mbox{Then (\ref{cbei1abl13}) for }\,\, 0<c<1 \, \,  \mbox{ is equivalent to } 
\frac{\exp(sa)-1}{\exp(s)-1} a \ge C(a, s) \, \mbox{ for } \, 1 - 85/(100Q) < a < 1 \nonumber \\
\mbox{ We show }(\exp(as)-1) \cdot a \ge  C(a, s) \cdot (\exp(s)-1).  \label{cbei1abl14}\\
\mbox{ For } a=1 \mbox{ both sides of (\ref{cbei1abl14}) are equal to } \exp(s)-1. \nonumber 
\end{eqnarray}
The left-hand-side of (\ref{cbei1abl14}) is convex. Therefore (\ref{cbei1abl14}) follows from

\begin{eqnarray}
\left( \frac{d}{da} (\exp(as)-1) \cdot a \right)_{|a=1} \, <\, \left(\frac{d}{da}C(a, s) \cdot (\exp(s)-1)\right)_{|a=1}. \label{cbei1abl15} \\
\frac{d}{da} (\exp(as)-1) \cdot a    \,= \, sa\exp(as)+ \exp(as)-1\,= \, (1+sa)\exp(as)-1. \nonumber \\
\frac{d}{da}C(a, s) \cdot (\exp(s)-1)\,= \,   \frac{100Q(\exp(s)-1)}{85}\,= \, \frac{100s(\exp(s)-1)^2}{85(\exp(s)-s-1)} \nonumber. \\
\mbox{Inequality (\ref{cbei1abl15}) follows from } (1+s)\exp(s)-1 \,< \, \frac{100s(\exp(s)-1)}{85} \nonumber \\
\Longleftrightarrow \frac{100s}{85} -1 < \exp(s) \left(\frac{15}{85}s-1\right). \label{cbei1abl16}
\end{eqnarray}
With $s=6$ inequality (\ref{cbei1abl16}) becomes $6.0 \dots \,< \, 23.7 \dots$ and 
then it holds for all $s\ge 6.$ 

\end{proof}